%% file: main.tex
\documentclass[aps,prb,reprint,showpacs,superscriptaddress,longbibliography]{revtex4-2}
\usepackage[english]{babel}
\usepackage{amsmath,amssymb,bbm,graphicx,color,comment,txfonts}
\usepackage[bookmarks=true,colorlinks,citecolor=blue,urlcolor=blue]{hyperref}
\usepackage{dsfont}
\usepackage[normalem]{ulem}
\usepackage{physics}

\usepackage{physics}
\usepackage{tikz}
\usepackage{mathdots}
\usepackage{yhmath}
\usepackage{cancel}
\usepackage{color}
\usepackage{siunitx}
\usepackage{array}
\usepackage{multirow}
\usepackage{amssymb}
\usepackage{gensymb}
\usepackage{tabularx}
\usepackage{extarrows}
\usepackage{booktabs}
\usetikzlibrary{fadings}
\usetikzlibrary{patterns}
\usetikzlibrary{shadows.blur}
\usetikzlibrary{shapes}
\usepackage{xurl}
\hypersetup{breaklinks=true}

\renewcommand{\Re}{\mathop{\mathrm{Re}}}
\renewcommand{\Im}{\mathop{\mathrm{Im}}}

\begin{document}

\date{\today}

\title{Monitored interacting Dirac fermions}
\author{T. M\"uller}
\affiliation{Institut f\"ur Theoretische Physik, Universit\"at zu K\"oln, D-50937 Cologne, Germany}
\affiliation{The Abdus Salam International Centre for Theoretical Physics (ICTP), Strada Costiera 11, 34151 Trieste, Italy}
\author{M. Buchhold}
\affiliation{Institut f\"ur Theoretische Physik, Universit\"at zu K\"oln, D-50937 Cologne, Germany}
\author{S. Diehl}
\affiliation{Institut f\"ur Theoretische Physik, Universit\"at zu K\"oln, D-50937 Cologne, Germany}

\begin{abstract}
We analytically study interacting Dirac fermions, described by the Thirring model, under weak local particle number measurements with monitoring rate $\gamma$. This system maps to a bosonic replica field theory, analyzed via the renormalization group. For a nonzero attractive interaction, a phase transition occurs at a critical measurement strength $\gamma_c$. When $\gamma>\gamma_c$, the system enters a localized phase characterized by exponentially decaying density-density correlations beyond a finite correlation length; for $\gamma<\gamma_c$, the correlations decay algebraically. The transition is of BKT-type, reflected by a characteristic scaling of the correlation length. In the non-interacting limit, $\gamma_c\to0$
 shifts to zero, reducing the algebraic phase to a single point in parameter space. This identifies weak measurements in the free case as an implicit double fine-tuning to the critical endpoint of the BKT phase transition. Along the non-interacting line, we compute the entanglement entropy from density-density correlation functions and find no entanglement transition at nonzero measurement strength in the thermodynamic limit.
\end{abstract}

\maketitle
\section{Introduction}
%
%
Quantum devices promise unique capabilities for many-body quantum dynamics, including the precise control of engineered unitary and non-unitary dynamics and the ability to perform mid-circuit measurements~\cite{Norcia_2023, Graham_2023, Koh_2023, DeCross_2023}. Such monitored systems can be conceptualized as driven open quantum systems, where measurements can be viewed as an effective environment. While under repeated measurements, the ensemble of monitored wave functions reaches a featureless infinite-temperature state~\cite{Wolff_2019, Wolff_2020, Poletti_2012, Pichler_2010}, recording measurement readouts for each experimental run provides access to quantum trajectories, revealing additional structure~\cite{Daley_2014}. Analyzing these trajectories using  theoretical tools from the realm of disordered quantum systems~\cite{Edwards_1975, Emery_1975, Grinstein_1976} uncovers nontrivial behavior beyond the infinite-temperature state~\cite{Buchhold_2021, Poboiko_2023, Poboiko_2024,Poboiko_2025, Bao_2020, Jian_2020, Skinner_2019,Fava_2023,Fava_2024,Guo_2024} 

A key diagnostic of such behavior is entanglement entropy~\cite{Skinner_2019, Li_2018, Li_2019,Choi_2020,Zabalo_2020,Agrawal_2022,Zabalo_2021}, which distinguishes phases in the stationary state of a monitored system. Local measurements disentangle the quantum state, while unitary dynamics entangle qubits. This competition may trigger a measurement-induced phase transition (MIPT), extensively studied in various models~\cite{Skinner_2019, Li_2018, Li_2019, Gullans_2020, Choi_2020, Jian_2020, Fan_2021, Li_2021, Nahum_2021, Fuji_2020, Jian_2021, Doggen_2022,Ippoliti_2021_duality,Ippoliti_2022,Biella_2021,Block_2022,Behrends_2024,Kelly_2025}. In these setups frequent measurements lead to the localization of quantum information, favoring an area-law phase, while rare measurements preserve an extensive growth of the entanglement entropy. This discovery sparked significant interest in exploring the possible phases and phase transitions in monitored quantum systems. For instance, models in higher dimensions~\cite{Sierant_2022,Turkeshi_2020,Lavasani_2021_2D,Lunt_2021}, have been studied, as well as the interplay of non-commuting measurements~\cite{Ippoliti_2021,Klocke_2023}. 

We focus on monitored quantum systems with global conservation laws~\cite{Cao_2019, Alberton_2021, Buchhold_2021, Poboiko_2023}, specifically fermions with particle number conservation~\cite{Chen_2020,Bernard_2019,Loio_2023,Soares_2024,Turkeshi_2024, Szyniszewski_2023, Jian_2022,Mueller_2022,Ladewig_2022,Barratt_2022,Agrawal_2022,Kells_2023}. Conservation laws translate to symmetries in the underlying dynamics and the effective action describing it. To explore the universal properties of monitored interacting fermions in one dimension, we adopt a Keldysh path integral approach~\cite{Buchhold_2021, Poboiko_2023}, combining the Keldysh formalism for open quantum systems~\cite{Sieberer_2016, Sieberer_2023} with the replica trick~\cite{Buchhold_2021, Poboiko_2023, Poboiko_2024,Poboiko_2025,Fava_2023,Fava_2024,Guo_2024,Starchl_2024} and renormalization group techniques.

\begin{figure}
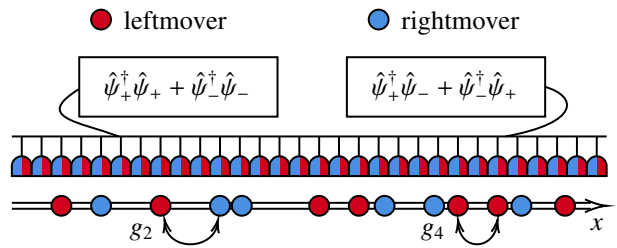

    \centering
    \include{Modelsketch}
    \caption{Sketch of the microscopic model: We consider chiral fermions moving continuously in one dimension, with a fixed directionality quantum number $d=\pm$. They are interacting locally and we monitor the two local Hermitian operators that are compatible with the conservation of total particle number and inversion symmetry. This is done continuously in both space and time.}
    \label{fig:modelsketch}
\end{figure}

We demonstrate this framework using a paradigmatic model of one-dimensional quantum matter in the spatial continuum: monitored interacting Dirac fermions (Fig.\ref{fig:modelsketch}), equivalently described as a monitored Luttinger liquid~\cite{Buchhold_2021}. This continuum model is particularly well suited to studying universal phenomena in measurement-induced dynamics due to its simplicity and potential to exhibit a quantum phase transition. It serves as a valuable foundation for understanding the mechanisms driving universal behavior in monitored quantum systems. \par
In lattice fermion systems~\cite{Cao_2019, Alberton_2021, Buchhold_2022, Mueller_2022, Minato_2022, Lumia_2024, Ladewig_2022, Szyniszewski_2023, Guo_2024}, measurements have been proposed to implement a U$(1)$ symmetry conserving the total number of fermions~\cite{Poboiko_2025, Poboiko_2024, Poboiko_2023, Soares_2024, Loio_2023, Zhang_2021, Fava_2023, Chatterjee_2024, Fava_2024, Xing_2024}. In contrast, in the continuum model of Dirac fermions that we propose, a U$(1)\times$U$(1)$ symmetry of left- and right-moving fermions is present for weak measurements, which is spontaneously broken for strong measurements. In this work, we analyze the monitored Dirac fermion model in detail. While a computation of the entanglement entropy in the interacting case remains an open problem, we characterize the non-trivial correlations in this model by means of the connected density-density correlation function, and provide results for the entanglement entropy in the absence of interactions. We then briefly discuss similarities and differences with lattice fermion models~\cite{Poboiko_2023, Poboiko_2024, Starchl_2024, Fava_2023, Poboiko_2025, Guo_2024, Fava_2024}. This comparison sheds light on the implications of symmetry and measurement dynamics in these systems.

\subsection{Synopsis}

We discuss the general framework for determining observables in continuous measurement protocols. We begin by introducing the rigorous replica quantum master equation, which closely resembles a Lindblad equation for Markovian open quantum systems (Sec.~\ref{sec:ReplicaMaster}), and which enables the analysis of interacting Dirac fermions under weak monitoring. Next, we define the Hamiltonian and measurement operators, guided by the underlying symmetries of the system (Sec.~\ref{sec:InteractingDirac}). Utilizing operator identities tied to the Dirac nature of the fermions, we bosonize the model and the relevant observables on the trajectory level (Sec.~\ref{sec:Observables}). Our primary focus is on the nonlinear two-point density-density correlation function, which captures correlations within the ensemble of quantum trajectories and serves as a sensitive probe of the system's monitored dynamics~\cite{Poboiko_2023,Poboiko_2024,Poboiko_2025,Buchhold_2021,Mueller_2022}. \par
In Sec.~\ref{sec:FeynmanConstruction}, we map the model onto a bosonic replica Keldysh path integral. The resulting action consists of a Gaussian term -- explicitly dependent on the measurement strength -- and a sine-Gordon nonlinearity controlled by the same parameter. A specific center-of-mass mode, which captures the system's heating to an infinite-temperature state, can be integrated out without affecting the nontrivial observables (Sec.~\ref{sec:FreeBosons}). This leads to an effective action for the relative modes in replica space, which decouples the two Keldysh contours. On each contour, the resulting model is a complex sine-Gordon theory. In the absence of nonlinearity, the path integral can be evaluated explicitly, revealing that the connected nonlinear density-density correlation decays algebraically, indicative of a critical phase (Sec.~\ref{sec:Fieldtheoryobservables}). \par
The stability of that critical phase is determined by the relevance of the non-linearity, which is studied using a second order perturbative RG calculation (Sec.~\ref{Sec:Nonlinearity}). A complex rescaling of space and time is needed at each step of the RG in order to preserve the exact symmetry of Hermiticity of the replicated density operator. \par
\begin{figure}
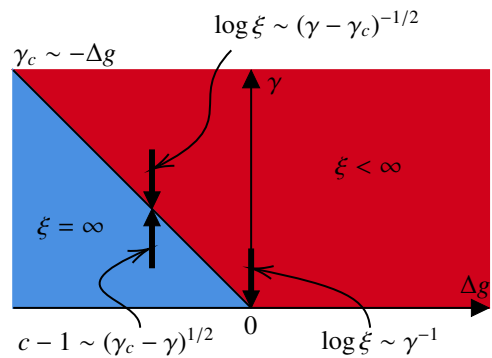

    \centering
    \include{Phasediagram}
    \caption{Schematic phase diagram of the monitored interacting Dirac fermion model obtained by a second order perturbative RG calculation. We find that attractive interactions $\Delta g < 0$, i.e. a Luttinger parameter $g=1+\Delta g<1$, stabilize a strongly correlated phase with continuously varying prefactor $c$ of the algebraically decaying correlations. Measurements localize the particles resulting in a finite correlation length $\xi$. The transition is of BKT nature. In the fine-tuned case of vanishing interactions $\Delta g=0$, we approach a critical endpoint of the phase transition line under $\gamma \to 0$ resulting in a correlation length divergence in agreement with the weak localization scenario.}
    \label{fig:PhaseDiagram}
\end{figure}
This analysis leads to the well-known flow equations of the sine-Gordon model (Sec.~\ref{sec:DiscussionOfAnalyticalApproach}), from which we derive the phase diagram shown in Fig.~\ref{fig:PhaseDiagram}. For attractive interactions, the system stabilizes a critical, Luttinger liquid-type phase characterized by an infinite correlation length and algebraic correlations for nonzero measurement rates $\gamma < \gamma_c$, where $\gamma_c$ depends on the interaction strength. At $\gamma=\gamma_c$, a Berezinskii-Kosterlitz-Thouless (BKT)~\cite{Berezinskii_1971, Berezinskii_1972, Kosterlitz_1973} phase transition occurs, leading to a localized, area law phase. In this regime, the sine-Gordon nonlinearity becomes RG relevant, resulting in a finite correlation length and exponentially decaying correlations. The correlation length diverges at the critical point following the BKT picture as 
\begin{align}
\log \xi \sim 1/\sqrt{\gamma-\gamma_c} \quad \text{ (interacting case)},
\end{align}
with nonvanishing $\gamma_c\neq 0$.

One central result of our work is that for vanishing interactions, the critical point shifts to $\gamma_c=0$. In this case, we can make a direct connection to the fermionic entanglement entropy and find that the system obeys area-law scaling for any non-zero measurement strength $\gamma>0$. The correlation length beyond which the area-law becomes visible now diverges exponentially as $\gamma \to 0$,
\begin{align}\label{eq:noint}
    \log \xi \sim 1/\gamma \quad \text{ (non-interacting case)}.
\end{align}
Interestingly, not only the critical point shifts to $\gamma_c = 0$, but we also find a \textit{modified}  scaling of the correlation length. This is due to the following finding: In the BKT  formula for the universal divergence of the correlation length as a function of the distance $\delta$ from the critical line, $\log \xi \sim 1/\sqrt{\delta}$, we find the form $\delta \sim (1-g)(\gamma - \gamma_c) +\mathcal{O} (\gamma^2)$, with Luttinger parameter $g$, 
characterizing the interactions. At the non-interacting point $g=1$, the next order in $\gamma$ becomes dominant, $\delta\sim \gamma^2$, which results in the scaling specified in Eq.~\eqref{eq:noint}. Phenomenologically, this behavior is also characteristic of weak localization~\cite{Poboiko_2023,Poboiko_2024} and indicates different universality classes for interacting and free Dirac fermions under monitoring.
\section{Model and theoretical framework} \label{sec:ModelandFramework}
\subsection{Replica master equation for weak measurements} \label{sec:ReplicaMaster}
We begin by introducing the continuous measurement protocol and discussing the structure of observables in the resulting ensemble of quantum trajectories. This discussion follows previous work~\cite{Nahum_2017,Fava_2023,Buchhold_2021,Buchhold_2022, Jian_2020} and sets the stage for the remainder of our work. Suppose that we continuously monitor the Hermitian operator $\hat{O}$ at a rate $\gamma$. In a single recording time step $\delta t$, the state of the system is updated to $\ket{\psi_t} \to \hat{P}(J_t) \ket{\psi_t}/\vert \vert \hat{P}(J_t) \ket{\psi_t} \vert \vert$ according to the generalized projector~\cite{Caves_1987,Gisin_1992,Wiseman_1993,deVega_2017,Strunz_1998,Percival_1998}
\begin{equation} 
\hat{P}(J_t) = \left( \frac{2\gamma \delta t}{\pi} \right)^{1/4} e^{-\gamma \delta t (\hat{O}-J_t)^2}.  \label{eq:genproj}
\end{equation}
Here, $J_t$ represents the measurement stream, a continuous random variable corresponding to the measurement outcomes. The probability for each outcome is given by the Born rule, $\mathcal{P}(J_t)=\bra{\psi_t} \hat{P}^\dagger(J_t) \hat{P}(J_t) \ket{\psi_t}$. In the limit of long times, $\gamma \delta t \to \infty$, this represents a standard projective measurement, while at short times $\gamma \delta t \to 0$, it leaves the state unchanged, $\hat{P}(J_t)\to \mathds{1}$, smoothly interpolating between the two limits. In addition to the continuous monitoring, a unitary evolution with a Hermitian Hamiltonian $\hat{H}$ is implemented by alternating measurement steps with the application of $e^{-i\hat{H} \delta t}$~\footnote{We set $\hbar=1$ throughout this work.}. This generates a stochastic evolution of a pure quantum state $\ket{\psi_t}$, i.e., a quantum trajectory. \par
The resulting evolution of $\ket{\psi_t}$ is stochastic due to the randomness in each measurement outcome. We introduce the notation $\overline{(\dots)}$ for the average over all possible measurement outcomes, i.e., all possible trajectories. Under a general monitoring scheme, the ensemble average $\hat{\rho}_t\equiv \overline{\ket{\psi_t}\bra{\psi_t}}$ over all trajectories reaches a featureless stationary state $\hat \rho_t \propto \mathds{1}$~\cite{Wolff_2019,Wolff_2020,Poletti_2012,Pichler_2010}. In order to capture the features of individual wave functions beyond that, one therefore needs to consider higher moments of the trajectory distribution. Such information is encoded in a replicated density matrix  $\hat{\rho}_{M,t}=\overline{\bigotimes_{r=1}^M \ket{\psi_t}\bra{\psi_t}}$. However,  for $M>1$  this object does not obey a closed equation due to the necessary normalization after each measurement. This obstacle is solved by applying a replica trick~\cite{Buchhold_2021,Poboiko_2023,Sieberer_2016,Buchhold_2021,Poboiko_2023,Sieberer_2023,Starchl_2024} (For details, see App.~\ref{App:replica}): One reformulates the evolution in terms of the un-normalized states $\ket{\tilde \psi_t}$ which evolve under the unitary evolution and projections $\ket{\tilde \psi_t} \to \hat{P}(J_t) \ket{\tilde \psi_t}$ such that for each realization of the measurement results $J_t$, $\ket{\psi_t}=\ket{\tilde \psi_t}/\vert \vert \ket{\tilde \psi_t} \vert \vert$, but $\ket{\tilde 
\psi_t}$ undergoes a linear evolution. Based on this, we define for all $R \in \mathbb{N}_+$ the un-normalized replica density operator
\begin{equation}
    \tilde \rho_{R,t} = \overline{\bigotimes_{r=1}^R \ket{\tilde \psi_t} \bra{\tilde \psi_t}}.
\end{equation}
This object is useful for two reasons: \par 
(i) It evolves according to a closed linear equation, a generalized replica quantum master equation~\cite{Chahine_2023,Poboiko_2023,Poboiko_2024} (see App.~\ref{App:replica}). If $\hat{O}^2=\hat{O}$ is a projector, it takes the simple form
\begin{multline}
    \partial_t \tilde{\rho}_{R,t} = -i [ \tilde{H}_R, \tilde{\rho}_{R,t}] - \frac{\gamma}{2} [ \tilde O_R ,[ \tilde O_R , \tilde{\rho}_{R,t} ]] \\  -\gamma \lbrace  \tilde O_R(1- \tilde O_R)  , \tilde{\rho}_{R,t}  \rbrace. \label{eq:LindbladGeneral}
\end{multline}
For a compact notation, we introduced replica averaged operators, i.e. $ \tilde O_R = \frac{1}{R} \sum_r \hat{O}^{(r)}$ and operators with upper index $(r)$ only act non-trivially on replica $r$. When multiple commuting operators are measured simultaneously, the corresponding terms can be simply added~\footnote{Non-commuting operators cannot be measured simultaneously. Therefore, it is important to specify the temporal order in which the respective generalized projections are applied at each time-step, and corresponding terms cannot simply be added.}. The anti-commutator term implies that the infinite temperature state is not a stationary solution for $R>1$, which results in nontrivial correlations of non-linear observables. This generalized Lindblad equation explicitly breaks trace preservation $\partial_t \Tr \tilde \rho_{R,t} = -2\gamma \Tr \tilde O_R(1-\tilde O_R) \tilde \rho_R$ due to the measurement that couples the different replicas. This reflects the missing normalization of the states. The additional anti-commutator term vanishes for $R = 1$, as $\tilde O_1^2=\tilde O_1$ is a projector, resulting in a trace preserving dynamics. In contrast, the replica density operator remains Hermitian under the evolution for arbitrary $R$, $\partial_t (\tilde \rho_{R,t}-\tilde \rho_{R,t}^\dagger)=0$ which relies on real $\gamma$ and Hermitian $\tilde O_R$.  \par
(ii) In the proper replica limit $R \to 1$, it yields the normalized $M$ replica density operator according to (see App.~\ref{App:replica})
\begin{align}
    \hat{\rho}_{M,t} &= \overline{\left(\Tr \ket{\tilde \psi_t} \bra{\tilde \psi_t}\right)^{1-M} \otimes_{r=1}^M \ket{\tilde \psi_t} \bra{\tilde \psi}} \\
    &= \lim_{R \to 1} \overline{\left(\Tr \ket{\tilde \psi_t} \bra{\tilde \psi_t}\right)^{R-M} \otimes_{r=1}^M \ket{\tilde \psi_t} \bra{\tilde \psi}} \\
    &= \lim_{R \to 1} \Tr_{r>M} \tilde \rho_{R,t},
\end{align}
required to compute $M$ replica observables. In the exponent $1-M$, the factor $-M$ provides normalization of the state, and the factor of unity weighs the trajectories according to Born's rule. This leads to the proper replica limit $R\to 1$ to be taken, as opposed to typical disorder problems, where one considers $R\to 0$. $\Tr_{r>M}$ denotes the trace over all replicas with a replica index $r$ larger than $M$. For instance, $\overline{\langle \hat{A} \rangle_t\langle \hat{B} \rangle_t}=\Tr \hat{A}^{(1)}\hat{B}^{(2)} \hat{\rho}_{2,t}$, where $\hat{A}$ and $\hat{B}$ are usual quantum mechanical observables, is a $M=2$ replica observable. For $M>1$, we interpret this as follows: First, we need to compute observables on the level of the un-normalized $R$ replica density operator for general $R \geq M$ and as a last step, we analytically continue the result to $R \in \mathbb{R}$ and take the limit $R \to 1$. While the replica limit procedure is not mathematically rigorous~\footnote{An alternative approach involves the introduction of auxiliary super-symmetric degrees of freedom to account for the normalization~\cite{Sedrakyan_2020,Efetov_1996,Verbaarschot_1985,Zirnbauer_2004}, which is subject of current research.}, it has proven useful in the physics of disordered systems~\cite{Edwards_1975,Emery_1975,Grinstein_1976}.
\subsection{Monitored interacting Dirac fermions} \label{sec:InteractingDirac}
\begin{figure}
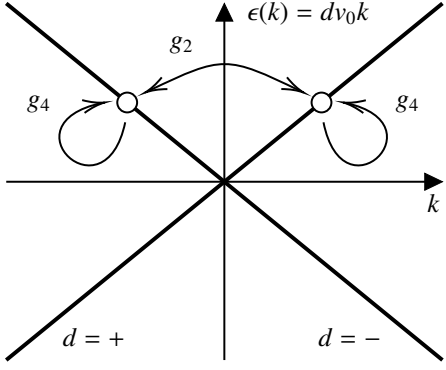

    \centering
\include{Momentumsketch}
    \caption{Sketch of the Hamiltonian processes in the interacting Dirac model. Both right ($d=+$) and left ($d=-$) moving particles have an unbound linear single particle dispersion relation $\epsilon(k)$. Interactions are local in space and consist of forward scattering $\sim g_4$ between fermions of the same flavor $d$ and backscattering $\sim g_2$ between the different kinds. Both these interactions are compatible with the chiral particle number $\mathrm{U}(1) \times \mathrm{U}(1)$ symmetry.}
    \label{fig:Dispersion}
\end{figure}

Specifically, we study a system of interacting massless Dirac fermions in one dimension under continuous monitoring of the local particle number (see Figs.~\ref{fig:modelsketch},\ref{fig:Dispersion}). 
To ensure a faithful and analytically tractable model, we define the microscopic model based on the following symmetries, and \textit{not} as an effective long wavelength description of an underlying lattice model:

(i) Number conservation (U(1) symmetry) -- The total fermion number is conserved, enforcing a strong U(1) symmetry on both the Hamiltonian and the measurement operators. (ii) Fermionic chirality -- Fermions are described by creation and annihilation operators $\hat{\psi}_d(x)$ and $\hat{\psi}_d^\dagger(x)$~\footnote{$\{ \hat{\psi}_d(x),\hat{\psi}_{d'}(x') \}=0$ and $\{ \hat{\psi}_d(x),\hat{\psi}^\dagger_{d'}(x') \}=\delta_{dd'}\delta(x-x')$.}, with $d=\pm$ for right- and left-moving fermions. The Hamiltonian conserves the number of right and left movers separately and exhibits a chiral U$(1) \times $U$(1)$ symmetry $\hat{\psi}_d(x) \to e^{ib_d} \hat{\psi}_d(x)$ where $b_+$ and $b_-$ can differ. The measurement operators conserve the total particle number -- they are symmetric under transformations where $b_+=b_-$ -- but break the chiral symmetry. (iii) The system is invariant under translations $\hat{\psi}_d(x) \to \hat{\psi}_d(x+\Delta)$ and inversions $\hat{\psi}_d(x) \to \hat{\psi}_{-d}(-x)$, with periodic boundary conditions. The randomness of measurements weakly breaks these symmetries in individual trajectories. (iv) The Hamiltonian is massless, implying gapless fermionic excitations with linear dispersion. (v) Locality -- The measurement process is local, with a uniform rate $\gamma \geq 0$ at each point in space, ensuring locality in the dynamics.

These properties set the leading order Hamiltonian~\cite{Luttinger_1963}, 
\begin{equation}
    \hat{H} = \int_{-L/2}^{L/2} dx \left( \sum_{d=\pm} \hat{\psi}^\dagger_d i d v_0  \partial_x \hat{\psi}_d + g_2 \hat{n}_+ \hat{n}_- \right).
\end{equation}
$\hat{n}_d = \hat{\psi}_d^\dagger \hat{\psi}_d$ are the densities of left and right moving particles, and $v_0, g_2 \in \mathbb{R}$ are the parameters of the model. The velocity $v_0$ characterizes the spectrum while $g_2$ parametrizes the symmetry-allowed local backscattering interaction. Relaxing the exact locality condition of the interactions allows to add forward scattering terms $\sim g_4 \hat{n}_d^2$. They do not change the results and for simplicity we set $g_4=0$. This model is also known as the  massless Thirring model~\cite{Thirring_1958} and its ground state represents a paradigmatic example for a solvable quantum field theory. \par
We define the measurement operator in accord with the above symmetries as
\begin{equation}
    \hat{O}(x) = \underbrace{\sum_{d=\pm} \hat{\psi}^\dagger_d(x) \hat{\psi}_d(x)}_{\hat{O}_{1}(x)} + \underbrace{m \sum_{d=\pm }\hat{\psi}^\dagger_d(x) \hat{\psi}_{-d}(x)}_{\hat{O}_{2}(x)}.
\end{equation}
This is unique up to a global prefactor that can be absorbed into the measurement rate $\gamma$~\footnote{The measurement rate here carries dimension $[\gamma]=1/\text{time} \times \text{length}$ (a density in space and time), as the measurement is applied on every point in a continuous fashion.}, and the parameter $m$ which we assume to be $\mathcal{O}(1)$. As $[\hat{O}_{1}(x),\hat{O}_{2}(x')]=0$, the monitoring of $\hat{O}(x)$ can also be understood as simultaneous monitoring of $\hat{O}_{1}(x)$ and $\hat{O}_{2}(x)$ with the same rate $\gamma$. $\hat{O}_{1}(x)$ is compatible with the chiral symmetry, while $\hat{O}_{2}(x)$ breaks it down to $\mathrm{U}(1)$ explicitly. $\hat{O}_{2}(x)$ is the only exactly local bilinear with that property compatible with inversion symmetry. \par
Both $\hat{H}$ and $\hat{O}(x)$ can be rewritten in terms of bosonic degrees of freedom by an exact operator mapping -- bosonization~\cite{vonDelft_1998,Luttinger_1963,Tomonaga_1950,Haldane_1981,Mattis_1965,Giamarchi_2003} -- in terms of the two Hermitian field operators $\hat{\phi}(x)$ and $\hat{\theta}(x)$, which satisfy $[\partial_x \hat{\theta}(x),\hat{\phi}(x')]=-i \pi \delta(x-x')$. For this operator mapping to work, it is required that the spectrum is unbounded which is ensured by chirality and the spatial continuum nature which are key ingredients to our model. The Hamiltonian can be written in terms of the bosonic operators as
\begin{equation} \label{eq:LuttingerLiquidHamiltonian}
    \hat{H} = \frac{v}{2\pi} \int dx \left[ g (\partial_x \hat{\phi})^2 + \frac{1}{g} (\partial_x \hat{\theta})^2 \right] .
\end{equation}
For the Hamiltonian, the parameters are exactly known as the group velocity $v=\sqrt{v_0^2-g_2^2}$ and the Luttinger parameter $g=\sqrt{\frac{v_0-g_2}{v_0+g_2}}$. $g=1$ therefore corresponds to the free case, $g_2=0$. We will be mainly interested in weak interactions and introduce the parameter $\Delta g=g-1$, where $\vert \Delta g \vert \ll 1$. $\Delta g < 0$ ($\Delta g>0$) corresponds to attractive (repulsive) interactions of the Dirac fermions. \par
Following the bosonization procedure~\cite{vonDelft_1998}, we also find an exact operator mapping for the monitored operators,
\begin{align}\label{eq:meas_boson}
    \hat{O}_{1}(x) &= -\frac{1}{\pi} \partial_x \hat{\phi}(x), & \hat{O}_{2}(x) &= \frac{m}{a} \cos 2 \hat{\phi}(x),
\end{align}
$a$ is a microscopic cutoff distance scale which we may absorb into the model parameter $m$ for brevity. Such a short distance cutoff -- or equivalently large momentum cutoff $\Lambda = \pi/a$ -- is required for a ultraviolet (UV) regularization of the theory.  
\subsection{Observables} \label{sec:Observables}
A convenient observable is the connected correlation function of the monitored operator itself~\cite{Buchhold_2021,Poboiko_2023,Mueller_2022},
\begin{equation}
    C(x-x') = \overline{\langle \hat{O}(x) \hat{O}(x') \rangle - \langle \hat{O}(x) \rangle \langle \hat{O}(x') \rangle}.
\end{equation}
It contains the term $\overline{\langle \hat{O}(x) \rangle\langle \hat{O}(x') \rangle}=\Tr \hat{O}(x) \otimes \hat{O}(x') \overline{\hat{\rho}\otimes \hat{\rho}}$. Therefore, it is sensitive to the second moment of the density operator $\overline{\hat{\rho} \otimes \hat{\rho}}$ and behaves non-trivially under measurement-induced dynamics. In terms of the replica density operator we find (for any $r \neq r'$)
\begin{equation}
    C(x-x') = \lim_{R \to 1}  \Tr \left( \hat{O}^{(r)}(x)(\hat{O}^{(r)}(x')-\hat{O}^{(r')}(x')) \tilde \rho_{R} \right)  . 
\end{equation}
Based on this, we find the second cumulant of particle number fluctuations in a subsystem $A$ as
\begin{equation}
    C_A^{(2)} = \int_A dx \int_A dx' C(x,x').
\end{equation}
For non-interacting Dirac fermions, there is a direct connection of this cumulant to the commonly studied von-Neumann entanglement entropy $S_\text{vN}(A)=\overline{-\Tr \hat{\rho}_A \log \hat{\rho}_A}$ where $\hat{\rho}_A=\Tr_B \ket{\psi_t}\bra{\psi_t}$ is the reduced density matrix of a subsystem $A$ and $B$ is the complement of $A$, via the Klich-Levitov relation~\cite{Klich_2009}
\begin{equation}
    S_\text{vN}(A) = \sum_{q=1}^\infty 2 \zeta(2q) C_A^{(2q)}.
\end{equation}
$C_A^{(2q)}$ is the $2q$-th particle number cumulant in $A$. For small measurement strength $\gamma\to 0$, the sum is dominated by the lowest order cumulants~\cite{Poboiko_2023}. In that regime, we may approximate  
\begin{equation}
    S_\text{vN}(A) \simeq \frac{\pi^2}{3} C_A^{(2)}.
\end{equation}
For interacting fermions, so far no such relation has been developed and the computation of the von Neumann entanglement entropy remains a formidable task~\cite{Bastianello_2019}. We therefore focus on the non-linear correlation function and discuss the entanglement entropy only in the non-interacting case. \par
\section{Path integral approach} \label{sec:PathIntegralApproach}
\subsection{Constructing the Feynman Keldysh path integral} \label{sec:FeynmanConstruction}
The dynamics of the bosonic model defined through Eqs.~\eqref{eq:LuttingerLiquidHamiltonian}, \eqref{eq:meas_boson} and the replica quantum master Eq.~\eqref{eq:LindbladGeneral} can be expressed in terms of a Feynman path integral. The corresponding replica action is (see App.~\ref{app:Feynman_Keldysh})
\begin{align} \label{eq:ActionFromLindblad}
    S_R &= \sum_r \left[ S_V^{(r)} + S_H^{(r)} + S_\mathcal{L}^{(r)} \right] + \sum_{r \neq r'} S_{\mathcal{M}}^{(r,r')},  
\end{align}
where
\begin{align} 
    S_V^{(r)} &= -\frac{1}{\pi} \int_{t,x} \left( \phi_+^{(r)} \partial_x \partial_t \theta_+^{(r)} - \phi_-^{(r)} \partial_x \partial_t \theta_-^{(r)} \right), \label{eq:ActionBothFields1} \\
    S_H^{(r)} &= - \int_t \left( H_+^{(r)} - H_-^{(r)} \right), \label{eq:ActionBothFields2} \\
    S_\mathcal{L}^{(r)} &= -i \gamma \sum_{n=1}^2 \int_{t,x} \left(  O_{n,+}^{(r)}O_{n,-}^{(r)} - \frac{1}{2} O^{(r)}_{n,+} O^{(r)}_{n,+} - \frac{1}{2} O^{(r)}_{n,-} O^{(r)}_{n,-} \right) ,  \label{eq:ActionBothFields3} \\
    S_\mathcal{M}^{(r,r')} &= -i \gamma \sum_{n=1}^2 \int_{t,x} \left( O_{n,+}^{(r)}O_{n,-}^{(r')} + \frac{1}{2} O^{(r)}_{n,+} O^{(r')}_{n,+} + \frac{1}{2} O^{(r)}_{n,-} O^{(r')}_{n,-} \right) . \label{eq:ActionBothFields4}
\end{align}
Here, $1\le r\le R$ denotes the replica index, $n=1,2$ stands for the different measurement operators $\hat{O}_1,\hat{O}_2$, and $\pm$ is the Keldysh index labeling fields on the $\pm$-contour. We consider the simultaneous monitoring of $\hat{O}_{1}(x)=-\frac{1}{\pi} \partial_x \hat{\phi}(x)$ and $\hat{O}_{2}(x)=m \cos(2 \hat{\phi}(x))$. The sum $S_V^{(r)} + S_H^{(r)} + S_\mathcal{L}^{(r)}$ describes $R$ independent copies of Luttinger liquids subject to dephasing. The term  $S_\mathcal{M}^{(r,r')}$ describes the measurement-induced inter-replica coupling, and it is responsible for both the evolution into a nontrivial stationary state as discussed in Sec.~\ref{sec:ReplicaMaster} and the violation of the normalization. Normalization is restored after taking the replica limit $R \to 1$. \par  
Correlation functions of normal ordered operators, such as, for instance, $\hat{O}^{(r)}(x) \hat{O}^{(r')}(x')$ at a fixed time $t$, are evaluated as expectation values of the corresponding 'classical' operators $O_{c}^{(r)} = (O_+^{(r)} + O_-^{(r)})/\sqrt{2}$. This definition applies to arbitrary operators, i.e. $\hat{O}=\hat{\phi},\hat{\theta}$, but also any nonlinear function thereof. It is different from the usual definition of classical and quantum fields in the Keldysh formalism, where it is only used in this form for linear combinations of field operators. The definition we use here makes sure that the correlation functions of arbitrary operators obey causality and reality (see App.~\ref{app:Feynman_Keldysh}). We find
\begin{multline}
    \Tr (\hat{O}^{(r)}(x) \hat{O}^{(r')}(x') \tilde \rho_R(t))=\langle O_c^{(r)}(x,t) O_c^{(r')}(x',t)\rangle  \\= \frac{1}{2} \int \mathcal{D} (\Phi_\pm, \Theta_\pm) O_c^{(r)}(x,t) O_c^{(r')}(x',t) e^{iS_R},
\end{multline}
and thus the correlation function 
\begin{equation}
    C(x,x') = \lim_{R \to 1} \langle O_c^{(r)}(x,t)(O_c^{(r)}(x',t) - O_c^{(r')}(x',t)) \rangle. \label{eq:FieldCdef}
\end{equation}
One also defines the corresponding quantum fields $O_{q}^{(r)} = (O_+^{(r)} - O_-^{(r)})/\sqrt{2}$. 
\subsection{Free Bosonic theory} \label{sec:FreeBosons}
Monitoring $\hat{O}_{1}(x)$ yields a contribution to the Gaussian part of the action, while monitoring $\hat{O}_{2}(x)$ yields a non-linear contribution to the replica Keldysh action. Let us consider first the case $m=0$ for which the model is Gaussian and thus exactly solvable. It is then useful to represent the action in Fourier space for position, time and the replica index: $\phi_{c/q}^{(r)}=\sum_{k=0}^{R-1}e^{-i(2\pi kr /R)} \phi_{c/q}^{(k)}$.
Integrating out the conjugate field $\hat \theta$ (see App.~\ref{App:IntegratingTheta}), which appears only up to quadratic order,  yields the Gaussian action~\footnote{$\int_{p,\omega}=\int_{-\infty}^\infty \frac{dp}{2\pi} \int_{-\infty}^\infty \frac{d\omega}{2\pi}$}
\begin{multline} 
    S_0[\phi] = \\ \frac{g}{2\pi v} \int_{p,\omega} \left( \begin{array}{cc}
        \phi_c^{(0)*} & \phi_q^{(0)*} 
    \end{array} \right) \left( \begin{array}{cc}
        0 & \omega^2 - v^2 p^2 \\
        \omega^2 - v^2 p^2 & \frac{2i\gamma v}{\pi g} p^2
    \end{array} \right) \left( \begin{array}{c}
        \phi_c^{(0)} \\ \phi_q^{(0)} 
    \end{array} \right) \\ \label{eq:bosogaussianreplica} 
    + \sum_{k>0} \frac{1}{2 \pi v} \sum_{\sigma} \int_{p,\omega} \phi_\sigma^{(k)*} \left( g \sigma(\omega^2- v^2  p^2) + \frac{2i \gamma v}{\pi} p^2 \right) \phi_\sigma^{(k)} . 
\end{multline}
The contour-coupling term thus only appears in the $k=0$ sector. This induces an unbounded growth of fluctuations of this center-of-mass mode, which reflects the heating towards an infinite temperature state. To demonstrate this, we compute the equal time correlation function $\left\langle \left( \phi^{(0)}_{c}(t,x) \right)^2 \right\rangle$. The $k=0$-mode is independent of the replica limit and we thus take $R \to 1$ before calculating correlation functions. This recovers the known properties of a regular Keldysh field theory including that the retarded and the advanced Green's function vanish at equal times. With an infinitesimal regularization $0^+>0$, the Keldysh Green's function is given by (see App.~\ref{app:infititecorrelations})
\begin{equation}
    \left\langle \left( \phi^{(0)}_{c}(t,x) \right)^2 \right\rangle = \int \frac{dp}{2\pi} \frac{\gamma}{4 g^2 (0^+)}. \label{eq:InfiniteKeldysh}
\end{equation}
This divergent, infinite temperature growth of the $k=0$-mode decouples from the remaining part of the Gaussian action and does not enter the correlation function $C(x)$. Thus, it has no influence on the evolution in the Gaussian limit $m\to0$. We will show in Sec.~\ref{Sec:Nonlinearity}, that even for $m\neq 0$ the divergent $k=0$-mode can be separated from the $k\neq0$-modes, yielding a well-defined field theory.
\subsection{Observables on the Gaussian level} \label{sec:Fieldtheoryobservables}
\begin{figure}
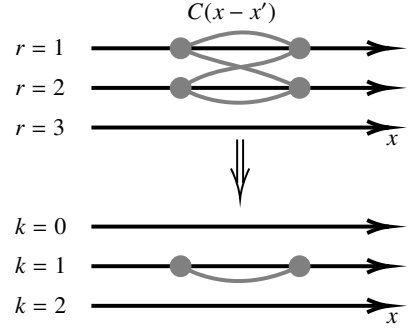

    \centering
    \include{Observablerotation}
    \caption{The correlation function $C(x,x')$ involves terms connecting two arbitrary replicas. The Fourier transformation in replica space allows us to write it in terms of a single arbitrary replica with $k>0$. As all $k>0$ modes decouple and have the same action, it is sufficient to solve the problem for a single replica in Fourier space with $k>0$.}
    \label{fig:observablesdecoupling}
\end{figure}

Within the Gaussian limit, correlation functions such as $C(x-x')$ can be computed exactly from the $k>0$ replica Fourier modes. The results obtained in this limit hold also for $m\neq 0$ on distances shorter than the measurement-induced correlation length.
Consider the Gaussian action
\begin{equation}
    S_0[\phi]= \sum_{\sigma=\pm} \frac{\sigma K_\sigma}{2\pi} \sum_{k>0} \int_{p, \omega} \phi_\sigma^{(k)*} \left( \frac{1}{\eta_\sigma} \omega^2 - \eta_\sigma p^2 \right) \phi_\sigma^{(k)} \label{eq:GaussianModel}
\end{equation}
with
\begin{align}\label{eq:Keta}
    K_\sigma &= g \sqrt{1-\frac{2i\sigma \gamma}{\pi v g}}, & \eta_\sigma&= \frac{v K_\sigma}{g}.
\end{align}
The Gaussian action decouples into independent contributions for all $k>0$, for which $K_\sigma, \eta_\sigma$ do not depend on $k$~\footnote{This property is preserved by the RG flow which justifies to keep these definitions for a re-scaled effective theory.}. The correlation function $C(x-x')$ defined in Eq.~\eqref{eq:FieldCdef} can be simplified by realizing that $\langle O_{2,\sigma}^{(r)}(x,t) O_{2,\sigma}^{(r')}(x',t) \rangle\sim e^{-4\langle \left( \phi_\sigma^{(r)}(x,t) \right)^2 \rangle}\sim(L/a)^{-\frac{2R}{K_\sigma}}$ vanishes in the thermodynamic limit $L\to\infty$, see App.~\ref{app:BosoObservables}. Cross terms $\langle O_{1,\sigma}^{(r)}(x,t) O_{2,\sigma}^{(r')}(x',t) \rangle$ need to vanish due to the inversion symmetry. Therefore, $\langle O_\sigma^{(r)}(x,t) O_\sigma^{(r')}(x',t) \rangle\to \langle O_{1,\sigma}^{(r)}(x,t) O_{1,\sigma}^{(r')}(x',t)\rangle$ and thus  
\begin{align}
    C(x-x') &= 
    \lim_{R \to 1}  \frac{1}{R-1} \frac{1}{\pi^2} \partial_x \partial_{x'} \sum_{k>0} \left\langle \phi_c^{(k)*}(x,t) \phi_c^{(k)}(x',t) \right\rangle \notag \\ &= \lim_{R \to 1} \frac{1}{\pi^2} \partial_x \partial_{x'} \left\langle \phi_c^{(k>0)*}(x,t) \phi_c^{(k>0)}(x',t) \right\rangle, \label{eq:Cofpdef}
\end{align}
for any $k>0$, since each of the $R-1$ correlation functions yields the same contribution, visualized in  Fig.~\ref{fig:observablesdecoupling}.
For the Fourier-transformed correlation function, we find (see App.~\ref{app:BosoObservables})
\begin{equation}
    C(p) \equiv \int_x e^{ipx} C(x) = \frac{c \vert p \vert}{2\pi}, \label{eq:Cofpres}
\end{equation}
where we introduced
\begin{equation}
    c=\Re 1/K_+.
\end{equation}
Based on this result, after a Fourier transformation we find (for $x$ much larger than the short distance cutoff $a$)
\begin{equation}
    C(x) \simeq - \frac{c}{2 \pi^2 x^2} . \label{eq:CritCorrf}
\end{equation}
The second cumulant of particle number fluctuations can be directly computed from the result in momentum space~\cite{Poboiko_2023,Poboiko_2024}
\begin{equation*}
    C_A^{(2)} = \int_0^l dx \int_0^l dx' C(x-x') = \frac{c}{\pi^2} \int_0^\infty dp \frac{1-\cos pl}{p}.
\end{equation*}
We regularize this divergent integral using the UV cutoff $\Lambda=\pi/a$ and obtain
\begin{equation*}
    C_A^{(2)} = \frac{c}{\pi^2} \int_0^{\pi l/a} ds \frac{1-\cos s}{s} = \frac{c}{\pi^2} \left( \Gamma + \log(\Lambda l) + \mathcal{O}((\Lambda l)^{-2}) \right) ,
\end{equation*}
where $\Gamma$ is the Euler-Mascheroni constant. We conclude that for $l \gg a$
\begin{equation*}
    C_A^{(2)} \simeq \frac{c \Gamma}{\pi^2} + \frac{c}{\pi^2} \log(\Lambda l) .
\end{equation*}
In the case of free Dirac fermions ($g=1$ i.e. $\Delta g=0$), we can connect this result quantitatively to the entanglement entropy via~\cite{Klich_2009} (see sec.~\ref{sec:Observables})
\begin{equation}
    S_\text{vN}(A) \simeq \frac{\pi^2}{3} C_A^{(2)} \simeq \frac{c \Gamma}{3} + \frac{c}{3} \log(\Lambda l). \label{eq:entanglement_scaling}
\end{equation}
In the limit of vanishing measurements $\gamma\to0$, the value for the ground state of the Dirac fermion Hamiltonian,  $c\to1$, is recovered. This differs from the monitored fermions on a one-dimensional tight-binding lattice, where $c$ appears to diverge in the limit $\gamma\to0$~\cite{Alberton_2021, Poboiko_2023}. The constant contribution $s_0=\frac{c \Gamma}{3}$ is non-universal and affected by the cutoff $\Lambda$. In the Gaussian limit, $m\to0$, all information about correlations on large distances is thus encoded in the parameter $c$, which is in turn determined by microscopic parameters.  
\subsection{Non-linearities and perturbative renormalization group} \label{Sec:Nonlinearity}
In the non-Gaussian case, $m\neq 0$, the measurement of $\hat{O}_{2}(x)$ introduces a nonlinear coupling between different replicas, including the central $k=0$ mode. In the following, we show how the RG flow of the theory reveals both a scale $\xi$ (the correlation length) beyond which the non-linearity is relevant, and a renormalization of the parameter $c$. As the observables are determined by correlations of the fields with replica momentum $k>0$, we integrate out the $k=0$ mode also in the presence of the non-linearity. Since the different $k$ modes are coupled non-linearly, this has to be done perturbatively (see App.~\ref{app:IntegratingOutK0}). However, due to the unbounded fluctuations of the $k=0$ mode induced by measurements of $\hat{O}_1(x)$ in the free theory, the perturbative expansion is exact including only the first term. We find that for $m\neq0$ -- after integrating out $\phi_\sigma^{(k=0)}$ -- the action consists of the Gaussian part $S_0$ and a nonlinear part $\Delta S$ (see App.~\ref{app:IntegratingOutK0}), i.e., $S = S_0 + \Delta S$, with
\begin{equation}
    \Delta S = -i m^2 \gamma \sum_{\sigma = \pm} \sum_{r \neq r'} \int_{x,t} \cos 2(\phi^{(r)}_\sigma-\phi^{(r')}_\sigma). \label{eq:relativereplicaaction}
\end{equation}
A similar Sine-Gordon non-linearity was found for the charge-sharpening transition of U$(1)$ symmetric monitored random circuits~\cite{Barratt_2022}. This term is independent of $\phi_\sigma^{(k=0)}$ as it has to be and we can see explicitly rewriting $\phi_\sigma^{(r)}-\phi_\sigma^{(r')} = \sum_{k=0}^{R-1} \left( e^{-i2\pi k r/R} - e^{-i2\pi k r'/R} \right) \phi_\sigma^{(k)}$. The prefactor of $\phi_\sigma^{(k=0)}$ vanishes for all $r,r'$. We find that all other $\phi_\sigma^{(k)}$ are coupled among each other, in contrast to the Gaussian case. However, the resulting action on the forward and backward contour still decouples $S[\phi_+,\phi_-]=S_+[\phi_+]+S_-[\phi_-]$. This provides an interpretation of the evolution in terms of a non-Hermitian Hamiltonian, acting individually on each contour, but coupling all fields $\phi_\sigma^{(k)}$ with $k\neq 0$ on a given contour. The non-Hermitian nature of the Hamiltonian results in a cooling of the system into a ground state on both contours. In order to understand the effect of the non-linearity on the correlations at large distances, we take an RG perspective and analyze the emerging replica sine-Gordon model. As the $\pm$ contours decouple, we can perform the perturbative RG calculation independently. The action can be reduced to
\begin{multline}
    S_\sigma[\phi_\sigma]= \frac{\sigma K_\sigma}{2\pi} \sum_{k>0} \int_{p, \omega} \phi_\sigma^{(k)*} \left( \frac{1}{\eta_\sigma} \omega^2 - \eta_\sigma p^2 \right) \phi_\sigma^{(k)} \\ - \lambda_\sigma \sum_{r \neq r'} \int_{x,t} \cos 2(\phi_\sigma^{(r)}-\phi_\sigma^{(r')}).  \label{eq:nearest_neighbor_action}
\end{multline}
The bare couplings can be read off as (see Eq.~\eqref{eq:Keta})
\begin{align}
    K_\sigma &= g\sqrt{1-\frac{2i\sigma \gamma}{\pi v g}}, & \eta_\sigma &= \frac{vK_\sigma}{g}, & \lambda_\sigma &= i m^2 \gamma.
\end{align}

The microscopic action obeys an important symmetry relating the forward and backward contour:
\begin{equation}
    S[\phi_+,\phi_-]=-S^*[\phi_-,\phi_+]. \label{eq:ActionSymmetry}
\end{equation}
This symmetry ensures Hermiticity of the density matrix $\tilde \rho_R$ and the reality of physical observables (see App.~\ref{app:Feynman_Keldysh}). It is an exact property of the generalized Lindblad generator in Eq.~\eqref{eq:LindbladGeneral}. Equation~\eqref{eq:ActionSymmetry} implies that 
\begin{align}
    K_\sigma&=K_{-\sigma}^*, &  \eta_\sigma&=\eta_{-\sigma}^*, & \lambda_\sigma &= -\lambda _{-\sigma}^*,
\end{align}
which is obeyed by the microscopic couplings. 

\begin{figure}
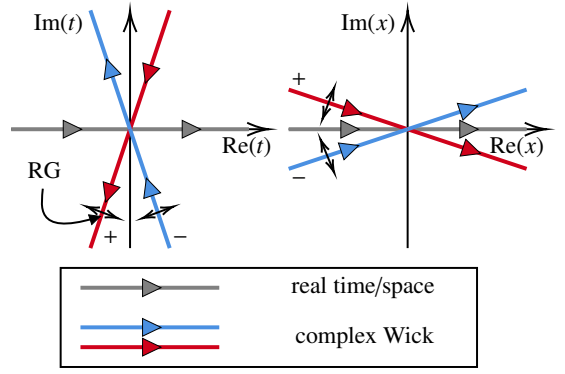

    \centering
    \include{ComplexWick}
    \caption{Illustration of the different complex Wick rotations from the original representation of space and time (grey) on the forward (red) and backward (blue) Keldysh contour. Through analytic continuation, we transform space and time to eliminate $\eta_\sigma$, obtaining the Euclidean form of the action. The complex phase of the velocity $\eta_\sigma$ causes the rotation angle to vary continuously, depending on $\gamma$ and differing between the forward and backward contours. Under the renormalization group (RG) flow, this angle adjusts to maintain the Hermiticity of the density operator. $\Im K_\sigma$, governs both the magnitude and direction of this scale-dependent rotation.
}
    \label{fig:ComplexWick}
\end{figure}

In the following, we will apply a perturbative RG scheme, which treats both contours individually. In order to preserve the above exact symmetry between the contours within this scheme, we implement a uniquely defined Wick-type space-time rotation, acting opposite on both contours. Starting from the microscopic action with a UV momentum cutoff $\Lambda=\pi/a$. We perform a Wick rotation $t \to i\sigma t$ and then eliminate $\eta_\sigma$ by rescaling space and time. Note that the rescaling includes an additional Wick rotation, which is different for both contours, see Fig.~\ref{fig:ComplexWick}, as it depends on the complex phase of $\eta_\sigma$.
We obtain on each contour the Euclidian action
\begin{multline}
    S_\sigma[\phi_\sigma]=\frac{K_\sigma}{2 \pi}  \sum_{k>0} \int_{\omega,p} \phi^{(k)*}_\sigma \left( p^2 +  \omega^2 \right)\phi^{(k)}_\sigma \\ + i \lambda_\sigma \int_{x,t} \sum_{r,r'}   \cos 2(\phi_\sigma^{(r)}-\phi_\sigma^{(r')}), \label{eq:EuclidianAction}
\end{multline}
We split the fields into slow and fast modes according to
\begin{align}
    \phi^{(k)}_{\sigma,<}(x,t) &= \int_{-b\Lambda}^{b \Lambda} \frac{d p}{2 \pi} e^{ipx} \int_{-\infty}^{\infty} \frac{d \omega}{2 \pi} e^{i\omega t} \phi_\sigma^{(k)}(p,\omega), \\ \phi^{(k)}_{\sigma,>}(x,t) &= \int_{b \Lambda < \vert p \vert <\Lambda} \frac{d p}{2 \pi} e^{ipx} \int_{-\infty}^{\infty} \frac{d \omega}{2 \pi} e^{i\omega t} \phi_\sigma^{(k)}(p,\omega),
\end{align}
where $b=e^{-s}$ and $0<s \ll 1$. We then integrate out the fast modes $\phi_>$ to obtain an action for the slow ones,
\begin{equation}
    e^{-S_{\sigma,<}[\phi_{\sigma}]} \equiv \langle e^{-S_\sigma[\phi_\sigma]} \rangle_{0,>}.
\end{equation}
Here $\langle \dots \rangle_{0,>}=\int \mathcal{D}\phi_> \dots e^{-S_{0}[\phi_{>}]}$ denotes the fast mode expectation value and $S=S_{0}+\Delta S$ where $S_0$ is the free part of the Euclidean action~\eqref{eq:EuclidianAction} and $\Delta S$ is the cosine non-linearity. 

The perturbative RG analysis provides faithful results in the free, Gaussian limit, i.e. for the Luttinger parameter $g\to 1$, i.e. $\Delta g\to 0$, and the monitoring strength $\gamma \to 0$. Up to second order 
\begin{align}
    S_< \simeq S_{0,<} &+ \langle \Delta S \rangle_{0,>} - \frac{1}{2} \left(  \langle ( \Delta S )^2 \rangle_{0,>} -  \langle \Delta S \rangle_{0,>}^2 \right)\nonumber\\
    \simeq S_{0,<} &+ i \sum_\sigma \lambda_\sigma \int_{x,t} \sum_{r,r'} \cos 2 (\phi^{(r')}_{\sigma,<} - \phi^{(r')}_{\sigma,<}) e^{-\frac{2s}{K_\sigma}}\nonumber\\
    & -  \sum_\sigma \frac{4\lambda_\sigma^2 I s}{\Lambda^4  K_\sigma} \sum_{k>0} \int_{\omega,p} \phi_{<,\sigma}^{(k)*} (p^2+\omega^2) \phi_{<,\sigma}^{(k)}, \label{eq:secondOrderPert}
\end{align}
see App.~\ref{app:RGNearestNeighbor}. 
Here, $I$ is a function of $K_\sigma$. An expansion around the free Gaussian value $K_\sigma = 1$ yields $I \approx 0.07805$. Rescaling space and time $x,t \to x/b,t/b$ and setting $s=0$ after taking a derivative with respect to $s$ yields the RG flow equations
\begin{align}
    \partial_s \lambda_\sigma &= \left(1-\frac{1}{K_\sigma}\right) \lambda_\sigma + \mathcal{O}(\gamma^3), \\
    \partial_s K_\sigma &= -\frac{4 I \lambda_\sigma^2}{\Lambda^4 K_\sigma} + \mathcal{O}(\gamma^3).
\end{align}
These flow equations take the conventional BKT form with, however, complex parameters. In the present form, they yield oscillatory behavior~\cite{Buchhold_2021} and break the above discussed symmetry $\lambda_\sigma \neq \lambda_{-\sigma}^*$ along the evolution, which is required for Hermiticity  $\tilde \rho_R = \tilde \rho_R^\dagger$. Therefore, we implement the above discussed, additional Wick-type rotation and rescale $(x,t) \to (x,t) e^{i\beta_\sigma}/b$ (see Fig.~\ref{fig:ComplexWick}). The choice $\beta_\sigma=s \Im \frac{1}{K_\sigma}$ ensures that Hermeticity, i.e., $\lambda_\sigma = \lambda_{-\sigma}^*$, is preserved on each RG step. The RG evolution of the phase $e^{i\beta_\sigma}$ does not need to be tracked: it guarantees at each RG step that physical observables remain real-valued and thus does not enter their computation (see Sec.~\ref{sec:Fieldtheoryobservables}). We then obtain the flow equations
\begin{align}
    \partial_s \lambda_\sigma &= \left( 1-\Re \frac{1}{K_\sigma} \right) \lambda_\sigma + \mathcal{O}(\gamma^3) ,\\
   \partial_s K_\sigma &= -\frac{4I\lambda_\sigma^2}{\Lambda^4 K_\sigma} + \mathcal{O}(\gamma^3).
\end{align}
We then define the real-valued couplings $X=1-1/\Re K_\sigma$ and $Y^2=-4I\lambda_\sigma^2/\Lambda^4$. Due to Hermeticity, they do not depend on $\sigma$. In the vicinity of the free limit $\Delta g=\gamma=0$, an expansion in both $X$ and $Y$ to leading order yields the closed equations
\begin{align}
    \partial_s Y &= 2X Y, &\partial_s X &=Y^2. \label{eq:complexRG}
\end{align}
These are precisely the well known BKT flow equations~\cite{Berezinskii_1971,Berezinskii_1972,Kosterlitz_1973}.
\subsection{Discussion of the flow equations and implications} \label{sec:DiscussionOfAnalyticalApproach}
\begin{figure}
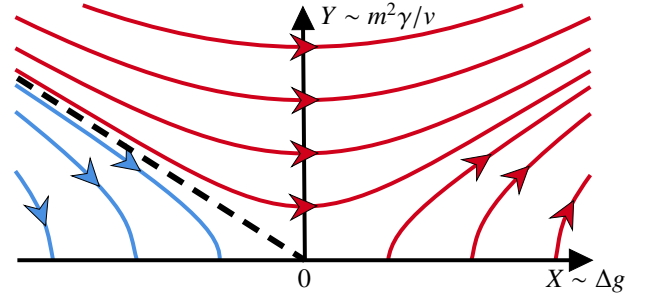

    \centering
    \include{RGflow}
    \caption{RG flow of the dimensionless parameters for the measurement strength of the chirality-breaking current observable $\hat{O}_{2}$, $Y$, and the effective interaction strength $X$. It is determined to leading order by the bare interaction parameter $X \sim \Delta g=g-1$, where $g$ is the Luttinger parameter. Sub-leading contributions $\sim \gamma^2$ come from the measurement of $\hat{O}_1$. The RG flow follows the BKT paradigm and we find that only for attractive interactions $\Delta g<0$, there is a parameter regime where the flow ends in a line of fixed points characterizing the free bosonic theory of a non-Hermitian Luttinger liquid. For vanishing interactions $\Delta g=0$, this regime reduces to the single point $\gamma \to 0$. The dashed line visualizes the phase transition where the monitoring of $\hat{O}_{2}$ becomes RG relevant.}
    \label{fig:RGflow}
\end{figure}

The flow equations \eqref{eq:complexRG} are real-valued and contour-independent, which both is a consequence of the Hermiticity symmetry. We will now discuss their solution to leading order and relate them to the behavior of the dimensionless interaction parameter $\Delta g$ from the Hamiltonian and the dimensionless measurement strength $m^2 \gamma/v$ of the chirality-breaking measurement of $\hat{O}_{2}(x)$ as shown in Fig.~\ref{fig:RGflow}. \par
Defining the quantity $\delta = Y^2-2X^2$, we can eliminate $Y$ and find new closed flow equations
\begin{align} 
    \partial_s X &= \delta + 2X^2, & \partial_s \delta = 0. \label{eq:DeltaRG}
\end{align}
Hence, $\delta$ is conserved which reduces the problem to a single parameter RG flow for $X$, determined by the bare values of $X$ and $\delta$. The perturbative RG result is valid in the regime of small interactions $\Delta g \to 0$ and measurement rates $\gamma\to 0$. In this regime, we find in a leading order expansion that~\footnote{We re-introduced the short distance cutoff $a$ in the parameter $m \to m/a$ to eliminate the cutoff dependence of the critical point.}, the 
\begin{equation}
    \delta \simeq \frac{8  m^4 I}{\pi^3} \frac{\gamma^2}{v^2} - 2\Delta g^2.
\end{equation}
Note that $X = \mathcal{O}(\gamma^2)$, as it depends only on the real part of $K_\sigma$, which appears only at second order in $\gamma$, i.e., $\Re K_\sigma \sim \gamma^2$. Consequently, the $\gamma$-dependence from $X^2 \sim \gamma^4$ to $\delta$ is subleading compared to the contribution from $Y^2 \sim \gamma^2$. Incorporating the effect of the measurement of $\hat{O}_1$ -- which enters through $X$ -- would require a systematic treatment up to the next order in perturbation theory in $\gamma$. However, this does not alter the universal behavior near $\Delta g = \gamma = 0$~\footnote{This power counting argument relies on $m=\mathcal{O}(1)$}. \par
The signs of both $\delta$ and $X$ determine the relevance of the non-linearity and therefore the underlying thermodynamic phase: \par
(i) For $X<0$, $\delta<0$ a stationary solution is approached at $X=-\sqrt{-\delta/2}$. This yields only small modifications to the Gaussian theory in Sec.~\ref{sec:Fieldtheoryobservables}, e.g., an infinitely large correlation length and only small corrections to the parameter $c$ in the correlation function in Eq.~\eqref{eq:CritCorrf}. To leading order, we find
\begin{equation}
    c-1 \sim \sqrt{\gamma_c-\gamma},
\end{equation}
where $\gamma_c \simeq -\frac{\Delta g\pi^{3/2}}{2 m^2 I}$. Here, the square root scaling in the vicinity of the critical point and the value $c=1$ right at the critical point are universal and do not depend on microscopic parameters. However, this scenario only turns into a robust phase for $\Delta g < 0$, i.e., the Gaussian model with all the features of a non-Hermitian Luttinger Liquid~\eqref{eq:GaussianModel}, can only be stabilized by attractive interactions for any nonzero measurement strength. \par
(ii) For $X<0$ and $\delta>0$, the parameter $c$ diverges at a parameter-dependent scale, which is commonly identified as the correlation length in the equilibrium BKT scenario~\cite{Berezinskii_1971,Berezinskii_1972,Kosterlitz_1973}. The flow Eqs.~\eqref{eq:DeltaRG} yield the explicit solution 
\begin{equation}
    X(s)=\sqrt{\frac{\delta}{2}} \tan(\sqrt{2\delta}s+\arctan(\sqrt{\frac{2}{\delta}}X(0))). \label{eq:FlowSolution}
\end{equation}
This expression diverges when the argument of the tangent function approaches the value $\pi/2$. For $\delta \to 0$ the $\arctan$ is negligible in the argument and the divergence occurs at the RG length scale $s_* \simeq \pi/\sqrt{2 \delta}$. The RG scale $s_*$ is then identified with the physical correlation length $\xi$ through $\log \xi \sim s_*$. On distances larger than $\xi$, correlation functions decay exponentially fast with the distance $C(x) \sim e^{-x/\xi}$ and the cumulant $C_A^{(2)} \sim O(A^0)$ is consistent with the entanglement entropy obeying an area law $S_\text{vN}(A) \sim A^0$. 

An important characteristic of the model is the scaling of the correlation length when approaching the critical point $\delta=0$. To leading order  we find
\begin{equation}
    \log \xi \sim \frac{1}{\sqrt{\gamma-\gamma_c}},
\end{equation}
which matches the BKT essential scaling. \par
(iii) For $X=0$, the parameter $\delta=\frac{8Im^4}{\pi^3}\frac{\gamma^2}{v^2}>0$ is always positive. This shifts the critical measurement strength $\gamma_c\to0$ in this limit, implying that the correlation length is always finite $\xi<\infty$ for any non-zero measurement strength $\gamma>0$. For $\gamma_c=0$ the scaling close to the non-monitored case is modified compared to case (ii), yielding
\begin{equation}
    \log \xi \sim \frac{1}{\sqrt{\delta}} \propto \frac{1}{\gamma}.
\end{equation}
This happens because the distance from criticality vanishes quadratically, $\delta \sim \gamma^2$ instead of linearly $\delta \sim \gamma-\gamma_c$. In this sense, one can identify the free theory ($\gamma=\Delta g=0$) as located at a critical endpoint of the BKT critical line from scenario (ii), and the essential scaling is modified towards a characteristic $1/\gamma$ divergence. This degree of divergence towards the free limit is phenomenologically consistent with the one predicted for monitored fermions undergoing weak localization~\cite{Poboiko_2023}. On length scales much smaller than the correlation length, the scaling of the entanglement entropy is consistent with the free model, $S_\text{vN} \simeq \frac{c}{3} \log L$, with a slightly renormalized prefactor $c$. Here we assume a large correlation length (i.e., small $\gamma$) such that $c$ remains close the free fermion result, $c \approx 1$.
 \par
(iv) If $X>0$, independently of $\gamma$, the correlation length is always finite and (for a fixed interaction strength $\Delta g$) its dependence on $\gamma$ is non-universal. \par
This discussion is summarized in Fig.~\ref{fig:PhaseDiagram}: Weak attractive interactions stabilize a phase with algebraic correlations and a divergent correlation length. The free case then represents a doubly fine-tuned scenario where the divergence of the correlation length is modified compared to the conventional BKT scenario. At this critical point, the system is non-interacting on the level of fermions, becomes conformally invariant and therefore has logarithmic scaling of the entanglement entropy and $c=1$. In this case, we can interpret $c$ as the central charge. 
\section{Discussion}
\subsection{Relation to previous works} \label{sec:RelationsPrevious}
Monitored fermions in one dimension have recently attracted significant interest across various contexts~\cite{Cao_2019, Alberton_2021, Buchhold_2021, Poboiko_2023, Starchl_2024, Coppola_2022, Guo_2024, Poboiko_2025, Chen_2020,Bernard_2019,Loio_2023,Soares_2024,Turkeshi_2024, Szyniszewski_2023, Jian_2022,Mueller_2022,Ladewig_2022,Fava_2023,Fava_2024,Xing_2024,Paviglianiti_2023, Turkeshi_2021, Malakar_2024, Tirrito_2023,Weinstein_2023}. In this section, we locate our work within this broader landscape, emphasizing the following aspects: the model and its symmetries, the analytical workflow and the physical implications of the results. \par
(i) \emph{Dirac fermions}: We focus on chiral Dirac fermions with a linear dispersion in the spatial continuum~\cite{Buchhold_2021}, characterized by a $\mathrm{U}(1) \times \mathrm{U}(1)$ symmetry that conserves left- and right-moving fermions independently. The model is amenable to exact bosonization through operator identities, making it a paradigmatic framework for analytically studying measurement-induced phase transitions in the perturbative regime of weak interactions and weak monitoring.

Dirac fermions naturally emerge as the low-energy degrees of freedom in one-dimensional lattice fermion models, such as those described by Luttinger liquid theory~\cite{Luttinger_1963}. However, in monitored systems, energy conservation is broken, and measurements can populate fermions across the entire band. Thus there is no direct, a priori link between the dynamics of monitored fermions on a tight-binding lattice~\cite{Cao_2019, Alberton_2021, Poboiko_2023, Starchl_2024, Coppola_2022, Guo_2024, Poboiko_2025} and monitored Dirac fermions. However, for lattice models with linear or approximately linear dispersion, we expect the results for monitored Dirac fermions to remain valid on long timescales determined by the inverse of the band curvature. This suggests that the critical behavior and universal properties identified in the continuum model may persist in lattice systems under suitable conditions. \par
(ii) \emph{Weak vs. projective measurements}: Here we consider weak-continuous measurements leading to the form of the replica quantum master equation in Eq.~\eqref{eq:LindbladGeneral}. Interpolating between infinitely weak, continuous measurements and projective measurements can be done by increasing the factor $\gamma\delta t$ in the generalized projector $\hat P(J_t)$ in Eq.~\eqref{eq:genproj}. For the present case of Dirac fermions we can rigorously argue that this does not modify the underlying universal behavior: expanding the exponential up to higher orders in the measurement operators $\hat O_1, \hat O_2$ produces only terms that are less relevant in the RG sense. For instance, the $l$-th order in $\hat O_2$ yields the higher harmonics $\cos 2l(\phi^{(r)}-\phi^{(r')})$ for $l>1$ which are less relevant than the $l=1$ term. While such terms may introduce non-universal corrections, e.g., to the precise location of the transition, they will not modify the scaling behavior at the BKT transition or at the critical point $\Delta g=\gamma=0$. For these reasons we only discuss weak measurements in this work. \par
(iii) \emph{Monitored operators vs. type of interactions}: In this work we emphasize the competition between measurements of the local operators $\hat O_2\sim \sum_{d=\pm}\hat\psi^\dagger_d\hat\psi^{\phantom{\dagger}}_{-d}$ and a Hamiltonian of interacting Dirac fermions. For this type of competition, an entangled phase with algebraic correlations is stabilized for attractive interactions, $\Delta g<0$, below a critical measurement rate $\gamma<\gamma_c$. In the bosonized framework, the present theory is dual to $g \to 1/ g$, i.e. $\Delta g \to -\Delta g$, and $\hat\theta\leftrightarrow\hat\phi$, which can be readily verified from Eqs.~\eqref{eq:ActionBothFields1}-\eqref{eq:ActionBothFields4}. This corresponds to a sign flip of the interactions and a change of the monitored operator $\hat O_2\to \hat O_3=\sum_{d=\pm} d\hat\psi^\dagger_d\partial_x\hat\psi^{\phantom{\dagger}}_d$, i.e., from monitoring the particle density to monitoring the current density. In this case, the entangled phase with algebraic correlations is stabilized for repulsive interactions, $\Delta g>0$ and weak measurements, as one readily verifies from the mentioned duality.

Consequently, a symmetric phase diagram, which is invariant under $\Delta g\to-\Delta g$ is obtained when both $\hat O_2$ and $\hat O_3$ are measured with equal rate. Such a scenario would align with the predicted shape of the phase diagram for interacting, monitored fermions on a tight-binding lattice~\cite{Poboiko_2025,Guo_2024}, for which a symmetric phase diagram has been obtained. The precise form of the phase diagram, and whether an entangled phase survives for arbitrarily weak interactions $|\Delta g|,\gamma\to0$, can, however, not readily be determined from our second order perturbative RG treatment. The competition between the non-commuting operators $\hat O_2, \hat O_3$ requires terms beyond second order perturbation theory. In thermal equilibrium, such scenarios have been studied using exact techniques for self-dual sine-Gordon models~\cite{Lecheminant_2002} or for interacting fermions with additional Rashba-type spin orbit coupling~\cite{Gritsev_2005}.

(iv) \emph{Entanglement entropy of interacting Dirac fermions}: In general, a direct computation of the entanglement entropy in bosonization is tedious in the interacting case due to the compact nature of the fields  -- even  in the ground state problem~\cite{Bastianello_2019}. However, we expect the finite correlation length to result in area law entanglement as the effects of compactness are overwritten by an emergent mass term. Therefore, we expect area-law entanglement in the weakly correlated phase shown in Fig.~\ref{fig:PhaseDiagram}. By our approach we do not find a phase with volume-law entanglement, in contrast to recent results for lattice fermions. There, a volume-law phase is predicted for sufficiently large measurement and interaction strength~\cite{Poboiko_2025,Guo_2024}.  \par
(v) \emph{Relation to previous works on monitored Dirac fermions}: Compared to previous work on monitored non-interacting Dirac fermions~\cite{Buchhold_2021}, here we derived the replica master equation and the replicated Keldysh action using the replica trick. This approach implements normalization of the density matrix by taking the replica limit $R\to1$ at the end of the computation. It is formally more rigorous than the approach used in Ref.~\cite{Buchhold_2021}, which kept the normalization at each level of the calculation -- leading to a hierarchy of master equations for larger and larger replica numbers, which was truncated based on perturbative arguments. For Dirac fermions, the replica limit $R\to1$ can be carried out explicitly due to the exact decoupling of the center of mass mode $k=0$ from the relative modes $k\neq0$, as we demonstrated. In particular, the RG flow is now tracked for general replica numbers $R$. This makes our approach more directly comparable to complementary work~\cite{Poboiko_2023, Poboiko_2024,Poboiko_2025,Fava_2023,Fava_2024,Guo_2024,Starchl_2024}. We remark, however, that the improvements in methodological rigor do not change the results compared to the previous work~\cite{Buchhold_2021} but rather confirm the finding of a BKT flow governing the universal behavior of the model. \par
(vi) \emph{RG flow equations}: The RG flow equations are of the BKT-type but with \textit{complex} flowing parameters. A priori, this increases the complexity from two to four flowing couplings, and to oscillatory behavior of the RG flow in the UV-regime, as presented in Ref.~\cite{Buchhold_2021}. In this previous work, it was found that asymptotically, the RG flow approaches that of a corresponding -- though not identical -- BKT problem at thermal equilibrium, which has a well-defined solution. Here, we argue that the oscillatory behavior is a consequence of truncating the RG flow equations at a finite order of perturbation theory. Implementing Hermiticity of the replicated density matrix, which is an exact symmetry, the oscillatory behavior is eliminated, yielding a more accurate solution of the RG flow equations starting from microscopic parameters. As a consequence, the present work refines the prediction from Ref.~\cite{Buchhold_2021} and clarifies that for non-interacting Dirac fermions, the critical point of the transition shifts towards zero measurement strength, $\gamma_c=0$. This doubly fine-tuned point obeys a divergence of the correlation length $\sim \exp(1/
\gamma)$. Both results are in agreement with the results predicted for fermions on tight-binding lattices discussed in Refs.~\cite{Poboiko_2023,Starchl_2024,Coppola_2022}.
\subsection{Concluding remarks}
We found that interacting monitored Dirac fermions host a BKT phase transition that can be diagnosed by studying the non-linear connected density-density correlation function. Attractive interactions stabilize a critical phase with algebraically decaying correlations in the presence of weak monitoring. Stronger measurements induce a BKT phase transition that can be seen in the emergence of a finite correlation length. In the presence of weak attractive interactions, the divergence of the correlation length at the critical point $\gamma_c$ follows the standard BKT result $\log \xi \sim 1/\sqrt{\gamma-\gamma_c}$. Interestingly, in the absence of interactions, the critical point moves to $\gamma_c=0$, and the scaling of the correlation length agrees with the weak localization result $\log \xi \sim 1/\gamma$~\cite{Altshuler_1980,Finkelstein_1984}, found in lattice fermion models in Refs.~\cite{Poboiko_2023, Poboiko_2024,Poboiko_2025,Fava_2023,Fava_2024,Guo_2024,Starchl_2024}.  

Interesting open questions are to both characterize the entanglement entropy in the interacting case analytically, and numerically confirm the existence of a critical phase in monitored interacting Dirac fermions. Both these questions are challenging: A direct computation of the entanglement entropy in the critical phase requires a thorough calculation that takes the compact nature of the fields into account. Numerical simulations on the level of the Dirac fermions require a spatial discretization of a manifestly continuous model, and interactions drastically limit the system size. Alternatively, a simulation of the resulting model after bosonization is worth considering. \par
Besides that, the formalism developed in this work paves the way up to understand related problems with the same (or similar) symmetries, such as models with long-ranged hoppings~\cite{Mueller_2022} or interactions, active feedback~\cite{Buchhold_2022}, impurities and others. More broadly, the replica Keldysh field theory approach for weak measurement protocols is applicable to other symmetries or higher dimensional systems, and it will be exciting to explore more of them in the future. 
\section*{Acknowledgements}
We thank Alex Altland, Romain Daviet, Johannes Lang, Carl Zelle and Martin Zirnbauer for fruitful discussions.
We acknowledge support from the Deutsche Forschungsgemeinschaft (DFG, German Research Foundation)
under Germany’s Excellence Strategy Cluster of
Excellence Matter and Light for Quantum Computing
(ML4Q) EXC 2004/1 390534769, and by the DFG
Collaborative Research Center (CRC) 183 Project
No. 277101999 -- project B02. 
TM acknowledges support from ERC under grant agreement n.101053159 (RAVE).
\bibliography{Bib}
\onecolumngrid
\appendix
\include{appendix}
\end{document}

%% file: Modelsketch.tex
  
\tikzset {_m063fkidr/.code = {\pgfsetadditionalshadetransform{ \pgftransformshift{\pgfpoint{0 bp } { 0 bp }  }  \pgftransformrotate{0 }  \pgftransformscale{2 }  }}}
\pgfdeclarehorizontalshading{_0xafkqvxj}{150bp}{rgb(0bp)=(0.29,0.56,0.89);
rgb(37.5bp)=(0.29,0.56,0.89);
rgb(49.75bp)=(0.29,0.56,0.89);
rgb(50bp)=(0.82,0.01,0.11);
rgb(62.5bp)=(0.82,0.01,0.11);
rgb(100bp)=(0.82,0.01,0.11)}

  
\tikzset {_ax93prglk/.code = {\pgfsetadditionalshadetransform{ \pgftransformshift{\pgfpoint{0 bp } { 0 bp }  }  \pgftransformrotate{0 }  \pgftransformscale{2 }  }}}
\pgfdeclarehorizontalshading{_n864rhe1y}{150bp}{rgb(0bp)=(0.29,0.56,0.89);
rgb(37.5bp)=(0.29,0.56,0.89);
rgb(49.75bp)=(0.29,0.56,0.89);
rgb(50bp)=(0.82,0.01,0.11);
rgb(62.5bp)=(0.82,0.01,0.11);
rgb(100bp)=(0.82,0.01,0.11)}

  
\tikzset {_0kh4bbm47/.code = {\pgfsetadditionalshadetransform{ \pgftransformshift{\pgfpoint{0 bp } { 0 bp }  }  \pgftransformrotate{0 }  \pgftransformscale{2 }  }}}
\pgfdeclarehorizontalshading{_z61uyweh5}{150bp}{rgb(0bp)=(0.29,0.56,0.89);
rgb(37.5bp)=(0.29,0.56,0.89);
rgb(49.75bp)=(0.29,0.56,0.89);
rgb(50bp)=(0.82,0.01,0.11);
rgb(62.5bp)=(0.82,0.01,0.11);
rgb(100bp)=(0.82,0.01,0.11)}

  
\tikzset {_ish55w2lu/.code = {\pgfsetadditionalshadetransform{ \pgftransformshift{\pgfpoint{0 bp } { 0 bp }  }  \pgftransformrotate{0 }  \pgftransformscale{2 }  }}}
\pgfdeclarehorizontalshading{_v47vt7j17}{150bp}{rgb(0bp)=(0.29,0.56,0.89);
rgb(37.5bp)=(0.29,0.56,0.89);
rgb(49.75bp)=(0.29,0.56,0.89);
rgb(50bp)=(0.82,0.01,0.11);
rgb(62.5bp)=(0.82,0.01,0.11);
rgb(100bp)=(0.82,0.01,0.11)}

  
\tikzset {_78tqxihig/.code = {\pgfsetadditionalshadetransform{ \pgftransformshift{\pgfpoint{0 bp } { 0 bp }  }  \pgftransformrotate{0 }  \pgftransformscale{2 }  }}}
\pgfdeclarehorizontalshading{_iw8jwknve}{150bp}{rgb(0bp)=(0.29,0.56,0.89);
rgb(37.5bp)=(0.29,0.56,0.89);
rgb(49.75bp)=(0.29,0.56,0.89);
rgb(50bp)=(0.82,0.01,0.11);
rgb(62.5bp)=(0.82,0.01,0.11);
rgb(100bp)=(0.82,0.01,0.11)}

  
\tikzset {_pucbrqgou/.code = {\pgfsetadditionalshadetransform{ \pgftransformshift{\pgfpoint{0 bp } { 0 bp }  }  \pgftransformrotate{0 }  \pgftransformscale{2 }  }}}
\pgfdeclarehorizontalshading{_nvopac2in}{150bp}{rgb(0bp)=(0.29,0.56,0.89);
rgb(37.5bp)=(0.29,0.56,0.89);
rgb(49.75bp)=(0.29,0.56,0.89);
rgb(50bp)=(0.82,0.01,0.11);
rgb(62.5bp)=(0.82,0.01,0.11);
rgb(100bp)=(0.82,0.01,0.11)}

  
\tikzset {_cj9esbe9r/.code = {\pgfsetadditionalshadetransform{ \pgftransformshift{\pgfpoint{0 bp } { 0 bp }  }  \pgftransformrotate{0 }  \pgftransformscale{2 }  }}}
\pgfdeclarehorizontalshading{_qbaxayynw}{150bp}{rgb(0bp)=(0.29,0.56,0.89);
rgb(37.5bp)=(0.29,0.56,0.89);
rgb(49.75bp)=(0.29,0.56,0.89);
rgb(50bp)=(0.82,0.01,0.11);
rgb(62.5bp)=(0.82,0.01,0.11);
rgb(100bp)=(0.82,0.01,0.11)}

  
\tikzset {_pj4bq2vq8/.code = {\pgfsetadditionalshadetransform{ \pgftransformshift{\pgfpoint{0 bp } { 0 bp }  }  \pgftransformrotate{0 }  \pgftransformscale{2 }  }}}
\pgfdeclarehorizontalshading{_qpdxhav6t}{150bp}{rgb(0bp)=(0.29,0.56,0.89);
rgb(37.5bp)=(0.29,0.56,0.89);
rgb(49.75bp)=(0.29,0.56,0.89);
rgb(50bp)=(0.82,0.01,0.11);
rgb(62.5bp)=(0.82,0.01,0.11);
rgb(100bp)=(0.82,0.01,0.11)}

  
\tikzset {_3pmu19gsq/.code = {\pgfsetadditionalshadetransform{ \pgftransformshift{\pgfpoint{0 bp } { 0 bp }  }  \pgftransformrotate{0 }  \pgftransformscale{2 }  }}}
\pgfdeclarehorizontalshading{_qox4ho64o}{150bp}{rgb(0bp)=(0.29,0.56,0.89);
rgb(37.5bp)=(0.29,0.56,0.89);
rgb(49.75bp)=(0.29,0.56,0.89);
rgb(50bp)=(0.82,0.01,0.11);
rgb(62.5bp)=(0.82,0.01,0.11);
rgb(100bp)=(0.82,0.01,0.11)}

  
\tikzset {_ye9hixqgo/.code = {\pgfsetadditionalshadetransform{ \pgftransformshift{\pgfpoint{0 bp } { 0 bp }  }  \pgftransformrotate{0 }  \pgftransformscale{2 }  }}}
\pgfdeclarehorizontalshading{_r02sv9rhy}{150bp}{rgb(0bp)=(0.29,0.56,0.89);
rgb(37.5bp)=(0.29,0.56,0.89);
rgb(49.75bp)=(0.29,0.56,0.89);
rgb(50bp)=(0.82,0.01,0.11);
rgb(62.5bp)=(0.82,0.01,0.11);
rgb(100bp)=(0.82,0.01,0.11)}

  
\tikzset {_ii27ucmlu/.code = {\pgfsetadditionalshadetransform{ \pgftransformshift{\pgfpoint{0 bp } { 0 bp }  }  \pgftransformrotate{0 }  \pgftransformscale{2 }  }}}
\pgfdeclarehorizontalshading{_58areozb5}{150bp}{rgb(0bp)=(0.29,0.56,0.89);
rgb(37.5bp)=(0.29,0.56,0.89);
rgb(49.75bp)=(0.29,0.56,0.89);
rgb(50bp)=(0.82,0.01,0.11);
rgb(62.5bp)=(0.82,0.01,0.11);
rgb(100bp)=(0.82,0.01,0.11)}

  
\tikzset {_dxw0rlgh4/.code = {\pgfsetadditionalshadetransform{ \pgftransformshift{\pgfpoint{0 bp } { 0 bp }  }  \pgftransformrotate{0 }  \pgftransformscale{2 }  }}}
\pgfdeclarehorizontalshading{_rkmoepgoz}{150bp}{rgb(0bp)=(0.29,0.56,0.89);
rgb(37.5bp)=(0.29,0.56,0.89);
rgb(49.75bp)=(0.29,0.56,0.89);
rgb(50bp)=(0.82,0.01,0.11);
rgb(62.5bp)=(0.82,0.01,0.11);
rgb(100bp)=(0.82,0.01,0.11)}

  
\tikzset {_b77hi1t3a/.code = {\pgfsetadditionalshadetransform{ \pgftransformshift{\pgfpoint{0 bp } { 0 bp }  }  \pgftransformrotate{0 }  \pgftransformscale{2 }  }}}
\pgfdeclarehorizontalshading{_nbjh7vnb0}{150bp}{rgb(0bp)=(0.29,0.56,0.89);
rgb(37.5bp)=(0.29,0.56,0.89);
rgb(49.75bp)=(0.29,0.56,0.89);
rgb(50bp)=(0.82,0.01,0.11);
rgb(62.5bp)=(0.82,0.01,0.11);
rgb(100bp)=(0.82,0.01,0.11)}

  
\tikzset {_rvgd7r7x9/.code = {\pgfsetadditionalshadetransform{ \pgftransformshift{\pgfpoint{0 bp } { 0 bp }  }  \pgftransformrotate{0 }  \pgftransformscale{2 }  }}}
\pgfdeclarehorizontalshading{_1uxpe801q}{150bp}{rgb(0bp)=(0.29,0.56,0.89);
rgb(37.5bp)=(0.29,0.56,0.89);
rgb(49.75bp)=(0.29,0.56,0.89);
rgb(50bp)=(0.82,0.01,0.11);
rgb(62.5bp)=(0.82,0.01,0.11);
rgb(100bp)=(0.82,0.01,0.11)}

  
\tikzset {_gxr821z3s/.code = {\pgfsetadditionalshadetransform{ \pgftransformshift{\pgfpoint{0 bp } { 0 bp }  }  \pgftransformrotate{0 }  \pgftransformscale{2 }  }}}
\pgfdeclarehorizontalshading{_vq09e9enk}{150bp}{rgb(0bp)=(0.29,0.56,0.89);
rgb(37.5bp)=(0.29,0.56,0.89);
rgb(49.75bp)=(0.29,0.56,0.89);
rgb(50bp)=(0.82,0.01,0.11);
rgb(62.5bp)=(0.82,0.01,0.11);
rgb(100bp)=(0.82,0.01,0.11)}

  
\tikzset {_teaqiz7dt/.code = {\pgfsetadditionalshadetransform{ \pgftransformshift{\pgfpoint{0 bp } { 0 bp }  }  \pgftransformrotate{0 }  \pgftransformscale{2 }  }}}
\pgfdeclarehorizontalshading{_88fxp42ey}{150bp}{rgb(0bp)=(0.29,0.56,0.89);
rgb(37.5bp)=(0.29,0.56,0.89);
rgb(49.75bp)=(0.29,0.56,0.89);
rgb(50bp)=(0.82,0.01,0.11);
rgb(62.5bp)=(0.82,0.01,0.11);
rgb(100bp)=(0.82,0.01,0.11)}

  
\tikzset {_6m755k16h/.code = {\pgfsetadditionalshadetransform{ \pgftransformshift{\pgfpoint{0 bp } { 0 bp }  }  \pgftransformrotate{0 }  \pgftransformscale{2 }  }}}
\pgfdeclarehorizontalshading{_c6t6012as}{150bp}{rgb(0bp)=(0.29,0.56,0.89);
rgb(37.5bp)=(0.29,0.56,0.89);
rgb(49.75bp)=(0.29,0.56,0.89);
rgb(50bp)=(0.82,0.01,0.11);
rgb(62.5bp)=(0.82,0.01,0.11);
rgb(100bp)=(0.82,0.01,0.11)}

  
\tikzset {_km5oetg9i/.code = {\pgfsetadditionalshadetransform{ \pgftransformshift{\pgfpoint{0 bp } { 0 bp }  }  \pgftransformrotate{0 }  \pgftransformscale{2 }  }}}
\pgfdeclarehorizontalshading{_8tgdt0fyc}{150bp}{rgb(0bp)=(0.29,0.56,0.89);
rgb(37.5bp)=(0.29,0.56,0.89);
rgb(49.75bp)=(0.29,0.56,0.89);
rgb(50bp)=(0.82,0.01,0.11);
rgb(62.5bp)=(0.82,0.01,0.11);
rgb(100bp)=(0.82,0.01,0.11)}

  
\tikzset {_4wx078wbt/.code = {\pgfsetadditionalshadetransform{ \pgftransformshift{\pgfpoint{0 bp } { 0 bp }  }  \pgftransformrotate{0 }  \pgftransformscale{2 }  }}}
\pgfdeclarehorizontalshading{_c687h7ytd}{150bp}{rgb(0bp)=(0.29,0.56,0.89);
rgb(37.5bp)=(0.29,0.56,0.89);
rgb(49.75bp)=(0.29,0.56,0.89);
rgb(50bp)=(0.82,0.01,0.11);
rgb(62.5bp)=(0.82,0.01,0.11);
rgb(100bp)=(0.82,0.01,0.11)}

  
\tikzset {_5yyt9o3f9/.code = {\pgfsetadditionalshadetransform{ \pgftransformshift{\pgfpoint{0 bp } { 0 bp }  }  \pgftransformrotate{0 }  \pgftransformscale{2 }  }}}
\pgfdeclarehorizontalshading{_61czy95vn}{150bp}{rgb(0bp)=(0.29,0.56,0.89);
rgb(37.5bp)=(0.29,0.56,0.89);
rgb(49.75bp)=(0.29,0.56,0.89);
rgb(50bp)=(0.82,0.01,0.11);
rgb(62.5bp)=(0.82,0.01,0.11);
rgb(100bp)=(0.82,0.01,0.11)}

  
\tikzset {_t0yi688qf/.code = {\pgfsetadditionalshadetransform{ \pgftransformshift{\pgfpoint{0 bp } { 0 bp }  }  \pgftransformrotate{0 }  \pgftransformscale{2 }  }}}
\pgfdeclarehorizontalshading{_oszzpkw1f}{150bp}{rgb(0bp)=(0.29,0.56,0.89);
rgb(37.5bp)=(0.29,0.56,0.89);
rgb(49.75bp)=(0.29,0.56,0.89);
rgb(50bp)=(0.82,0.01,0.11);
rgb(62.5bp)=(0.82,0.01,0.11);
rgb(100bp)=(0.82,0.01,0.11)}

  
\tikzset {_pnh4bc92y/.code = {\pgfsetadditionalshadetransform{ \pgftransformshift{\pgfpoint{0 bp } { 0 bp }  }  \pgftransformrotate{0 }  \pgftransformscale{2 }  }}}
\pgfdeclarehorizontalshading{_64jrfhql7}{150bp}{rgb(0bp)=(0.29,0.56,0.89);
rgb(37.5bp)=(0.29,0.56,0.89);
rgb(49.75bp)=(0.29,0.56,0.89);
rgb(50bp)=(0.82,0.01,0.11);
rgb(62.5bp)=(0.82,0.01,0.11);
rgb(100bp)=(0.82,0.01,0.11)}

  
\tikzset {_065mtx5sv/.code = {\pgfsetadditionalshadetransform{ \pgftransformshift{\pgfpoint{0 bp } { 0 bp }  }  \pgftransformrotate{0 }  \pgftransformscale{2 }  }}}
\pgfdeclarehorizontalshading{_u8305ya2u}{150bp}{rgb(0bp)=(0.29,0.56,0.89);
rgb(37.5bp)=(0.29,0.56,0.89);
rgb(49.75bp)=(0.29,0.56,0.89);
rgb(50bp)=(0.82,0.01,0.11);
rgb(62.5bp)=(0.82,0.01,0.11);
rgb(100bp)=(0.82,0.01,0.11)}

  
\tikzset {_uyffxq5ch/.code = {\pgfsetadditionalshadetransform{ \pgftransformshift{\pgfpoint{0 bp } { 0 bp }  }  \pgftransformrotate{0 }  \pgftransformscale{2 }  }}}
\pgfdeclarehorizontalshading{_30tzevrxf}{150bp}{rgb(0bp)=(0.29,0.56,0.89);
rgb(37.5bp)=(0.29,0.56,0.89);
rgb(49.75bp)=(0.29,0.56,0.89);
rgb(50bp)=(0.82,0.01,0.11);
rgb(62.5bp)=(0.82,0.01,0.11);
rgb(100bp)=(0.82,0.01,0.11)}

  
\tikzset {_og8jei06z/.code = {\pgfsetadditionalshadetransform{ \pgftransformshift{\pgfpoint{0 bp } { 0 bp }  }  \pgftransformrotate{0 }  \pgftransformscale{2 }  }}}
\pgfdeclarehorizontalshading{_dgefh48te}{150bp}{rgb(0bp)=(0.29,0.56,0.89);
rgb(37.5bp)=(0.29,0.56,0.89);
rgb(49.75bp)=(0.29,0.56,0.89);
rgb(50bp)=(0.82,0.01,0.11);
rgb(62.5bp)=(0.82,0.01,0.11);
rgb(100bp)=(0.82,0.01,0.11)}

  
\tikzset {_hzo6ber61/.code = {\pgfsetadditionalshadetransform{ \pgftransformshift{\pgfpoint{0 bp } { 0 bp }  }  \pgftransformrotate{0 }  \pgftransformscale{2 }  }}}
\pgfdeclarehorizontalshading{_pszdrkfgn}{150bp}{rgb(0bp)=(0.29,0.56,0.89);
rgb(37.5bp)=(0.29,0.56,0.89);
rgb(49.75bp)=(0.29,0.56,0.89);
rgb(50bp)=(0.82,0.01,0.11);
rgb(62.5bp)=(0.82,0.01,0.11);
rgb(100bp)=(0.82,0.01,0.11)}

  
\tikzset {_po66myqwb/.code = {\pgfsetadditionalshadetransform{ \pgftransformshift{\pgfpoint{0 bp } { 0 bp }  }  \pgftransformrotate{0 }  \pgftransformscale{2 }  }}}
\pgfdeclarehorizontalshading{_qff0shcbe}{150bp}{rgb(0bp)=(0.29,0.56,0.89);
rgb(37.5bp)=(0.29,0.56,0.89);
rgb(49.75bp)=(0.29,0.56,0.89);
rgb(50bp)=(0.82,0.01,0.11);
rgb(62.5bp)=(0.82,0.01,0.11);
rgb(100bp)=(0.82,0.01,0.11)}

  
\tikzset {_d7ad8xqy8/.code = {\pgfsetadditionalshadetransform{ \pgftransformshift{\pgfpoint{0 bp } { 0 bp }  }  \pgftransformrotate{0 }  \pgftransformscale{2 }  }}}
\pgfdeclarehorizontalshading{_5kqdlzf47}{150bp}{rgb(0bp)=(0.29,0.56,0.89);
rgb(37.5bp)=(0.29,0.56,0.89);
rgb(49.75bp)=(0.29,0.56,0.89);
rgb(50bp)=(0.82,0.01,0.11);
rgb(62.5bp)=(0.82,0.01,0.11);
rgb(100bp)=(0.82,0.01,0.11)}

  
\tikzset {_gmlhf5744/.code = {\pgfsetadditionalshadetransform{ \pgftransformshift{\pgfpoint{0 bp } { 0 bp }  }  \pgftransformrotate{0 }  \pgftransformscale{2 }  }}}
\pgfdeclarehorizontalshading{_bf61zpe1k}{150bp}{rgb(0bp)=(0.29,0.56,0.89);
rgb(37.5bp)=(0.29,0.56,0.89);
rgb(49.75bp)=(0.29,0.56,0.89);
rgb(50bp)=(0.82,0.01,0.11);
rgb(62.5bp)=(0.82,0.01,0.11);
rgb(100bp)=(0.82,0.01,0.11)}

  
\tikzset {_ycqkwssw6/.code = {\pgfsetadditionalshadetransform{ \pgftransformshift{\pgfpoint{0 bp } { 0 bp }  }  \pgftransformrotate{0 }  \pgftransformscale{2 }  }}}
\pgfdeclarehorizontalshading{_7ot6fj95x}{150bp}{rgb(0bp)=(0.29,0.56,0.89);
rgb(37.5bp)=(0.29,0.56,0.89);
rgb(49.75bp)=(0.29,0.56,0.89);
rgb(50bp)=(0.82,0.01,0.11);
rgb(62.5bp)=(0.82,0.01,0.11);
rgb(100bp)=(0.82,0.01,0.11)}
\tikzset{every picture/.style={line width=0.75pt}} 

\begin{tikzpicture}[x=0.75pt,y=0.75pt,yscale=-1,xscale=1]

\draw    (130,103.5) -- (422,103.5)(130,106.5) -- (422,106.5) ;
\draw [shift={(430,105)}, rotate = 180] [color={rgb, 255:red, 0; green, 0; blue, 0 }  ][line width=0.75]    (10.93,-3.29) .. controls (6.95,-1.4) and (3.31,-0.3) .. (0,0) .. controls (3.31,0.3) and (6.95,1.4) .. (10.93,3.29)   ;
\draw  [fill={rgb, 255:red, 208; green, 2; blue, 27 }  ,fill opacity=1 ] (150,105) .. controls (150,102.24) and (152.24,100) .. (155,100) .. controls (157.76,100) and (160,102.24) .. (160,105) .. controls (160,107.76) and (157.76,110) .. (155,110) .. controls (152.24,110) and (150,107.76) .. (150,105) -- cycle ;
\draw  [fill={rgb, 255:red, 74; green, 144; blue, 226 }  ,fill opacity=1 ] (170,105) .. controls (170,102.24) and (172.24,100) .. (175,100) .. controls (177.76,100) and (180,102.24) .. (180,105) .. controls (180,107.76) and (177.76,110) .. (175,110) .. controls (172.24,110) and (170,107.76) .. (170,105) -- cycle ;
\draw  [fill={rgb, 255:red, 208; green, 2; blue, 27 }  ,fill opacity=1 ] (200,105) .. controls (200,102.24) and (202.24,100) .. (205,100) .. controls (207.76,100) and (210,102.24) .. (210,105) .. controls (210,107.76) and (207.76,110) .. (205,110) .. controls (202.24,110) and (200,107.76) .. (200,105) -- cycle ;
\draw  [fill={rgb, 255:red, 74; green, 144; blue, 226 }  ,fill opacity=1 ] (241,105) .. controls (241,102.24) and (243.24,100) .. (246,100) .. controls (248.76,100) and (251,102.24) .. (251,105) .. controls (251,107.76) and (248.76,110) .. (246,110) .. controls (243.24,110) and (241,107.76) .. (241,105) -- cycle ;
\draw  [fill={rgb, 255:red, 74; green, 144; blue, 226 }  ,fill opacity=1 ] (230,105) .. controls (230,102.24) and (232.24,100) .. (235,100) .. controls (237.76,100) and (240,102.24) .. (240,105) .. controls (240,107.76) and (237.76,110) .. (235,110) .. controls (232.24,110) and (230,107.76) .. (230,105) -- cycle ;
\draw  [fill={rgb, 255:red, 208; green, 2; blue, 27 }  ,fill opacity=1 ] (280,105) .. controls (280,102.24) and (282.24,100) .. (285,100) .. controls (287.76,100) and (290,102.24) .. (290,105) .. controls (290,107.76) and (287.76,110) .. (285,110) .. controls (282.24,110) and (280,107.76) .. (280,105) -- cycle ;
\draw  [fill={rgb, 255:red, 208; green, 2; blue, 27 }  ,fill opacity=1 ] (300,105) .. controls (300,102.24) and (302.24,100) .. (305,100) .. controls (307.76,100) and (310,102.24) .. (310,105) .. controls (310,107.76) and (307.76,110) .. (305,110) .. controls (302.24,110) and (300,107.76) .. (300,105) -- cycle ;
\draw  [fill={rgb, 255:red, 74; green, 144; blue, 226 }  ,fill opacity=1 ] (313,105) .. controls (313,102.24) and (315.24,100) .. (318,100) .. controls (320.76,100) and (323,102.24) .. (323,105) .. controls (323,107.76) and (320.76,110) .. (318,110) .. controls (315.24,110) and (313,107.76) .. (313,105) -- cycle ;
\path  [shading=_0xafkqvxj,_m063fkidr] (130,85) .. controls (130,82.24) and (132.24,80) .. (135,80) -- (135,80) .. controls (137.76,80) and (140,82.24) .. (140,85) -- (140,90) .. controls (140,90) and (140,90) .. (140,90) -- (130,90) .. controls (130,90) and (130,90) .. (130,90) -- cycle ; 
 \draw   (130,85) .. controls (130,82.24) and (132.24,80) .. (135,80) -- (135,80) .. controls (137.76,80) and (140,82.24) .. (140,85) -- (140,90) .. controls (140,90) and (140,90) .. (140,90) -- (130,90) .. controls (130,90) and (130,90) .. (130,90) -- cycle ; 

\draw    (135,80) -- (135,70) ;
\path  [shading=_n864rhe1y,_ax93prglk] (140,85) .. controls (140,82.24) and (142.24,80) .. (145,80) -- (145,80) .. controls (147.76,80) and (150,82.24) .. (150,85) -- (150,90) .. controls (150,90) and (150,90) .. (150,90) -- (140,90) .. controls (140,90) and (140,90) .. (140,90) -- cycle ; 
 \draw   (140,85) .. controls (140,82.24) and (142.24,80) .. (145,80) -- (145,80) .. controls (147.76,80) and (150,82.24) .. (150,85) -- (150,90) .. controls (150,90) and (150,90) .. (150,90) -- (140,90) .. controls (140,90) and (140,90) .. (140,90) -- cycle ; 

\draw    (145,80) -- (145,70) ;
\path  [shading=_z61uyweh5,_0kh4bbm47] (150,85) .. controls (150,82.24) and (152.24,80) .. (155,80) -- (155,80) .. controls (157.76,80) and (160,82.24) .. (160,85) -- (160,90) .. controls (160,90) and (160,90) .. (160,90) -- (150,90) .. controls (150,90) and (150,90) .. (150,90) -- cycle ; 
 \draw   (150,85) .. controls (150,82.24) and (152.24,80) .. (155,80) -- (155,80) .. controls (157.76,80) and (160,82.24) .. (160,85) -- (160,90) .. controls (160,90) and (160,90) .. (160,90) -- (150,90) .. controls (150,90) and (150,90) .. (150,90) -- cycle ; 

\draw    (155,80) -- (155,70) ;
\path  [shading=_v47vt7j17,_ish55w2lu] (160,85) .. controls (160,82.24) and (162.24,80) .. (165,80) -- (165,80) .. controls (167.76,80) and (170,82.24) .. (170,85) -- (170,90) .. controls (170,90) and (170,90) .. (170,90) -- (160,90) .. controls (160,90) and (160,90) .. (160,90) -- cycle ; 
 \draw   (160,85) .. controls (160,82.24) and (162.24,80) .. (165,80) -- (165,80) .. controls (167.76,80) and (170,82.24) .. (170,85) -- (170,90) .. controls (170,90) and (170,90) .. (170,90) -- (160,90) .. controls (160,90) and (160,90) .. (160,90) -- cycle ; 

\draw    (165,80) -- (165,70) ;
\path  [shading=_iw8jwknve,_78tqxihig] (170,85) .. controls (170,82.24) and (172.24,80) .. (175,80) -- (175,80) .. controls (177.76,80) and (180,82.24) .. (180,85) -- (180,90) .. controls (180,90) and (180,90) .. (180,90) -- (170,90) .. controls (170,90) and (170,90) .. (170,90) -- cycle ; 
 \draw   (170,85) .. controls (170,82.24) and (172.24,80) .. (175,80) -- (175,80) .. controls (177.76,80) and (180,82.24) .. (180,85) -- (180,90) .. controls (180,90) and (180,90) .. (180,90) -- (170,90) .. controls (170,90) and (170,90) .. (170,90) -- cycle ; 

\draw    (175,80) -- (175,70) ;
\path  [shading=_nvopac2in,_pucbrqgou] (180,85) .. controls (180,82.24) and (182.24,80) .. (185,80) -- (185,80) .. controls (187.76,80) and (190,82.24) .. (190,85) -- (190,90) .. controls (190,90) and (190,90) .. (190,90) -- (180,90) .. controls (180,90) and (180,90) .. (180,90) -- cycle ; 
 \draw   (180,85) .. controls (180,82.24) and (182.24,80) .. (185,80) -- (185,80) .. controls (187.76,80) and (190,82.24) .. (190,85) -- (190,90) .. controls (190,90) and (190,90) .. (190,90) -- (180,90) .. controls (180,90) and (180,90) .. (180,90) -- cycle ; 

\draw    (185,80) -- (185,70) ;
\path  [shading=_qbaxayynw,_cj9esbe9r] (190,85) .. controls (190,82.24) and (192.24,80) .. (195,80) -- (195,80) .. controls (197.76,80) and (200,82.24) .. (200,85) -- (200,90) .. controls (200,90) and (200,90) .. (200,90) -- (190,90) .. controls (190,90) and (190,90) .. (190,90) -- cycle ; 
 \draw   (190,85) .. controls (190,82.24) and (192.24,80) .. (195,80) -- (195,80) .. controls (197.76,80) and (200,82.24) .. (200,85) -- (200,90) .. controls (200,90) and (200,90) .. (200,90) -- (190,90) .. controls (190,90) and (190,90) .. (190,90) -- cycle ; 

\draw    (195,80) -- (195,70) ;
\path  [shading=_qpdxhav6t,_pj4bq2vq8] (200,85) .. controls (200,82.24) and (202.24,80) .. (205,80) -- (205,80) .. controls (207.76,80) and (210,82.24) .. (210,85) -- (210,90) .. controls (210,90) and (210,90) .. (210,90) -- (200,90) .. controls (200,90) and (200,90) .. (200,90) -- cycle ; 
 \draw   (200,85) .. controls (200,82.24) and (202.24,80) .. (205,80) -- (205,80) .. controls (207.76,80) and (210,82.24) .. (210,85) -- (210,90) .. controls (210,90) and (210,90) .. (210,90) -- (200,90) .. controls (200,90) and (200,90) .. (200,90) -- cycle ; 

\draw    (205,80) -- (205,70) ;
\path  [shading=_qox4ho64o,_3pmu19gsq] (210,85) .. controls (210,82.24) and (212.24,80) .. (215,80) -- (215,80) .. controls (217.76,80) and (220,82.24) .. (220,85) -- (220,90) .. controls (220,90) and (220,90) .. (220,90) -- (210,90) .. controls (210,90) and (210,90) .. (210,90) -- cycle ; 
 \draw   (210,85) .. controls (210,82.24) and (212.24,80) .. (215,80) -- (215,80) .. controls (217.76,80) and (220,82.24) .. (220,85) -- (220,90) .. controls (220,90) and (220,90) .. (220,90) -- (210,90) .. controls (210,90) and (210,90) .. (210,90) -- cycle ; 

\draw    (215,80) -- (215,70) ;
\path  [shading=_r02sv9rhy,_ye9hixqgo] (220,85) .. controls (220,82.24) and (222.24,80) .. (225,80) -- (225,80) .. controls (227.76,80) and (230,82.24) .. (230,85) -- (230,90) .. controls (230,90) and (230,90) .. (230,90) -- (220,90) .. controls (220,90) and (220,90) .. (220,90) -- cycle ; 
 \draw   (220,85) .. controls (220,82.24) and (222.24,80) .. (225,80) -- (225,80) .. controls (227.76,80) and (230,82.24) .. (230,85) -- (230,90) .. controls (230,90) and (230,90) .. (230,90) -- (220,90) .. controls (220,90) and (220,90) .. (220,90) -- cycle ; 

\draw    (225,80) -- (225,70) ;
\path  [shading=_58areozb5,_ii27ucmlu] (230,85) .. controls (230,82.24) and (232.24,80) .. (235,80) -- (235,80) .. controls (237.76,80) and (240,82.24) .. (240,85) -- (240,90) .. controls (240,90) and (240,90) .. (240,90) -- (230,90) .. controls (230,90) and (230,90) .. (230,90) -- cycle ; 
 \draw   (230,85) .. controls (230,82.24) and (232.24,80) .. (235,80) -- (235,80) .. controls (237.76,80) and (240,82.24) .. (240,85) -- (240,90) .. controls (240,90) and (240,90) .. (240,90) -- (230,90) .. controls (230,90) and (230,90) .. (230,90) -- cycle ; 

\draw    (235,80) -- (235,70) ;
\path  [shading=_rkmoepgoz,_dxw0rlgh4] (240,85) .. controls (240,82.24) and (242.24,80) .. (245,80) -- (245,80) .. controls (247.76,80) and (250,82.24) .. (250,85) -- (250,90) .. controls (250,90) and (250,90) .. (250,90) -- (240,90) .. controls (240,90) and (240,90) .. (240,90) -- cycle ; 
 \draw   (240,85) .. controls (240,82.24) and (242.24,80) .. (245,80) -- (245,80) .. controls (247.76,80) and (250,82.24) .. (250,85) -- (250,90) .. controls (250,90) and (250,90) .. (250,90) -- (240,90) .. controls (240,90) and (240,90) .. (240,90) -- cycle ; 

\draw    (245,80) -- (245,70) ;
\path  [shading=_nbjh7vnb0,_b77hi1t3a] (250,85) .. controls (250,82.24) and (252.24,80) .. (255,80) -- (255,80) .. controls (257.76,80) and (260,82.24) .. (260,85) -- (260,90) .. controls (260,90) and (260,90) .. (260,90) -- (250,90) .. controls (250,90) and (250,90) .. (250,90) -- cycle ; 
 \draw   (250,85) .. controls (250,82.24) and (252.24,80) .. (255,80) -- (255,80) .. controls (257.76,80) and (260,82.24) .. (260,85) -- (260,90) .. controls (260,90) and (260,90) .. (260,90) -- (250,90) .. controls (250,90) and (250,90) .. (250,90) -- cycle ; 

\draw    (255,80) -- (255,70) ;
\path  [shading=_1uxpe801q,_rvgd7r7x9] (260,85) .. controls (260,82.24) and (262.24,80) .. (265,80) -- (265,80) .. controls (267.76,80) and (270,82.24) .. (270,85) -- (270,90) .. controls (270,90) and (270,90) .. (270,90) -- (260,90) .. controls (260,90) and (260,90) .. (260,90) -- cycle ; 
 \draw   (260,85) .. controls (260,82.24) and (262.24,80) .. (265,80) -- (265,80) .. controls (267.76,80) and (270,82.24) .. (270,85) -- (270,90) .. controls (270,90) and (270,90) .. (270,90) -- (260,90) .. controls (260,90) and (260,90) .. (260,90) -- cycle ; 

\draw    (265,80) -- (265,70) ;
\path  [shading=_vq09e9enk,_gxr821z3s] (270,85) .. controls (270,82.24) and (272.24,80) .. (275,80) -- (275,80) .. controls (277.76,80) and (280,82.24) .. (280,85) -- (280,90) .. controls (280,90) and (280,90) .. (280,90) -- (270,90) .. controls (270,90) and (270,90) .. (270,90) -- cycle ; 
 \draw   (270,85) .. controls (270,82.24) and (272.24,80) .. (275,80) -- (275,80) .. controls (277.76,80) and (280,82.24) .. (280,85) -- (280,90) .. controls (280,90) and (280,90) .. (280,90) -- (270,90) .. controls (270,90) and (270,90) .. (270,90) -- cycle ; 

\draw    (275,80) -- (275,70) ;
\path  [shading=_88fxp42ey,_teaqiz7dt] (280,85) .. controls (280,82.24) and (282.24,80) .. (285,80) -- (285,80) .. controls (287.76,80) and (290,82.24) .. (290,85) -- (290,90) .. controls (290,90) and (290,90) .. (290,90) -- (280,90) .. controls (280,90) and (280,90) .. (280,90) -- cycle ; 
 \draw   (280,85) .. controls (280,82.24) and (282.24,80) .. (285,80) -- (285,80) .. controls (287.76,80) and (290,82.24) .. (290,85) -- (290,90) .. controls (290,90) and (290,90) .. (290,90) -- (280,90) .. controls (280,90) and (280,90) .. (280,90) -- cycle ; 

\draw    (285,80) -- (285,70) ;
\path  [shading=_c6t6012as,_6m755k16h] (290,85) .. controls (290,82.24) and (292.24,80) .. (295,80) -- (295,80) .. controls (297.76,80) and (300,82.24) .. (300,85) -- (300,90) .. controls (300,90) and (300,90) .. (300,90) -- (290,90) .. controls (290,90) and (290,90) .. (290,90) -- cycle ; 
 \draw   (290,85) .. controls (290,82.24) and (292.24,80) .. (295,80) -- (295,80) .. controls (297.76,80) and (300,82.24) .. (300,85) -- (300,90) .. controls (300,90) and (300,90) .. (300,90) -- (290,90) .. controls (290,90) and (290,90) .. (290,90) -- cycle ; 

\draw    (295,80) -- (295,70) ;
\path  [shading=_8tgdt0fyc,_km5oetg9i] (300,85) .. controls (300,82.24) and (302.24,80) .. (305,80) -- (305,80) .. controls (307.76,80) and (310,82.24) .. (310,85) -- (310,90) .. controls (310,90) and (310,90) .. (310,90) -- (300,90) .. controls (300,90) and (300,90) .. (300,90) -- cycle ; 
 \draw   (300,85) .. controls (300,82.24) and (302.24,80) .. (305,80) -- (305,80) .. controls (307.76,80) and (310,82.24) .. (310,85) -- (310,90) .. controls (310,90) and (310,90) .. (310,90) -- (300,90) .. controls (300,90) and (300,90) .. (300,90) -- cycle ; 

\draw    (305,80) -- (305,70) ;
\path  [shading=_c687h7ytd,_4wx078wbt] (310,85) .. controls (310,82.24) and (312.24,80) .. (315,80) -- (315,80) .. controls (317.76,80) and (320,82.24) .. (320,85) -- (320,90) .. controls (320,90) and (320,90) .. (320,90) -- (310,90) .. controls (310,90) and (310,90) .. (310,90) -- cycle ; 
 \draw   (310,85) .. controls (310,82.24) and (312.24,80) .. (315,80) -- (315,80) .. controls (317.76,80) and (320,82.24) .. (320,85) -- (320,90) .. controls (320,90) and (320,90) .. (320,90) -- (310,90) .. controls (310,90) and (310,90) .. (310,90) -- cycle ; 

\draw    (315,80) -- (315,70) ;
\path  [shading=_61czy95vn,_5yyt9o3f9] (320,85) .. controls (320,82.24) and (322.24,80) .. (325,80) -- (325,80) .. controls (327.76,80) and (330,82.24) .. (330,85) -- (330,90) .. controls (330,90) and (330,90) .. (330,90) -- (320,90) .. controls (320,90) and (320,90) .. (320,90) -- cycle ; 
 \draw   (320,85) .. controls (320,82.24) and (322.24,80) .. (325,80) -- (325,80) .. controls (327.76,80) and (330,82.24) .. (330,85) -- (330,90) .. controls (330,90) and (330,90) .. (330,90) -- (320,90) .. controls (320,90) and (320,90) .. (320,90) -- cycle ; 

\draw    (325,80) -- (325,70) ;
\draw    (130,70) -- (430,70) ;
\draw    (184,70) .. controls (163.29,62.43) and (140.71,60.43) .. (164,46) ;
\draw   (164,30) -- (264,30) -- (264,60) -- (164,60) -- cycle ;
\draw    (378.98,70) .. controls (417.41,63) and (415.12,49.57) .. (398.98,45) ;
\draw   (298.98,30) -- (398.98,30) -- (398.98,60) -- (298.98,60) -- cycle ;
\draw    (207.12,114.19) .. controls (209.01,128.3) and (231.79,127.51) .. (233.85,113.78) ;
\draw [shift={(234,112)}, rotate = 103.41] [color={rgb, 255:red, 0; green, 0; blue, 0 }  ][line width=0.75]    (6.56,-1.97) .. controls (4.17,-0.84) and (1.99,-0.18) .. (0,0) .. controls (1.99,0.18) and (4.17,0.84) .. (6.56,1.97)   ;
\draw [shift={(207,112)}, rotate = 78.33] [color={rgb, 255:red, 0; green, 0; blue, 0 }  ][line width=0.75]    (6.56,-1.97) .. controls (4.17,-0.84) and (1.99,-0.18) .. (0,0) .. controls (1.99,0.18) and (4.17,0.84) .. (6.56,1.97)   ;
\draw  [fill={rgb, 255:red, 208; green, 2; blue, 27 }  ,fill opacity=1 ] (170,11) .. controls (170,8.24) and (172.24,6) .. (175,6) .. controls (177.76,6) and (180,8.24) .. (180,11) .. controls (180,13.76) and (177.76,16) .. (175,16) .. controls (172.24,16) and (170,13.76) .. (170,11) -- cycle ;
\draw  [fill={rgb, 255:red, 74; green, 144; blue, 226 }  ,fill opacity=1 ] (310,11) .. controls (310,8.24) and (312.24,6) .. (315,6) .. controls (317.76,6) and (320,8.24) .. (320,11) .. controls (320,13.76) and (317.76,16) .. (315,16) .. controls (312.24,16) and (310,13.76) .. (310,11) -- cycle ;
\path  [shading=_oszzpkw1f,_t0yi688qf] (330,85) .. controls (330,82.24) and (332.24,80) .. (335,80) -- (335,80) .. controls (337.76,80) and (340,82.24) .. (340,85) -- (340,90) .. controls (340,90) and (340,90) .. (340,90) -- (330,90) .. controls (330,90) and (330,90) .. (330,90) -- cycle ; 
 \draw   (330,85) .. controls (330,82.24) and (332.24,80) .. (335,80) -- (335,80) .. controls (337.76,80) and (340,82.24) .. (340,85) -- (340,90) .. controls (340,90) and (340,90) .. (340,90) -- (330,90) .. controls (330,90) and (330,90) .. (330,90) -- cycle ; 

\draw    (335,80) -- (335,70) ;
\path  [shading=_64jrfhql7,_pnh4bc92y] (340,85) .. controls (340,82.24) and (342.24,80) .. (345,80) -- (345,80) .. controls (347.76,80) and (350,82.24) .. (350,85) -- (350,90) .. controls (350,90) and (350,90) .. (350,90) -- (340,90) .. controls (340,90) and (340,90) .. (340,90) -- cycle ; 
 \draw   (340,85) .. controls (340,82.24) and (342.24,80) .. (345,80) -- (345,80) .. controls (347.76,80) and (350,82.24) .. (350,85) -- (350,90) .. controls (350,90) and (350,90) .. (350,90) -- (340,90) .. controls (340,90) and (340,90) .. (340,90) -- cycle ; 

\draw    (345,80) -- (345,70) ;
\path  [shading=_u8305ya2u,_065mtx5sv] (350,85) .. controls (350,82.24) and (352.24,80) .. (355,80) -- (355,80) .. controls (357.76,80) and (360,82.24) .. (360,85) -- (360,90) .. controls (360,90) and (360,90) .. (360,90) -- (350,90) .. controls (350,90) and (350,90) .. (350,90) -- cycle ; 
 \draw   (350,85) .. controls (350,82.24) and (352.24,80) .. (355,80) -- (355,80) .. controls (357.76,80) and (360,82.24) .. (360,85) -- (360,90) .. controls (360,90) and (360,90) .. (360,90) -- (350,90) .. controls (350,90) and (350,90) .. (350,90) -- cycle ; 

\draw    (355,80) -- (355,70) ;
\path  [shading=_30tzevrxf,_uyffxq5ch] (360,85) .. controls (360,82.24) and (362.24,80) .. (365,80) -- (365,80) .. controls (367.76,80) and (370,82.24) .. (370,85) -- (370,90) .. controls (370,90) and (370,90) .. (370,90) -- (360,90) .. controls (360,90) and (360,90) .. (360,90) -- cycle ; 
 \draw   (360,85) .. controls (360,82.24) and (362.24,80) .. (365,80) -- (365,80) .. controls (367.76,80) and (370,82.24) .. (370,85) -- (370,90) .. controls (370,90) and (370,90) .. (370,90) -- (360,90) .. controls (360,90) and (360,90) .. (360,90) -- cycle ; 

\draw    (365,80) -- (365,70) ;
\path  [shading=_dgefh48te,_og8jei06z] (370,85) .. controls (370,82.24) and (372.24,80) .. (375,80) -- (375,80) .. controls (377.76,80) and (380,82.24) .. (380,85) -- (380,90) .. controls (380,90) and (380,90) .. (380,90) -- (370,90) .. controls (370,90) and (370,90) .. (370,90) -- cycle ; 
 \draw   (370,85) .. controls (370,82.24) and (372.24,80) .. (375,80) -- (375,80) .. controls (377.76,80) and (380,82.24) .. (380,85) -- (380,90) .. controls (380,90) and (380,90) .. (380,90) -- (370,90) .. controls (370,90) and (370,90) .. (370,90) -- cycle ; 

\draw    (375,80) -- (375,70) ;
\path  [shading=_pszdrkfgn,_hzo6ber61] (380,85) .. controls (380,82.24) and (382.24,80) .. (385,80) -- (385,80) .. controls (387.76,80) and (390,82.24) .. (390,85) -- (390,90) .. controls (390,90) and (390,90) .. (390,90) -- (380,90) .. controls (380,90) and (380,90) .. (380,90) -- cycle ; 
 \draw   (380,85) .. controls (380,82.24) and (382.24,80) .. (385,80) -- (385,80) .. controls (387.76,80) and (390,82.24) .. (390,85) -- (390,90) .. controls (390,90) and (390,90) .. (390,90) -- (380,90) .. controls (380,90) and (380,90) .. (380,90) -- cycle ; 

\draw    (385,80) -- (385,70) ;
\path  [shading=_qff0shcbe,_po66myqwb] (390,85) .. controls (390,82.24) and (392.24,80) .. (395,80) -- (395,80) .. controls (397.76,80) and (400,82.24) .. (400,85) -- (400,90) .. controls (400,90) and (400,90) .. (400,90) -- (390,90) .. controls (390,90) and (390,90) .. (390,90) -- cycle ; 
 \draw   (390,85) .. controls (390,82.24) and (392.24,80) .. (395,80) -- (395,80) .. controls (397.76,80) and (400,82.24) .. (400,85) -- (400,90) .. controls (400,90) and (400,90) .. (400,90) -- (390,90) .. controls (390,90) and (390,90) .. (390,90) -- cycle ; 

\draw    (395,80) -- (395,70) ;
\path  [shading=_5kqdlzf47,_d7ad8xqy8] (400,85) .. controls (400,82.24) and (402.24,80) .. (405,80) -- (405,80) .. controls (407.76,80) and (410,82.24) .. (410,85) -- (410,90) .. controls (410,90) and (410,90) .. (410,90) -- (400,90) .. controls (400,90) and (400,90) .. (400,90) -- cycle ; 
 \draw   (400,85) .. controls (400,82.24) and (402.24,80) .. (405,80) -- (405,80) .. controls (407.76,80) and (410,82.24) .. (410,85) -- (410,90) .. controls (410,90) and (410,90) .. (410,90) -- (400,90) .. controls (400,90) and (400,90) .. (400,90) -- cycle ; 

\draw    (405,80) -- (405,70) ;
\path  [shading=_bf61zpe1k,_gmlhf5744] (410,85) .. controls (410,82.24) and (412.24,80) .. (415,80) -- (415,80) .. controls (417.76,80) and (420,82.24) .. (420,85) -- (420,90) .. controls (420,90) and (420,90) .. (420,90) -- (410,90) .. controls (410,90) and (410,90) .. (410,90) -- cycle ; 
 \draw   (410,85) .. controls (410,82.24) and (412.24,80) .. (415,80) -- (415,80) .. controls (417.76,80) and (420,82.24) .. (420,85) -- (420,90) .. controls (420,90) and (420,90) .. (420,90) -- (410,90) .. controls (410,90) and (410,90) .. (410,90) -- cycle ; 

\draw    (415,80) -- (415,70) ;
\draw  [fill={rgb, 255:red, 74; green, 144; blue, 226 }  ,fill opacity=1 ] (338,105) .. controls (338,102.24) and (340.24,100) .. (343,100) .. controls (345.76,100) and (348,102.24) .. (348,105) .. controls (348,107.76) and (345.76,110) .. (343,110) .. controls (340.24,110) and (338,107.76) .. (338,105) -- cycle ;
\draw  [fill={rgb, 255:red, 74; green, 144; blue, 226 }  ,fill opacity=1 ] (382,105) .. controls (382,102.24) and (384.24,100) .. (387,100) .. controls (389.76,100) and (392,102.24) .. (392,105) .. controls (392,107.76) and (389.76,110) .. (387,110) .. controls (384.24,110) and (382,107.76) .. (382,105) -- cycle ;
\draw  [fill={rgb, 255:red, 208; green, 2; blue, 27 }  ,fill opacity=1 ] (350,105) .. controls (350,102.24) and (352.24,100) .. (355,100) .. controls (357.76,100) and (360,102.24) .. (360,105) .. controls (360,107.76) and (357.76,110) .. (355,110) .. controls (352.24,110) and (350,107.76) .. (350,105) -- cycle ;
\draw  [fill={rgb, 255:red, 208; green, 2; blue, 27 }  ,fill opacity=1 ] (370,105) .. controls (370,102.24) and (372.24,100) .. (375,100) .. controls (377.76,100) and (380,102.24) .. (380,105) .. controls (380,107.76) and (377.76,110) .. (375,110) .. controls (372.24,110) and (370,107.76) .. (370,105) -- cycle ;
\draw  [fill={rgb, 255:red, 208; green, 2; blue, 27 }  ,fill opacity=1 ] (404,105) .. controls (404,102.24) and (406.24,100) .. (409,100) .. controls (411.76,100) and (414,102.24) .. (414,105) .. controls (414,107.76) and (411.76,110) .. (409,110) .. controls (406.24,110) and (404,107.76) .. (404,105) -- cycle ;
\draw    (355.08,114.19) .. controls (356.42,128.3) and (373.29,127.51) .. (374.88,113.78) ;
\draw [shift={(375,112)}, rotate = 100.27] [color={rgb, 255:red, 0; green, 0; blue, 0 }  ][line width=0.75]    (6.56,-1.97) .. controls (4.17,-0.84) and (1.99,-0.18) .. (0,0) .. controls (1.99,0.18) and (4.17,0.84) .. (6.56,1.97)   ;
\draw [shift={(355,112)}, rotate = 81.55] [color={rgb, 255:red, 0; green, 0; blue, 0 }  ][line width=0.75]    (6.56,-1.97) .. controls (4.17,-0.84) and (1.99,-0.18) .. (0,0) .. controls (1.99,0.18) and (4.17,0.84) .. (6.56,1.97)   ;
\path  [shading=_7ot6fj95x,_ycqkwssw6] (420,85) .. controls (420,82.24) and (422.24,80) .. (425,80) -- (425,80) .. controls (427.76,80) and (430,82.24) .. (430,85) -- (430,90) .. controls (430,90) and (430,90) .. (430,90) -- (420,90) .. controls (420,90) and (420,90) .. (420,90) -- cycle ; 
 \draw   (420,85) .. controls (420,82.24) and (422.24,80) .. (425,80) -- (425,80) .. controls (427.76,80) and (430,82.24) .. (430,85) -- (430,90) .. controls (430,90) and (430,90) .. (430,90) -- (420,90) .. controls (420,90) and (420,90) .. (420,90) -- cycle ; 

\draw    (425,80) -- (425,70) ;

\draw (175,35) node [anchor=north west][inner sep=0.75pt]  [font=\normalsize]  {$\hat{\psi }_{+}^{\dagger }\hat{\psi }_{+} +\hat{\psi }_{-}^{\dagger }\hat{\psi }_{-}$};
\draw (310,35) node [anchor=north west][inner sep=0.75pt]  [font=\normalsize]  {$\hat{\psi }_{+}^{\dagger }\hat{\psi }_{-} +\hat{\psi }_{-}^{\dagger }\hat{\psi }_{+}$};
\draw (188,112.4) node [anchor=north west][inner sep=0.75pt]  [font=\normalsize]   {$g_{2}$};
\draw (185,5) node [anchor=north west][inner sep=0.75pt] [font=\normalsize]  [align=left] {leftmover};
\draw (325,5) node [anchor=north west][inner sep=0.75pt]  [font=\normalsize]  [align=left] {rightmover};
\draw (335,112.4) node [anchor=north west][inner sep=0.75pt]  [font=\normalsize]   {$g_{4}$};

\draw (420,110) node [anchor=north west][inner sep=0.75pt]  [font=\normalsize]   {$x$};

\end{tikzpicture}

%% file: Phasediagram.tex
\tikzset{every picture/.style={line width=0.75pt}} 

\begin{tikzpicture}[x=0.75pt,y=0.75pt,yscale=-1,xscale=1]

\draw  [color={rgb, 255:red, 208; green, 2; blue, 27 }  ,draw opacity=0.5 ][fill={rgb, 255:red, 208; green, 2; blue, 27 }  ,fill opacity=0.5 ] (350,170) -- (230,170) -- (110,50) -- (350,50) -- cycle ;
\draw  [color={rgb, 255:red, 74; green, 144; blue, 226 }  ,draw opacity=0.5 ][fill={rgb, 255:red, 74; green, 144; blue, 226 }  ,fill opacity=0.5 ] (110,50) -- (230,170) -- (110,170) -- cycle ;
\draw    (230,170) -- (110,50) ;
\draw    (110,170) -- (347,170) ;
\draw [shift={(350,170)}, rotate = 180] [fill={rgb, 255:red, 0; green, 0; blue, 0 }  ][line width=0.08]  [draw opacity=0] (8.93,-4.29) -- (0,0) -- (8.93,4.29) -- cycle    ;
\draw    (230,170) -- (230,53) ;
\draw [shift={(230,50)}, rotate = 90] [fill={rgb, 255:red, 0; green, 0; blue, 0 }  ][line width=0.08]  [draw opacity=0] (8.93,-4.29) -- (0,0) -- (8.93,4.29) -- cycle    ;
\draw    (290,180) .. controls (288.68,165.31) and (276.58,159.29) .. (231.38,150.27) ;
\draw [shift={(230,150)}, rotate = 11.19] [color={rgb, 255:red, 0; green, 0; blue, 0 }  ][line width=0.75]    (10.93,-3.29) .. controls (6.95,-1.4) and (3.31,-0.3) .. (0,0) .. controls (3.31,0.3) and (6.95,1.4) .. (10.93,3.29)   ;
\draw [color={rgb, 255:red, 0; green, 0; blue, 0 }  ,draw opacity=1 ][line width=2.25]    (230,140) -- (230,165) ;
\draw [shift={(230,170)}, rotate = 270] [fill={rgb, 255:red, 0; green, 0; blue, 0 }  ,fill opacity=1 ][line width=0.08]  [draw opacity=0] (8.57,-4.12) -- (0,0) -- (8.57,4.12) -- cycle    ;
\draw [color={rgb, 255:red, 0; green, 0; blue, 0 }  ,draw opacity=1 ][line width=2.25]    (180,90) -- (180,115) ;
\draw [shift={(180,120)}, rotate = 270] [fill={rgb, 255:red, 0; green, 0; blue, 0 }  ,fill opacity=1 ][line width=0.08]  [draw opacity=0] (8.57,-4.12) -- (0,0) -- (8.57,4.12) -- cycle    ;
\draw    (220,40) .. controls (195.25,65.25) and (228.65,76.28) .. (181.46,99.3) ;
\draw [shift={(180,100)}, rotate = 334.53] [color={rgb, 255:red, 0; green, 0; blue, 0 }  ][line width=0.75]    (10.93,-3.29) .. controls (6.95,-1.4) and (3.31,-0.3) .. (0,0) .. controls (3.31,0.3) and (6.95,1.4) .. (10.93,3.29)   ;
\draw [color={rgb, 255:red, 0; green, 0; blue, 0 }  ,draw opacity=1 ][line width=2.25]    (180,150) -- (180,125) ;
\draw [shift={(180,120)}, rotate = 90] [fill={rgb, 255:red, 0; green, 0; blue, 0 }  ,fill opacity=1 ][line width=0.08]  [draw opacity=0] (8.57,-4.12) -- (0,0) -- (8.57,4.12) -- cycle    ;
\draw    (174,180) .. controls (174.42,164.02) and (125.66,162.86) .. (177.4,140.68) ;
\draw [shift={(179,140)}, rotate = 157.21] [color={rgb, 255:red, 0; green, 0; blue, 0 }  ][line width=0.75]    (10.93,-3.29) .. controls (6.95,-1.4) and (3.31,-0.3) .. (0,0) .. controls (3.31,0.3) and (6.95,1.4) .. (10.93,3.29)   ;

\draw (237,52) node [anchor=north west][inner sep=0.75pt]  [font=\normalsize]  {$\gamma $};
\draw (334,152) node [anchor=north west][inner sep=0.75pt]  [font=\normalsize]  {$\Delta g $};
\draw (264,178.4) node [anchor=north west][inner sep=0.75pt]  [font=\normalsize]  {$\log \xi \sim \gamma ^{-1}$};
\draw (210,18.4) node [anchor=north west][inner sep=0.75pt]  [font=\normalsize]  {$\log \xi \sim ( \gamma -\gamma _{c})^{-1/2}$};
\draw (121,122.4) node [anchor=north west][inner sep=0.75pt]  [font=\normalsize]  {$\xi =\infty $};
\draw (271,92.4) node [anchor=north west][inner sep=0.75pt]  [font=\normalsize]  {$\xi < \infty $};
\draw (109,35) node [anchor=north west][inner sep=0.75pt]  [font=\normalsize]  {$\gamma _{c} \sim -\Delta g $};
\draw (111,178.4) node [anchor=north west][inner sep=0.75pt]   [font=\normalsize]  {$c-1\sim ( \gamma _{c} -\gamma )^{1/2}$};

\draw (225,173) node [anchor=north west][inner sep=0.75pt]   [font=\normalsize]  {$0$};

\end{tikzpicture}

%% file: Momentumsketch.tex
    

\tikzset{every picture/.style={line width=0.75pt}} 

\begin{tikzpicture}[x=0.75pt,y=0.75pt,yscale=-1,xscale=1]

\draw    (150,120) -- (367,120) ;
\draw [shift={(370,120)}, rotate = 180] [fill={rgb, 255:red, 0; green, 0; blue, 0 }  ][line width=0.08]  [draw opacity=0] (8.93,-4.29) -- (0,0) -- (8.93,4.29) -- cycle    ;
\draw    (260,210) -- (260,33) ;
\draw [shift={(260,30)}, rotate = 90] [fill={rgb, 255:red, 0; green, 0; blue, 0 }  ][line width=0.08]  [draw opacity=0] (8.93,-4.29) -- (0,0) -- (8.93,4.29) -- cycle    ;
\draw [line width=1.5]    (150,210) -- (370,30) ;
\draw [line width=1.5]    (370,210) -- (150,30) ;
\draw  [fill={rgb, 255:red, 255; green, 255; blue, 255 }  ,fill opacity=1 ] (304,80) .. controls (304,77.24) and (306.24,75) .. (309,75) .. controls (311.76,75) and (314,77.24) .. (314,80) .. controls (314,82.76) and (311.76,85) .. (309,85) .. controls (306.24,85) and (304,82.76) .. (304,80) -- cycle ;
\draw  [fill={rgb, 255:red, 255; green, 255; blue, 255 }  ,fill opacity=1 ] (206,80) .. controls (206,77.24) and (208.24,75) .. (211,75) .. controls (213.76,75) and (216,77.24) .. (216,80) .. controls (216,82.76) and (213.76,85) .. (211,85) .. controls (208.24,85) and (206,82.76) .. (206,80) -- cycle ;
\draw    (222.35,73.87) .. controls (259.46,56.11) and (261.09,56.21) .. (298.28,74.17) ;
\draw [shift={(300,75)}, rotate = 205.79] [color={rgb, 255:red, 0; green, 0; blue, 0 }  ][line width=0.75]    (10.93,-3.29) .. controls (6.95,-1.4) and (3.31,-0.3) .. (0,0) .. controls (3.31,0.3) and (6.95,1.4) .. (10.93,3.29)   ;
\draw [shift={(220,75)}, rotate = 334.4] [color={rgb, 255:red, 0; green, 0; blue, 0 }  ][line width=0.75]    (10.93,-3.29) .. controls (6.95,-1.4) and (3.31,-0.3) .. (0,0) .. controls (3.31,0.3) and (6.95,1.4) .. (10.93,3.29)   ;
\draw    (310,90) .. controls (319.9,139.34) and (369.33,90.82) .. (321.48,80.31) ;
\draw [shift={(320,80)}, rotate = 11.05] [color={rgb, 255:red, 0; green, 0; blue, 0 }  ][line width=0.75]    (10.93,-3.29) .. controls (6.95,-1.4) and (3.31,-0.3) .. (0,0) .. controls (3.31,0.3) and (6.95,1.4) .. (10.93,3.29)   ;
\draw    (210,90) .. controls (199.73,139.34) and (148.47,90.82) .. (198.09,80.31) ;
\draw [shift={(199.63,80)}, rotate = 169.33] [color={rgb, 255:red, 0; green, 0; blue, 0 }  ][line width=0.75]    (10.93,-3.29) .. controls (6.95,-1.4) and (3.31,-0.3) .. (0,0) .. controls (3.31,0.3) and (6.95,1.4) .. (10.93,3.29)   ;

\draw (361,125) node [anchor=north west][inner sep=0.75pt]   [font=\normalsize]  {$k$};
\draw (270,28) node [anchor=north west][inner sep=0.75pt]  [font=\normalsize]   {$\epsilon (k) =dv_{0} k$};
\draw (231,45) node [anchor=north west][inner sep=0.75pt]   [font=\normalsize]  {$g_{2}$};
\draw (345,75) node [anchor=north west][inner sep=0.75pt]  [font=\normalsize]   {$g_{4}$};
\draw (159,75) node [anchor=north west][inner sep=0.75pt]   [font=\normalsize]  {$g_{4}$};
\draw (177,192.4) node [anchor=north west][inner sep=0.75pt]   [font=\normalsize]  {$d=+$};
\draw (306,192.4) node [anchor=north west][inner sep=0.75pt]   [font=\normalsize]  {$d=-$};

\end{tikzpicture}

%% file: Observablerotation.tex
\tikzset{every picture/.style={line width=0.75pt}} 

\begin{tikzpicture}[x=0.75pt,y=0.75pt,yscale=-1,xscale=1]

\draw [color={rgb, 255:red, 128; green, 128; blue, 128  }  ,draw opacity=1 ][line width=1.5]    (255,160) .. controls (228.6,147.8) and (212.6,148.4) .. (195,140) ;
\draw [color={rgb, 255:red, 128; green, 128; blue, 128  }  ,draw opacity=1 ][line width=1.5]    (255,140) .. controls (234.6,152.8) and (216.2,145.8) .. (195,160) ;
\draw [color={rgb, 255:red, 128; green, 128; blue, 128  }  ,draw opacity=1 ][line width=1.5]    (195,140) .. controls (218.2,131.2) and (229.66,127.6) .. (255,140) ;
\draw  [line width=0.75]  (226.5,186) -- (226.5,208)(223.5,186) -- (223.5,208) ;
\draw [shift={(225,216)}, rotate = 270] [color={rgb, 255:red, 0; green, 0; blue, 0 }  ][line width=0.75]    (15.3,-4.61) .. controls (9.73,-1.96) and (4.63,-0.42) .. (0,0) .. controls (4.63,0.42) and (9.73,1.96) .. (15.3,4.61)   ;
\draw [line width=1.5]    (150,140) -- (298,140) ;
\draw [shift={(300,140)}, rotate = 180] [color={rgb, 255:red, 0; green, 0; blue, 0 }  ][line width=1.5]    (10.93,-3.29) .. controls (6.95,-1.4) and (3.31,-0.3) .. (0,0) .. controls (3.31,0.3) and (6.95,1.4) .. (10.93,3.29)   ;
\draw [line width=1.5]    (150,160) -- (298,160) ;
\draw [shift={(300,160)}, rotate = 180] [color={rgb, 255:red, 0; green, 0; blue, 0 }  ][line width=1.5]    (10.93,-3.29) .. controls (6.95,-1.4) and (3.31,-0.3) .. (0,0) .. controls (3.31,0.3) and (6.95,1.4) .. (10.93,3.29)   ;
\draw [line width=1.5]    (150,180) -- (298,180) ;
\draw [shift={(300,180)}, rotate = 180] [color={rgb, 255:red, 0; green, 0; blue, 0 }  ][line width=1.5]    (10.93,-3.29) .. controls (6.95,-1.4) and (3.31,-0.3) .. (0,0) .. controls (3.31,0.3) and (6.95,1.4) .. (10.93,3.29)   ;
\draw  [color={rgb, 255:red, 128; green, 128; blue, 128  }  ,draw opacity=1 ][fill={rgb, 255:red, 128; green, 128; blue, 128  }  ,fill opacity=1 ] (250,140) .. controls (250,137.24) and (252.24,135) .. (255,135) .. controls (257.76,135) and (260,137.24) .. (260,140) .. controls (260,142.76) and (257.76,145) .. (255,145) .. controls (252.24,145) and (250,142.76) .. (250,140) -- cycle ;
\draw  [color={rgb,255:red, 128; green, 128; blue, 128  }  ,draw opacity=1 ][fill={rgb, 255:red, 128; green, 128; blue, 128  }  ,fill opacity=1 ] (190,140) .. controls (190,137.24) and (192.24,135) .. (195,135) .. controls (197.76,135) and (200,137.24) .. (200,140) .. controls (200,142.76) and (197.76,145) .. (195,145) .. controls (192.24,145) and (190,142.76) .. (190,140) -- cycle ;
\draw  [color={rgb, 255:red, 128; green, 128; blue, 128  }  ,draw opacity=1 ][fill={rgb,255:red, 128; green, 128; blue, 128  }  ,fill opacity=1 ] (190,160) .. controls (190,157.24) and (192.24,155) .. (195,155) .. controls (197.76,155) and (200,157.24) .. (200,160) .. controls (200,162.76) and (197.76,165) .. (195,165) .. controls (192.24,165) and (190,162.76) .. (190,160) -- cycle ;
\draw  [color={rgb, 255:red, 128; green, 128; blue, 128  }  ,draw opacity=1 ][fill={rgb, 255:red, 128; green, 128; blue, 128  }  ,fill opacity=1 ] (250,160) .. controls (250,157.24) and (252.24,155) .. (255,155) .. controls (257.76,155) and (260,157.24) .. (260,160) .. controls (260,162.76) and (257.76,165) .. (255,165) .. controls (252.24,165) and (250,162.76) .. (250,160) -- cycle ;
\draw [color={rgb, 255:red, 128; green, 128; blue, 128  }  ,draw opacity=1 ][line width=1.5]    (195,160) .. controls (207.8,167) and (229,172.6) .. (255,160) ;
\draw [line width=1.5]    (150,230) -- (298,230) ;
\draw [shift={(300,230)}, rotate = 180] [color={rgb, 255:red, 0; green, 0; blue, 0 }  ][line width=1.5]    (10.93,-3.29) .. controls (6.95,-1.4) and (3.31,-0.3) .. (0,0) .. controls (3.31,0.3) and (6.95,1.4) .. (10.93,3.29)   ;
\draw [line width=1.5]    (150,250) -- (298,250) ;
\draw [shift={(300,250)}, rotate = 180] [color={rgb, 255:red, 0; green, 0; blue, 0 }  ][line width=1.5]    (10.93,-3.29) .. controls (6.95,-1.4) and (3.31,-0.3) .. (0,0) .. controls (3.31,0.3) and (6.95,1.4) .. (10.93,3.29)   ;
\draw [line width=1.5]    (150,270) -- (298,270) ;
\draw [shift={(300,270)}, rotate = 180] [color={rgb, 255:red, 0; green, 0; blue, 0 }  ][line width=1.5]    (10.93,-3.29) .. controls (6.95,-1.4) and (3.31,-0.3) .. (0,0) .. controls (3.31,0.3) and (6.95,1.4) .. (10.93,3.29)   ;
\draw  [color={rgb, 255:red, 128; green, 128; blue, 128  }  ,draw opacity=1 ][fill={rgb, 255:red, 128; green, 128; blue, 128 }  ,fill opacity=1 ] (190,250) .. controls (190,247.24) and (192.24,245) .. (195,245) .. controls (197.76,245) and (200,247.24) .. (200,250) .. controls (200,252.76) and (197.76,255) .. (195,255) .. controls (192.24,255) and (190,252.76) .. (190,250) -- cycle ;
\draw  [color={rgb, 255:red, 128; green, 128; blue, 128  }  ,draw opacity=1 ][fill={rgb,255:red, 128; green, 128; blue, 128  }  ,fill opacity=1 ] (250,250) .. controls (250,247.24) and (252.24,245) .. (255,245) .. controls (257.76,245) and (260,247.24) .. (260,250) .. controls (260,252.76) and (257.76,255) .. (255,255) .. controls (252.24,255) and (250,252.76) .. (250,250) -- cycle ;
\draw [color={rgb, 255:red, 128; green, 128; blue, 128  }  ,draw opacity=1 ][line width=1.5]    (195,250) .. controls (207.8,257) and (229,262.6) .. (255,250) ;

\draw (110,134) node [anchor=north west][inner sep=0.75pt]    {$r=1$};
\draw (110,154) node [anchor=north west][inner sep=0.75pt]    {$r=2$};
\draw (110,174) node [anchor=north west][inner sep=0.75pt]    {$r=3$};
\draw (197,115.4) node [anchor=north west][inner sep=0.75pt]    {$C( x-x')$};
\draw (297,182) node [anchor=north west][inner sep=0.75pt]    {$x$};
\draw (110,224) node [anchor=north west][inner sep=0.75pt]    {$k=0$};
\draw (110,244) node [anchor=north west][inner sep=0.75pt]    {$k=1$};
\draw (110,264) node [anchor=north west][inner sep=0.75pt]    {$k=2$};
\draw (297,272) node [anchor=north west][inner sep=0.75pt]    {$x$};

\end{tikzpicture}

%% file: ComplexWick.tex
\tikzset{every picture/.style={line width=0.75pt}} 

\begin{tikzpicture}[x=0.75pt,y=0.75pt,yscale=-1,xscale=1]

\draw    (130,140) -- (258,140) ;
\draw [shift={(260,140)}, rotate = 180] [color={rgb, 255:red, 0; green, 0; blue, 0 }  ][line width=0.75]    (10.93,-3.29) .. controls (6.95,-1.4) and (3.31,-0.3) .. (0,0) .. controls (3.31,0.3) and (6.95,1.4) .. (10.93,3.29)   ;
\draw    (190,200) -- (190,82) ;
\draw [shift={(190,80)}, rotate = 90] [color={rgb, 255:red, 0; green, 0; blue, 0 }  ][line width=0.75]    (10.93,-3.29) .. controls (6.95,-1.4) and (3.31,-0.3) .. (0,0) .. controls (3.31,0.3) and (6.95,1.4) .. (10.93,3.29)   ;
\draw [color={rgb, 255:red, 128; green, 128; blue, 128 }  ,draw opacity=1 ][line width=1.5]    (130,140) -- (190,140) ;
\draw [shift={(165.6,140)}, rotate = 180] [fill={rgb, 255:red, 128; green, 128; blue, 128 }  ,fill opacity=1 ][line width=0.08]  [draw opacity=0] (9.29,-4.46) -- (0,0) -- (9.29,4.46) -- cycle    ;
\draw [color={rgb, 255:red, 128; green, 128; blue, 128 }  ,draw opacity=1 ][fill={rgb, 255:red, 155; green, 155; blue, 155 }  ,fill opacity=1 ][line width=1.5]    (190,140) -- (250,140) ;
\draw [shift={(225.6,140)}, rotate = 180] [fill={rgb, 255:red, 128; green, 128; blue, 128 }  ,fill opacity=1 ][line width=0.08]  [draw opacity=0] (9.29,-4.46) -- (0,0) -- (9.29,4.46) -- cycle    ;
\draw [color={rgb, 255:red, 208; green, 2; blue, 27 }  ,draw opacity=1 ][line width=1.5]    (190,140) -- (210,80) ;
\draw [shift={(197.75,116.74)}, rotate = 288.43] [fill={rgb, 255:red, 208; green, 2; blue, 27 }  ,fill opacity=1 ][line width=0.08]  [draw opacity=0] (9.29,-4.46) -- (0,0) -- (9.29,4.46) -- cycle    ;
\draw [color={rgb, 255:red, 208; green, 2; blue, 27 }  ,draw opacity=1 ][line width=1.5]    (170,200) -- (190,140) ;
\draw [shift={(177.75,176.74)}, rotate = 288.43] [fill={rgb, 255:red, 208; green, 2; blue, 27 }  ,fill opacity=1 ][line width=0.08]  [draw opacity=0] (9.29,-4.46) -- (0,0) -- (9.29,4.46) -- cycle    ;
\draw [color={rgb, 255:red, 74; green, 144; blue, 226 }  ,draw opacity=1 ][line width=1.5]    (210,200) -- (190,140) ;
\draw [shift={(198.23,164.69)}, rotate = 71.57] [fill={rgb, 255:red, 74; green, 144; blue, 226 }  ,fill opacity=1 ][line width=0.08]  [draw opacity=0] (9.29,-4.46) -- (0,0) -- (9.29,4.46) -- cycle    ;
\draw [color={rgb, 255:red, 74; green, 144; blue, 226 }  ,draw opacity=1 ][line width=1.5]    (190,140) -- (170,80) ;
\draw [shift={(178.23,104.69)}, rotate = 71.57] [fill={rgb, 255:red, 74; green, 144; blue, 226 }  ,fill opacity=1 ][line width=0.08]  [draw opacity=0] (9.29,-4.46) -- (0,0) -- (9.29,4.46) -- cycle    ;
\draw    (210.13,179.7) -- (197.87,184.3) ;
\draw [shift={(196,185)}, rotate = 339.44] [color={rgb, 255:red, 0; green, 0; blue, 0 }  ][line width=0.75]    (6.56,-1.97) .. controls (4.17,-0.84) and (1.99,-0.18) .. (0,0) .. controls (1.99,0.18) and (4.17,0.84) .. (6.56,1.97)   ;
\draw [shift={(212,179)}, rotate = 159.44] [color={rgb, 255:red, 0; green, 0; blue, 0 }  ][line width=0.75]    (6.56,-1.97) .. controls (4.17,-0.84) and (1.99,-0.18) .. (0,0) .. controls (1.99,0.18) and (4.17,0.84) .. (6.56,1.97)   ;
\draw    (182.13,184.3) -- (169.87,179.7) ;
\draw [shift={(168,179)}, rotate = 20.56] [color={rgb, 255:red, 0; green, 0; blue, 0 }  ][line width=0.75]    (6.56,-1.97) .. controls (4.17,-0.84) and (1.99,-0.18) .. (0,0) .. controls (1.99,0.18) and (4.17,0.84) .. (6.56,1.97)   ;
\draw [shift={(184,185)}, rotate = 200.56] [color={rgb, 255:red, 0; green, 0; blue, 0 }  ][line width=0.75]    (6.56,-1.97) .. controls (4.17,-0.84) and (1.99,-0.18) .. (0,0) .. controls (1.99,0.18) and (4.17,0.84) .. (6.56,1.97)   ;
\draw [color={rgb, 255:red, 128; green, 128; blue, 128 }  ,draw opacity=1 ][line width=1.5]    (165,220) -- (235,220) ;
\draw [shift={(205.6,220)}, rotate = 180] [fill={rgb, 255:red, 128; green, 128; blue, 128 }  ,fill opacity=1 ][line width=0.08]  [draw opacity=0] (9.29,-4.46) -- (0,0) -- (9.29,4.46) -- cycle    ;
\draw    (270,140) -- (398,140) ;
\draw [shift={(400,140)}, rotate = 180] [color={rgb, 255:red, 0; green, 0; blue, 0 }  ][line width=0.75]    (10.93,-3.29) .. controls (6.95,-1.4) and (3.31,-0.3) .. (0,0) .. controls (3.31,0.3) and (6.95,1.4) .. (10.93,3.29)   ;
\draw    (330,200) -- (330,82) ;
\draw [shift={(330,80)}, rotate = 90] [color={rgb, 255:red, 0; green, 0; blue, 0 }  ][line width=0.75]    (10.93,-3.29) .. controls (6.95,-1.4) and (3.31,-0.3) .. (0,0) .. controls (3.31,0.3) and (6.95,1.4) .. (10.93,3.29)   ;
\draw [color={rgb, 255:red, 128; green, 128; blue, 128 }  ,draw opacity=1 ][line width=1.5]    (270,140) -- (330,140) ;
\draw [shift={(305.6,140)}, rotate = 180] [fill={rgb, 255:red, 128; green, 128; blue, 128 }  ,fill opacity=1 ][line width=0.08]  [draw opacity=0] (9.29,-4.46) -- (0,0) -- (9.29,4.46) -- cycle    ;
\draw [color={rgb, 255:red, 128; green, 128; blue, 128 }  ,draw opacity=1 ][fill={rgb, 255:red, 155; green, 155; blue, 155 }  ,fill opacity=1 ][line width=1.5]    (330,140) -- (390,140) ;
\draw [shift={(365.6,140)}, rotate = 180] [fill={rgb, 255:red, 128; green, 128; blue, 128 }  ,fill opacity=1 ][line width=0.08]  [draw opacity=0] (9.29,-4.46) -- (0,0) -- (9.29,4.46) -- cycle    ;
\draw [color={rgb, 255:red, 208; green, 2; blue, 27 }  ,draw opacity=1 ][line width=1.5]    (330,140) -- (270,120) ;
\draw [shift={(306.74,132.25)}, rotate = 198.43] [fill={rgb, 255:red, 208; green, 2; blue, 27 }  ,fill opacity=1 ][line width=0.08]  [draw opacity=0] (9.29,-4.46) -- (0,0) -- (9.29,4.46) -- cycle    ;
\draw [color={rgb, 255:red, 208; green, 2; blue, 27 }  ,draw opacity=1 ][line width=1.5]    (390,160) -- (330,140) ;
\draw [shift={(366.74,152.25)}, rotate = 198.43] [fill={rgb, 255:red, 208; green, 2; blue, 27 }  ,fill opacity=1 ][line width=0.08]  [draw opacity=0] (9.29,-4.46) -- (0,0) -- (9.29,4.46) -- cycle    ;
\draw [color={rgb, 255:red, 74; green, 144; blue, 226 }  ,draw opacity=1 ][line width=1.5]    (270,160) -- (330,140) ;
\draw [shift={(305.31,148.23)}, rotate = 161.57] [fill={rgb, 255:red, 74; green, 144; blue, 226 }  ,fill opacity=1 ][line width=0.08]  [draw opacity=0] (9.29,-4.46) -- (0,0) -- (9.29,4.46) -- cycle    ;
\draw [color={rgb, 255:red, 74; green, 144; blue, 226 }  ,draw opacity=1 ][line width=1.5]    (330,140) -- (390,120) ;
\draw [shift={(365.31,128.23)}, rotate = 161.57] [fill={rgb, 255:red, 74; green, 144; blue, 226 }  ,fill opacity=1 ][line width=0.08]  [draw opacity=0] (9.29,-4.46) -- (0,0) -- (9.29,4.46) -- cycle    ;
\draw    (292.3,120.87) -- (287.7,133.13) ;
\draw [shift={(287,135)}, rotate = 290.56] [color={rgb, 255:red, 0; green, 0; blue, 0 }  ][line width=0.75]    (6.56,-1.97) .. controls (4.17,-0.84) and (1.99,-0.18) .. (0,0) .. controls (1.99,0.18) and (4.17,0.84) .. (6.56,1.97)   ;
\draw [shift={(293,119)}, rotate = 110.56] [color={rgb, 255:red, 0; green, 0; blue, 0 }  ][line width=0.75]    (6.56,-1.97) .. controls (4.17,-0.84) and (1.99,-0.18) .. (0,0) .. controls (1.99,0.18) and (4.17,0.84) .. (6.56,1.97)   ;
\draw    (292.24,160.15) -- (286.76,146.85) ;
\draw [shift={(286,145)}, rotate = 67.62] [color={rgb, 255:red, 0; green, 0; blue, 0 }  ][line width=0.75]    (6.56,-1.97) .. controls (4.17,-0.84) and (1.99,-0.18) .. (0,0) .. controls (1.99,0.18) and (4.17,0.84) .. (6.56,1.97)   ;
\draw [shift={(293,162)}, rotate = 247.62] [color={rgb, 255:red, 0; green, 0; blue, 0 }  ][line width=0.75]    (6.56,-1.97) .. controls (4.17,-0.84) and (1.99,-0.18) .. (0,0) .. controls (1.99,0.18) and (4.17,0.84) .. (6.56,1.97)   ;
\draw [color={rgb, 255:red, 74; green, 144; blue, 226 }  ,draw opacity=1 ][line width=1.5]    (165,240) -- (235,240) ;
\draw [shift={(205.6,240)}, rotate = 180] [fill={rgb, 255:red, 74; green, 144; blue, 226 }  ,fill opacity=1 ][line width=0.08]  [draw opacity=0] (9.29,-4.46) -- (0,0) -- (9.29,4.46) -- cycle    ;
\draw [color={rgb, 255:red, 208; green, 2; blue, 27 }  ,draw opacity=1 ][line width=1.5]    (165,250) -- (235,250) ;
\draw [shift={(205.6,250)}, rotate = 180] [fill={rgb, 255:red, 208; green, 2; blue, 27 }  ,fill opacity=1 ][line width=0.08]  [draw opacity=0] (9.29,-4.46) -- (0,0) -- (9.29,4.46) -- cycle    ;
\draw   (155,210) -- (365,210) -- (365,260) -- (155,260) -- cycle ;
\draw    (146,170) .. controls (145.49,177.82) and (141.2,194.17) .. (172.52,184.77) ;
\draw [shift={(175,184)}, rotate = 162.06] [fill={rgb, 255:red, 0; green, 0; blue, 0 }  ][line width=0.08]  [draw opacity=0] (5.36,-2.57) -- (0,0) -- (5.36,2.57) -- cycle    ;

\draw (235,142.4) node [anchor=north west][inner sep=0.75pt]    {$\text{Re}( t)$};
\draw (140,80.4) node [anchor=north west][inner sep=0.75pt]    {$\text{Im}( t)$};
\draw (175,190) node [anchor=north west][inner sep=0.75pt]    {$+$};
\draw (210,190) node [anchor=north west][inner sep=0.75pt]    {$-$};
\draw (270,213) node [anchor=north west][inner sep=0.75pt]  [align=left] {real time/space};
\draw (370,142.4) node [anchor=north west][inner sep=0.75pt]    {$\text{Re}( x)$};
\draw (295,80.4) node [anchor=north west][inner sep=0.75pt]    {$\text{Im}( x)$};
\draw (270,110) node [anchor=north west][inner sep=0.75pt]    {$+$};
\draw (270,160) node [anchor=north west][inner sep=0.75pt]    {$-$};
\draw (272,238) node [anchor=north west][inner sep=0.75pt]   [align=left] {complex Wick};
\draw (137,155) node [anchor=north west][inner sep=0.75pt]   [align=left] {RG};

\end{tikzpicture}

%% file: RGflow.tex
\tikzset{every picture/.style={line width=0.75pt}} 

\begin{tikzpicture}[x=0.75pt,y=0.75pt,yscale=-1,xscale=1]

\draw [color={rgb, 255:red, 208; green, 2; blue, 27 }  ,draw opacity=1 ][line width=1.5]    (489,177) .. controls (472.33,200.11) and (473.89,207) .. (471.4,220) ;
\draw [shift={(480.56,189.88)}, rotate = 117.9] [fill={rgb, 255:red, 208; green, 2; blue, 27 }  ,fill opacity=1 ][line width=0.08]  [draw opacity=0] (13.4,-6.43) -- (0,0) -- (13.4,6.44) -- (8.9,0) -- cycle    ;
\draw [color={rgb, 255:red, 208; green, 2; blue, 27 }  ,draw opacity=1 ][line width=1.5]    (489.13,146) .. controls (428.11,193.89) and (432.11,208.56) .. (429,220) ;
\draw [shift={(458.07,172.58)}, rotate = 135.43] [fill={rgb, 255:red, 208; green, 2; blue, 27 }  ,fill opacity=1 ][line width=0.08]  [draw opacity=0] (13.4,-6.43) -- (0,0) -- (13.4,6.44) -- (8.9,0) -- cycle    ;
\draw [color={rgb, 255:red, 208; green, 2; blue, 27 }  ,draw opacity=1 ][line width=1.5]    (489,132.5) .. controls (407,186.72) and (389.22,203) .. (387,220) ;
\draw [shift={(438.78,166.66)}, rotate = 144.37] [fill={rgb, 255:red, 208; green, 2; blue, 27 }  ,fill opacity=1 ][line width=0.08]  [draw opacity=0] (13.4,-6.43) -- (0,0) -- (13.4,6.44) -- (8.9,0) -- cycle    ;
\draw [color={rgb, 255:red, 74; green, 144; blue, 226 }  ,draw opacity=1 ][line width=1.5]    (199,131.5) .. controls (275,182) and (303.2,203.7) .. (302,220) ;
\draw [shift={(262.02,174.94)}, rotate = 216.27] [fill={rgb, 255:red, 74; green, 144; blue, 226 }  ,fill opacity=1 ][line width=0.08]  [draw opacity=0] (13.4,-6.43) -- (0,0) -- (13.4,6.44) -- (8.9,0) -- cycle    ;
\draw [color={rgb, 255:red, 74; green, 144; blue, 226 }  ,draw opacity=1 ][line width=1.5]    (199,145) .. controls (261.4,194.7) and (255.6,203.7) .. (260.3,220) ;
\draw [shift={(241.32,182.63)}, rotate = 226.21] [fill={rgb, 255:red, 74; green, 144; blue, 226 }  ,fill opacity=1 ][line width=0.08]  [draw opacity=0] (13.4,-6.43) -- (0,0) -- (13.4,6.44) -- (8.9,0) -- cycle    ;
\draw [color={rgb, 255:red, 74; green, 144; blue, 226 }  ,draw opacity=1 ][line width=1.5]    (199,175) .. controls (215,196.7) and (215.6,207.1) .. (218,220) ;
\draw [shift={(214.22,202.64)}, rotate = 249.58] [fill={rgb, 255:red, 74; green, 144; blue, 226 }  ,fill opacity=1 ][line width=0.08]  [draw opacity=0] (13.4,-6.43) -- (0,0) -- (13.4,6.44) -- (8.9,0) -- cycle    ;
\draw [line width=1.5]    (200,220) -- (486,220) ;
\draw [shift={(490,220)}, rotate = 180] [fill={rgb, 255:red, 0; green, 0; blue, 0 }  ][line width=0.08]  [draw opacity=0] (11.61,-5.58) -- (0,0) -- (11.61,5.58) -- cycle    ;
\draw [line width=1.5]    (345,220) -- (344.03,96) ;
\draw [shift={(344,92)}, rotate = 89.55] [fill={rgb, 255:red, 0; green, 0; blue, 0 }  ][line width=0.08]  [draw opacity=0] (11.61,-5.58) -- (0,0) -- (11.61,5.58) -- cycle    ;
\draw [color={rgb, 255:red, 208; green, 2; blue, 27 }  ,draw opacity=1 ][line width=1.5]    (233,91.3) .. controls (309.46,118.38) and (375,120.38) .. (455,91.3) ;
\draw [shift={(351.31,112.29)}, rotate = 179.28] [fill={rgb, 255:red, 208; green, 2; blue, 27 }  ,fill opacity=1 ][line width=0.08]  [draw opacity=0] (13.4,-6.43) -- (0,0) -- (13.4,6.44) -- (8.9,0) -- cycle    ;
\draw [color={rgb, 255:red, 208; green, 2; blue, 27 }  ,draw opacity=1 ][line width=1.5]    (199,97.3) .. controls (317.8,155.1) and (377.6,151.1) .. (489,97.6) ;
\draw [shift={(352.02,139.07)}, rotate = 178.85] [fill={rgb, 255:red, 208; green, 2; blue, 27 }  ,fill opacity=1 ][line width=0.08]  [draw opacity=0] (13.4,-6.43) -- (0,0) -- (13.4,6.44) -- (8.9,0) -- cycle    ;
\draw [color={rgb, 255:red, 208; green, 2; blue, 27 }  ,draw opacity=1 ][line width=1.5]    (199,113) .. controls (328.2,183.3) and (360.6,184.1) .. (489,113.5) ;
\draw [shift={(351.56,165.93)}, rotate = 178.48] [fill={rgb, 255:red, 208; green, 2; blue, 27 }  ,fill opacity=1 ][line width=0.08]  [draw opacity=0] (13.4,-6.43) -- (0,0) -- (13.4,6.44) -- (8.9,0) -- cycle    ;
\draw [color={rgb, 255:red, 208; green, 2; blue, 27 }  ,draw opacity=1 ][line width=1.5]    (199,123.8) .. controls (356.2,217.1) and (334.4,214.5) .. (489,124.5) ;
\draw [shift={(351.02,192.6)}, rotate = 177.25] [fill={rgb, 255:red, 208; green, 2; blue, 27 }  ,fill opacity=1 ][line width=0.08]  [draw opacity=0] (13.4,-6.43) -- (0,0) -- (13.4,6.44) -- (8.9,0) -- cycle    ;
\draw [line width=2.25]  [dash pattern={on 6.75pt off 4.5pt}]  (200,128) -- (345,220) ;

\draw (465,223) node [anchor=north west][inner sep=0.75pt]  [font=\normalsize]  {$X \sim \Delta g$};
\draw (351,88.4) node [anchor=north west][inner sep=0.75pt]  [font=\normalsize]   {$Y \sim m^2 \gamma/v$};
\draw (340,223) node [anchor=north west][inner sep=0.75pt]  [font=\normalsize]   {$0$};

\end{tikzpicture}

%% file: appendix.tex
\section{Construction of the replica master equation} \label{App:replica}
We intend to compute the replica density operator $\hat{\rho}_M$ for $M \geq 2$. Suppose we drop the normalization condition for quantum states $\ket{\tilde \psi_t}$, just initializing it at a time $t_0$ to be normalized $\ket{\tilde \psi_{t_0}}=\ket{\psi_{t_0}}=\ket{\psi_0}$ and non-random. The state update for the non-normalized states is
\begin{equation}
    \ket{\tilde \psi_{t+dt}} = e^{-i \hat{H} dt} \hat P(J_{t}) \ket{\tilde \psi_t}. \label{eq:non-normalized}
\end{equation}
Hence, for any finite measurement strengths $\gamma$, $\ket{\tilde \psi_t}$ depends on the sequence of measurement outcomes $\{J_{\tau}\}_{t_0 \leq \tau <t}$. We do not write out this dependence for simplicity, but we note that the probability of finding a measurement sequence leading to that specific state is given by the norm of the state $\Tr \ket{\tilde \psi_t} \bra{\tilde \psi_t}$. Therefore, the probability of finding the system in a state $\ket{\psi_t}=\ket{\tilde \psi_t}/\abs{\abs{\ket{\tilde \psi_t}}}$ at time $t$, or equivalently finding the non-normalized state $\ket{\tilde \psi_t}$ is equal to the probability of measuring the corresponding sequence of measurement outcomes $\{J_{\tau}\}_{t_0 \leq \tau <t}$.
\begin{equation}
    \mathcal{P}(\{J_{\tau}\}_{t_0 \leq \tau <t}) = \abs{\abs{\ket{\tilde \psi_t}}}^2 = \Tr \ket{\tilde \psi_t} \bra{\tilde \psi_t}.
\end{equation}
We realize that at a given time $t$, the statistical average over realizations amounts to an average over all possible trajectories of measurement outcomes $\{J_{\tau}\}_{t_0 \leq \tau <t}$. Therefore, we make the average over trajectories more explicit, obtaining
\begin{equation}
    \overline{(\dots)} = \int \mathcal D J \mathcal{P}(\{ J_{\tau} \}_{t_0 \leq \tau <t}) (\dots), 
\end{equation}
where $\int \mathcal D J$ denotes the integral over all $J_{t_0 \leq \tau <t} \in \mathbb{R}$. This allows to rewrite the replica density operator as
\begin{align}
    \hat{\rho}_{M,t} &= \overline{\otimes_{r=1}^M \ket{\psi_t}\bra{\psi_t}} \notag \\ &= \int \mathcal D J p(\{ J_{\tau} \}_{t_0 \leq \tau <t}) \otimes_{r=1}^M \ket{\psi_t}\bra{\psi_t} \notag \\
    &= \int \mathcal D J \left( \Tr \ket{\tilde \psi_t} \bra{\tilde \psi_t} \right)^{1-M} \otimes_{r=1}^M \ket{\tilde \psi_t}\bra{\tilde \psi_t}.
\end{align}
Note the dependence of the states $\ket{\psi_t}$ and $\ket{\tilde \psi_t}$ on the measurement record $\{J_{\tau}\}_{t_0 \leq \tau <t}$. The division by the trace for each replica causes problems in obtaining a closed expression for the time-evolution expression of the replica density operator. Therefore, we employ the following replica trick introducing
\begin{equation}
    \hat{\rho}_{R,M,t} = \overline{\left(\Tr \ket{\tilde \psi_t} \bra{\tilde \psi_t}\right)^{R-M} \otimes_{r=1}^M \ket{\tilde \psi_t}\bra{ \tilde \psi_t}}.
\end{equation}
For $M \leq R$, this may be written as
\begin{align}
    \hat{\rho}_{R,M,t} &= \Tr_{r>M} \tilde \rho_{R,t}, & \tilde \rho_{R,t} &= \overline{\ket{\tilde \psi_t} \bra{\tilde \psi_t}}=\int \mathcal D J \otimes_{r=1}^R \ket{\tilde \psi_t}\bra{ \tilde \psi_t}. \label{eq:def_replica_density_operator}
\end{align}
This expression allows us to formulate a linear generalized Lindblad quantum master equation describing the time-evolution of for a general replica number $R$ in absence of normalization in every time-step. Applying Eq.~\eqref{eq:non-normalized}, the time-evolution of the replicated density operator is
\begin{equation}
    \tilde \rho_{R,t+dt} = \left( e^{-i \hat{H}dt} \right)^{\otimes R} \overline{\hat{P}^{\otimes 2R}} ( t, \tilde \rho_{R,t} ) \left( e^{i\hat{H}dt} \right)^{\otimes R}.
\end{equation}
Here,
\begin{align}
    \overline{\hat{P}^{\otimes 2R}} (t, \tilde \rho_{R,t}) = \int dJ_t \left( \otimes_{r=1}^R \hat{P}(J_t) \right) \tilde \rho_{R,t} \left( \otimes_{r=1}^R \hat{P}(J_t) \right),
\end{align}
is acting on the density-operator from both sides and the integral runs over $J_{t}$ at a fixed time $t$ describing the update from time $t \to t+dt$ due to the measurement at that time step. Since the generalized projectors are Gaussian in the measurement results $J_{l,t}$, the integral may be evaluated exactly, which allows us to derive a closed differential equation for the time-evolution of the un-normalized replica density operator. It is convenient to denote operators acting only on a single replica as $\hat{A}^{(r)}=\hat{1}^{\otimes r-1} \otimes \hat{A} \otimes \hat{1}^{\otimes R-r}$, i.e. $\hat{P}^{(r)}(J_{t}) = \left( \frac{2 \gamma dt}{\pi} \right)^{1/4} e^{-\gamma dt(J_t-\hat{O}^{(r)})^2}$. It follows that $\otimes_{r=1}^R \hat{P}(J_t) = \prod_r \hat{P}^{(r)}(J_t)$. Note that while the operator $\hat{O}^{(r)}$ depends on the chosen replica, the observed current does not, which induces a coupling between the different replicas in the time-evolution. The problem then simplifies to the average
\begin{align}
    \int dJ \left( \prod_{r=1}^R \hat{P}^{(r)}(J) \right) \tilde \rho_R \left( \prod_{r=1}^R \hat{P}^{(r)}(J) \right) \notag 
    &= \left( \frac{2 \gamma dt}{\pi} \right)^{R/2} \int dJ e^{-\gamma dt \sum_{r=1}^R (J-\hat{O}^{(r)})^2} \tilde \rho_R e^{-\gamma dt \sum_{r=1}^R (J-\hat{O}^{(r)})^2} \notag \\
    &= \left( \frac{2 \gamma dt}{\pi} \right)^{R/2} \int dJ e^{-\gamma dt \sum_{r=1}^R \left[ (J-\hat{O}^{(r)}_-)^2 + (J-\hat{O}^{(r)}_+)^2\right]} \tilde \rho_R,
\end{align}
where the index $\pm$ indicates that the operator acts on $\tilde \rho_R$ from either side. This notation corresponds to the forward and backward Keldysh contour in the path integral formalism. Introducing the replica-summed operator $\overline{O}^{R}= \sum_r \hat{O}^{(r)}$ we find
\begin{equation}
    \left( \frac{2 \gamma dt}{\pi} \right)^{R/2} \int dJ e^{-\gamma dt \left( 2J^2 - 2J \left( \overline{O_+}^R + \overline{O_-}^R\right) + \left( \overline{O_+^2}^R + \overline{O_-^2}^R\right) \right)} \tilde \rho_R \notag 
    = \frac{1}{\sqrt{R}} \left( \frac{2\gamma dt}{\pi} \right)^{\frac{R-1}{2}} e^{\frac{1}{2} \gamma dt \left[ \left( \overline{O_+}^R + \overline{O_-}^R \right)^2 - 2 \left( \overline{O^2_+}^R + \overline{O^2_-}^R \right) \right]} \tilde \rho_R.
\end{equation}
Ultimately, we are only interested in the limit $R \to 1$. While retaining the $R$ copies of all operators and states, the explicit presence of $R$ in the normalization does not affect the structure of the master equation. Therefore, we take the limit directly at this stage. This step is necessary to ensure differentiability in $dt$; otherwise, the result would be ill-defined due to the lack of proper normalization in the state update. Applying the correct replica limit then allows us to expand in small $dt$, leading to
\begin{align}
    \tilde \rho_R + dt \gamma \left[ \overline{O}^R \tilde \rho_R \overline{O}^R - \frac{1}{2} \left\{ 2 \overline{O^2}^R - \left(\overline{O}^R\right)^2, \tilde \rho_R \right\} \right] + \mathcal{O}(dt^2).
\end{align}
Adding the Hamiltonian contribution yields the replica master equation
\begin{align}
    &\partial_t \tilde \rho_R = -i[\overline{H}^R , \tilde \rho_R] \notag + \gamma \left[ \overline{O}^R \tilde \rho_R \overline{O}^R - \frac{1}{2} \left\{ 2 \overline{O^2}^R - \left(\overline{O}^R\right)^2, \tilde \rho_R \right\} \right].
\end{align}
In this result, we not assume any additional properties about the observables which makes it $\hat{O}$ very general. For $R=1$, we may use $\overline{O^2}^R=\hat O^2 = \left( \overline{O}^R \right)^2$ such that we obtain a simplification in the anti-commutator revealing the ordinary Lindblad quantum master equation. For $R>1$ however, we find that $\left( \overline{O}^R \right)^2$ contains products of operators acting on different replicas, introducing a coupling between the replicas. Taking the replica limit $R\to1$ wherever $R$ appears explicitly as a prefactor -- but keeping all operators -- yields the generalized Lindblad equation
\begin{align}
    \partial_t \tilde \rho_{R,t} &= \mathcal{L}_R \tilde \rho_{R,t} = \sum_{r=1}^R \mathcal{L}^{(r)} \tilde \rho_{R,t} + \sum_{r,r', r \neq r'} \mathcal{M}^{(r,r')} \tilde \rho_{R,t}, \label{eq:replicaME} \\
    \mathcal{L}^{(r)} ( . ) &= - i [\hat{H}^{(r)},( . )] - \frac{1}{2} \gamma \left[ \hat{O}^{(r)} , \left[ \hat{O}^{(r)} , ( . ) \right] \right], \\
    \mathcal{M}^{(r,r')} ( . ) &= \frac{1}{2}  \gamma \left\{ \hat{O}^{(r)} , \left\{ \hat{O}^{(r')} , ( . ) \right\} \right\} .  
\end{align}
Note that if we take $R=1$ everywhere, i.e. set all operators $\hat{O}^{(r>1)}=$, this yields the Lindblad quantum master equation. Otherwise, besides just copying the dynamics, we obtain a term $\mathcal{M}^{(r,r')}$ pairwise coupling the replicas. On every individual replica, we obtain the same dynamics as we would get for the linear average, i.e. heating to an infinite temperature state $\hat \rho \sim \hat{1}$ while this is not a solution for the inter-replica coupling terms due to the replacement of commutators $[,]$ by anti-commutators $\{,\}$. Note as well that $R$ does not appear as a prefactor at all but we had to set $R\to 1$ in order to obtain a proper differential equation in the first place. \par
\section{Feynman Keldysh path-integral construction for the correlation function}
\label{app:Feynman_Keldysh}
We start from the bosonized form of the monitored Dirac fermion model. It can straightforwardly be put on the replica level, resulting in the time evolution of the un-normalized replica density operator $\partial_t \tilde \rho_{R,t}=\mathcal{L}_R \tilde \rho_{R,t}$ according to the derivation in App.~\ref{App:replica}. The operators $\hat{H}^{(r)}$ and $\hat{O}_{1}^{(r)}(x),\hat{O}_{2}^{(r)}(x)$ appearing in the superoperator $\mathcal{L}_R$ can be all represented in terms of the Hermitian operators $\hat{\phi}^{(r)}(x)$ and $\hat{\theta}^{(r)}(x)$ with $[\hat{\phi}^{(r)}(x),\partial_{x'} \hat{\theta}^{(r')}(x')]=i \pi \delta(x-x') \delta_{r,r'}$. Introducing $\hat{\pi}^{(r)}(x)=\frac{1}{\pi}\partial_{x} \hat{\theta}^{(r)}(x)$ we find that $\hat{\phi}^{(r)}(x)$ and $\hat{\pi}^{(r)}(x)$ are canonically conjugate variables. Hence, we may use the Feynman Keldysh path integral construction, using 'position' and 'momentum' eigenstates on all replicas and positions.
\begin{align}
    \hat{\phi}^{(r)}(x)\ket{\{\phi^{(r)}(x)\}_{r,x}} &= \phi^{(r)}(x) \ket{\{\phi^{(r)}(x)\}_{r,x}}, & \hat{\pi}^{(r)}(x) \ket{\{\pi^{(r)}(x)\}_{r,x}} &= \pi^{(r)}(x) \ket{\{\pi^{(r)}(x)\}_{r,x}} ,
\end{align}
with the overlap
\begin{equation}
    \braket{\{\phi^{(r)}(x)\}_{r,x}}{\{\pi^{(r)}(x)\}_{r,x}} = \braket{\{\pi^{(r)}(x)\}_{r,x}}{\{\phi^{(r)}(x)\}_{r,x}}^* = e^{i \sum_{r=1}^{R} \int dx \phi^{(r)}(x) \pi^{(r)}(x)}.
\end{equation}
By the notation $\{ \phi(a,b) \}_a$ we mean the collection of fields with all possible indices $a$ for a given external index $b$. This is used throughout the following derivation with various indices, like space, time, contour index, and replica number. We find the resolutions of unity
\begin{align}
    \hat{1} &=\int d\{\phi^{(r)}(x)\}_{r,x} \ket{\{\phi^{(r)}(x)\}_{r,x}}\bra{\{\phi^{(r)}(x)\}_{r,x}}, & \hat{1}&= \int d\{\pi^{(r)}(x)\}_{r,x} \ket{\{\pi^{(r)}(x)\}_{r,x}}\bra{\{\pi^{(r)}(x)\}_{r,x}}.
\end{align}
All of them are treated as discrete variables. Based on these eigenstates, we now construct the Keldysh path integral for the correlations of the variable $\hat{O}^{(r)}(x)$ (which only depends locally on $\hat{\phi}^{(r)}(x)$ and its derivatives) at a given time $t$, i.e.
\begin{equation}
    \langle \hat{O}^{(r)}(x) \hat{O}^{(r')}(x') \rangle_t \equiv \lim_{R \to 1}  \Tr \hat{O}^{(r)}(x) \hat{O}^{(r')}(x') \tilde \rho_R(t) = \lim_{R \to 1}  \Tr \hat{O}^{(r)}(x) \hat{O}^{(r')}(x') U_{t-t_0} (\tilde \rho_R(t_0) ).
\end{equation}
where $U_{t-t_0}(\tilde \rho_R(t_0))=e^{(t-t_0)\mathcal{L}_R} \tilde{\rho}_R(t_0)$ is the (non-unitary) superoperator that generates the time evolution. In usual Lindblad dynamics, the time-evolution operator is trace preserving, i.e. $\Tr U_t(\hat{A})=\Tr \hat{A}$ for arbitrary operators $\hat{A}$. If we ignore the coupling between replicas induced by measurements, the trace is preserved and therefore
\begin{equation}
    \langle \hat{O}^{(r)}(x) \hat{O}^{(r')}(x') \rangle_t = \lim_{R \to 1}  \Tr ( U_{t_f-t} ( \hat{O}^{(r)}(x) \hat{O}^{(r')}(x') U_{t-t_0}(\tilde \rho_R(t_0)))).
\end{equation}
This gives the insertions of the field correlation function a physical meaning at each point in time, earlier than $t_f$. In particular, the cyclic property of the trace ensures that at any point in time earlier than $t_f$, we find
\begin{equation}
    \lim_{R \to 1}  \Tr ( U_{t_f-t} ( \hat{O}^{(r)}(x) \hat{O}^{(r')}(x') U_{t-t_0}(\tilde \rho_R(t_0))))=\lim_{R \to 1}  \Tr ( U_{t_f-t} ( U_{t-t_0}(\tilde \rho_R(t_0))\hat{O}^{(r)}(x) \hat{O}^{(r')}(x'))). 
\end{equation}
In the replica formalism, this property is only recovered for $R=1$, in which case the correlation function is only computable for $r=r'$. For the other correlation functions, the extension of the time evolution changes the result as the order of the evaluation matters. For that reason, we keep $R>1$ arbitrary. Therefore, we cannot directly extend time to $t_f$ and need to regularize the theory accordingly. An natural way of doing so is to extend the time by a modified time-evolution -- that preserves the trace -- after the time $t$ where the correlation function is computed. This can for example be realized by assuming that the generator of the dynamics, i.e. measurement rate $\gamma$ obtains an explicit time dependence and is switched off after the evaluation of the path integral. This means that $U_{t_f-t}$ preserves the trace and can be added without changing the result at time $t$. Under this assumption, we now construct the path integral. \par
We perform the usual Trotter decomposition for a Keldysh path integral, i.e.
\begin{equation}
    U_{t_f-t_0} = \prod_{n=0}^{{t_f}/\delta_t-1} \left(1+\delta t \mathcal{L}_R(\gamma(\tau_n))\right)
\end{equation}
where $\delta t \to 0$ and $\tau_n=t_0+n\delta t$. We have introduced a time-dependent generator of the dynamics via a time-dependent measurement strength, $\gamma(\tau_n)=\gamma$ for $\tau_n<t$ and $\gamma(\tau_n)=0$ else. The action of a single time-step on the initial state reads
\begin{multline}
    (1+\delta t \mathcal{L}_R(t_0)) \tilde \rho_R(t_0) = \int d\{ \phi_{\sigma}^{(r)}(x,\tau_0),\pi_{\sigma}^{(r)}(x,\tau_0) \}_{x,r,\sigma}  (1+\delta t \mathcal{L}_R(\tau_0))   \ket{\{\phi_+^{(r)}(x,\tau_0) \}_{x,r}} \bra{\{\phi_+^{(r)}(x,\tau_0) \}_{x,r}} \ket{\{\pi_+^{(r)}(x,\tau_0) \}_{x,r}} \\ \bra{\{\pi_+^{(r)}(x,\tau_0) \}_{x,r}} \tilde \rho_R(t_0) \ket{\{\phi_-^{(r)}(x,\tau_0) \}_{x,r}} \bra{\{\phi_-^{(r)}(x,\tau_0) \}_{x,r}} \ket{\{\pi_-^{(r)}(x,\tau_0) \}_{x,r}} \bra{\{\pi_-^{(r)}(x,\tau_0) \}_{x,r}} ,
\end{multline}
where $\sigma=\pm$ is the Keldysh contour index and $\tau_0$ is an auxiliary label. Evaluating the overlaps and reordering yields
\begin{multline}
    (1+\delta t \mathcal{L}_R(t_0)) \tilde \rho_R(t_0) = \int d\{ \phi_{\sigma}^{(r)}(x,\tau_0),\pi_{\sigma}^{(r)}(x,\tau_0) \}_{x,r,\sigma} e^{i\sum_{\sigma,r}\int_x \phi^{(r)}_\sigma(x,\tau_0)\pi^{(r)}_\sigma(x,\tau_0)} \bra{\{\pi_+^{(r)}(x,\tau_0) \}_{x,r}} \tilde \rho_R(t_0) \ket{\{\phi_-^{(r)}(x,\tau_0) \}_{x,r}} \\ (1+\delta t \mathcal{L}_R(\tau_0)) \ket{\{ \phi^{(r)}_+(x,\tau_0) \}_{x,r}} \bra{\{\pi^{(r)}_{-}(x,\tau_0)\}_{x,r}}.
\end{multline}
Repeating this $N=(t_f-t_0)/\delta t$ times and closing the trace yields
\begin{equation}
    \Tr \tilde \rho_{R}(t_f) = \lim_{\delta t \to 0} \int d\{ \phi_{\sigma}^{(r)}(x,\tau_n),\pi_{\sigma}^{(r)}(x,\tau_n) \}_{x,r,\sigma,n} e^{i S(\{ \phi_{\sigma}^{(r)}(x,\tau_n),\pi_{\sigma}^{(r)}(x,\tau_n) \}_{x,r,\sigma,n})} 
\end{equation}
where $\tau_n=t_0+n\delta t$ and $n$ runs from $0$ to $N$ and the action reads
\begin{multline}
    S(\{\phi_\sigma^{(r)}(x,\tau_n),\pi_\sigma^{(r)}(x,\tau_n)\}_{r,\sigma,x,n}) \\=  \sum_{\sigma,\sigma'=\pm} \sum_{n,n'=0}^{N} \int_x \sum_{r} \phi^{(r)}_\sigma(x,\tau_{n}) (\delta_{\sigma,\sigma'}(\delta_{n,n'}-\delta_{n,n'+\sigma}) - \delta_{\sigma,+}\delta_{\sigma',-} \delta_{n,N} \delta_{n',N} - \delta_{\sigma,-}\delta_{\sigma',+} \delta_{n,0} \delta_{n',0}) \pi^{(r)}_{\sigma'}(x,\tau_{n'})  \\ -i\delta t \sum_{n=1}^N \frac{\bra{\{ \pi_+^{(r)}(x,\tau_n) \}_{r,x}} \mathcal{L}_R(\gamma(\tau_{n-1}),\ket{\{\phi_+^{(r)}(x,\tau_{n-1})\}_{x,r}}\bra{\{\pi_-^{(r)}(x,\tau_{n-1})\}_{x,r}}) \ket{\{\phi_-^{(r)}(x,\tau_{n})\}_{x,r}}}{\bra{\{ \pi_+^{(r)}(x,\tau_n) \}_{r,x}} \ket{\{\phi_+^{(r)}(x,\tau_{n-1})\}_{x,r}}\bra{\{\pi_-^{(r)}(x,\tau_{n-1})\}_{x,r}} \ket{\{\phi_-^{(r)}(x,\tau_{n})\}_{x,r}}} .
\end{multline}
For the closing at $\tau_0$ we used that $\tilde \rho_R(t_0)\sim 1$, i.e., we initialize the system in an infinite temperature state independently on each replica. This is motivated by the fact that the average dynamics on a single replica approaches this state under the measurement dynamics. However, the boundary conditions are not important as we extend the initial and final time, where boundary conditions enter the action, to $\pm \infty$. Therefore, they are irrelevant for the observables in the stationary state that we intend to compute. \par
The correlation function can be written in terms of this path integral as,
\begin{align}
    \langle \hat{O}^{(r)}(x) \hat{O}^{(r')}(x') \rangle_t &= \lim_{R \to 1} \lim_{\delta t \to 0} \int d\{ \phi_{\sigma}^{(r)}(x,\tau_n),\pi_{\sigma}^{(r)}(x,\tau_n) \}_{x,r,\sigma,n}  O_+^{(r)}(x,t) O_+^{(r')}(x',t) e^{i S(\{ \phi_{\sigma}^{(r)}(x,\tau_n),\pi_{\sigma}^{(r)}(x,\tau_n) \}_{x,r,\sigma,n})} \notag \\ &= \lim_{R \to 1} \lim_{\delta t \to 0} \int d\{ \phi_{\sigma}^{(r)}(x,\tau_n),\pi_{\sigma}^{(r)}(x,\tau_n) \}_{x,r,\sigma,n}  O_-^{(r)}(x,t+\delta t) O_-^{(r')}(x',t+\delta t) e^{i S(\{ \phi_{\sigma}^{(r)}(x,\tau_n),\pi_{\sigma}^{(r)}(x,\tau_n) \}_{x,r,\sigma,n})} .
\end{align}
depending on where the fields are inserted prior to the evaluation of the time evolution from $t$ to $t_f$. Based on the construction, both representations have to be equal if the measurement is switched off after the time $t-\delta t$. In the path integral setting, physically meaningful observables cannot depend on the dynamics that happens after the time at which they are evaluated (by causality), such that we may just as well let $\gamma$ be constant for all times, also after $t$, in order to recover time-translation symmetry of the action, crucial for the RG procedure. However, this breaks the equality of the $++$ and $--$ correlation function as we can see by solving the Gaussian integral, ignoring the interacting measurement, see App.~\ref{app:BosoObservables}. Therefore, we need to regularize the path integral representation of the correlation function. As $\hat{\phi}^{(r)}(x)$ is Hermitian and commutes with itself, we know that the correlation function has to be real. Therefore, we make it manifestly real according to
\begin{equation}
    \langle \hat{O}^{(r)}(x) \hat{O}^{(r')}(x') \rangle_t = \frac{1}{2}\Tr U_{t_{f}-t}(\lbrace \hat{O}^{(r)}(x) \hat{O}^{(r')}(x') , U_{t-t_0}(\tilde \rho_R(t_0)) \rbrace).
\end{equation}
In this representation, we make sure that the result is real independently of the question whether $\gamma=0$ or $\gamma>0$ for $\tau\geq t$, as long as the generator of the dynamics preserves Hermeticity, which is a fundamental symmetry of the generalized Lindblad quantum master equation for measurements. For that reason, the causal path integral representation of the correlation function reads
\begin{equation}
    \langle \hat{O}^{(r)}(x) \hat{O}^{(r')}(x') \rangle_t = \frac{1}{2} \langle O_+^{(r)}(x,t) O_+^{(r')}(x',t)  \rangle + \frac{1}{2} \langle O_-^{(r)}(x,t) O_-^{(r')}(x',t)  \rangle = \langle O_c^{(r)}(x,t) O_c^{(r')}(x',t) \rangle.
\end{equation}
where we use $O_{c/q}^{(r)}(x,t)=(O_{+}^{(r)}(x,t)\pm O_{-}^{(r)}(x,t))/\sqrt{2}$. In the usual Keldysh formalism, this definition is only used for field operators. Here, however, the correlation functions of arbitrary observables are governed by the respective 'classical' fields as defined above. Additionally, in contrast to the usual Keldysh formalism, there emerges a finite $qq$ correlation function between the fields $\phi,\theta$ if the time evolution is continued with a finite measurement strength, which means that it violates causality, but this object does not have a physical meaning and the relevant observables retain a causal structure. \par
This construction implies that $\Im \langle O_c^{(r)}(x,t) O_c^{(r')}(x',t) \rangle=0$ which means that $\langle O_+^{(r)}(x,t) O_+^{(r')}(x',t) \rangle=\langle O_-^{(r)}(x,t) O_-^{(r')}(x',t) \rangle^*$. Hence, the path integral must be symmetric under exchange of $+$ and $-$ fields and simultaneous complex conjugation. For the action, this implies (ignoring the dependence on $\pi_\sigma^{(r)}(x,t)$ which is integrated out, taking the temporal continuum limit and suppressing all indices besides the contour index)
\begin{equation}
    S[\phi_+,\phi_-] = -S^*[\phi_-,\phi_+].
\end{equation}
In particular, this symmetry must be preserved by the RG because a breaking would imply that the correlation function is no longer real which is un-physical. 
\section{Integrating out $\theta$}
\label{App:IntegratingTheta}
The field $\theta$ does not appear in the measurement operator and the Hamiltonian of the nearest neighbor hopping model is simply given by
\begin{equation}
    \hat{H} = \frac{v}{2\pi} \int_x \left[ g(\partial_x \hat{\phi})^2 + \frac{1}{g}(\partial_x \hat{\theta})^2 \right].
\end{equation}
It is therefore quadratic in $\theta$ and we may integrate out $\theta$ exactly from the general $R$-replica Keldysh action without even considering the measurement terms. In this case, the action is found using Eq.~\eqref{eq:ActionFromLindblad} to be
\begin{align}
    S_{0,H} = -\frac{1}{2\pi} \sum_r \sum_{\sigma=\pm} \sigma \int_{t,x} \left[ \phi_\sigma^{(r)} \partial_x \partial_t \theta_\sigma^{(r)} + \theta_\sigma^{(r)} \partial_x \partial_t \phi_\sigma^{(r)} - gv \phi_\sigma^{(r)} \partial_x^2 \phi_\sigma^{(r)} - \frac{v}{g} \theta_\sigma^{(r)} \partial_x^2 \theta_\sigma^{(r)} \right] .
\end{align}
The action decouples the different $\sigma$ and $r$ components. Integrating out $\theta_\sigma^{(r)}$ is therefore done for all components individually (all fields have the contour index $\sigma$ and the replica index $r$, suppressed for short notation)
\begin{align}
    \int \mathcal{D}\theta e^{iS} &= \int \mathcal{D}\theta e^{\frac{i\sigma v}{2\pi g} \int_{x,t} \left[ \theta \partial_x^2 \theta - \frac{g}{v} \theta \partial_x \partial_t \phi - \frac{g}{v} \phi \partial_x \partial_t \theta + g^2 \phi \partial_x^2 \phi \right]} \notag  \\
    &= \int \mathcal{D}\theta e^{\frac{i\sigma v}{2\pi g} \int_{x,t} \left[ (\theta-f) \partial_x^2 (\theta-f) + f \partial_x^2 \theta + \theta \partial_x^2 f - f \partial_x^2 f - \frac{g}{v} \theta \partial_x \partial_t \phi - \frac{g}{v} \phi \partial_x \partial_t \theta + g^2 \phi \partial_x^2 \phi \right]} \notag \\
    &= \int \mathcal{D}\theta e^{\frac{i\sigma v}{2\pi g} \int_{x,t} \left[ (\theta-f) \partial_x^2 (\theta-f)   + g^2 \phi \partial_x^2 \phi - f \partial_x^2 f \right]}
\end{align}
We now fix $f$ such that $\partial_x f = \frac{g}{v} \partial_t \phi$ and shift the fields $\theta(x,t)=\tilde \theta(x,t)+ \frac{g}{v} \int_{x_0}^x dy \partial_t \tilde \phi(y,t)$ and $\phi(x,t)=\tilde \phi(x,t)$. Note that $\frac{\delta \phi(x,t)}{\delta \tilde \theta(x',t')}=0$. Therefore, the Jacobi matrix of the transformation is of block-triangular form. The blocks on the diagonal are unit matrices if one discretizes space such that the determinant is one and therefore we find $\int \mathcal{D} \phi \mathcal{D} \theta=\int \mathcal{D} \tilde \phi \mathcal{D} \tilde \theta$ for this transformation and we may equivalently just shift $f$ away from the $\theta$ term. This means
\begin{equation}
    \int \mathcal{D}\theta e^{iS} = \int \mathcal{D}\theta e^{\frac{i\sigma v}{2\pi g} \int_{x,t} \left[ \theta \partial_x^2 \theta + g^2 \phi \partial_x^2 \phi - \frac{g^2}{v^2} \phi \partial_t^2 \phi \right]}
\end{equation}
Therefore, the integral over $\theta$ can just be absorbed into the normalization of the integral and we obtain the action in terms of $\phi$ alone.
\section{Fluctuations in the infinite temperature state} \label{app:infititecorrelations}
Here we compute the correlation functions of $k=0$ fields using the Gaussian measurement action~\eqref{eq:bosogaussianreplica}. This is done in real space to make use of the fact that we integrate over real fields. $\phi^{(k=0)}_{c/q}(x,t)= \frac{1}{\sqrt{R}} \sum_r \phi^{(r)}_{c/q}(x,t)$ is real because all fields $\phi^{(r)}_{c/q}(x,t)$ are real. On the Gaussian level, the different $k$ sectors decouple and therefore the relevant Keldysh path integral has the structure ($k=0$ is implicit for brief notation)
\begin{align}
    \mathcal{Z} &= \int \mathcal{D}\phi e^{-\frac{1}{2} \int_{p,\omega} \phi^* (-iG^{-1}) \phi} = \int \mathcal{D}\phi e^{-\frac{1}{2} \int_{x,t} \phi (-iG^{-1}) \phi}, \\
   G^{-1} &= \frac{g}{\pi v} \left( \begin{array}{cc}
        0 & \omega^2 - v^2 p^2 \\
        \omega^2 - v^2 p^2 & \frac{2i\gamma v}{\pi g} p^2
    \end{array} \right) \to  \frac{g}{\pi v} \left( \begin{array}{cc}
        0 & -\partial_t^2 + v^2 \partial_x^2 \\
        -\partial_t^2 + v^2 \partial_x^2 & \frac{2i\gamma v}{\pi g} \partial_x^2
    \end{array} \right).
\end{align}
Next, we add a real source term $J(x,t)$ and use that the partition function is normalized in the replica limit $R \to 1$ which allows us to solve the Gaussian integral
\begin{equation}
    \mathcal{Z}[J] = \int \mathcal{D}\phi e^{-\frac{1}{2} \int_{x,t} \phi (-iG^{-1}) \phi + \int_{x,t} J^T \phi} = e^{\frac{1}{2} J^T iG J}.  
\end{equation}
Now, we can take functional derivatives w.r.t. $J$ and generate the correlation functions. After returning to momentum space, we can perform the inversion and we find
\begin{equation}
    \langle \phi^{(k=0)*}_{c/q}(\omega,p) \phi^{(k=0)}_{c/q}(\omega',p') \rangle = iG = i 2\pi \delta(p-p') 2\pi \delta(\omega-\omega') \frac{-\pi v/g}{(\omega^2-v^2 p^2)^2} \left( \begin{array}{cc}
        \frac{2i\gamma v }{\pi g} p^2 & -(\omega^2-v^2 p^2) \\
        -(\omega^2-v^2 p^2) & 0
    \end{array} \right).
\end{equation}
To get the occupation number of the modes, we are interested in the result at equal times, i.e.
\begin{equation}
    \langle \phi^{(k=0)*}_{c/q}(t,p) \phi^{(k=0)}_{c/q}(t,p') \rangle = 2\pi \delta(p-p') \int \frac{d\omega}{2\pi} \frac{-\pi i v/g}{(\omega^2-v^2 p^2)^2}  \left( \begin{array}{cc}
        \frac{2i\gamma v }{\pi g} p^2 & -(\omega^2-v^2 p^2) \\
        -(\omega^2-v^2 p^2) & 0
    \end{array} \right).
\end{equation}
All components of the integral are not convergent. To regularize, we may use the usual $\omega \to \omega \pm i0^+$ in the retarded/advanced sector introducing an infinitesimal dissipation $0^+>0$. This renders the frequency-integral finite and we obtain
\begin{multline}
    \langle \phi^{(k=0)*}_{c/q}(t,p) \phi^{(k=0)}_{c/q}(t,p') \rangle = 2\pi \delta(p-p') \int \frac{d\omega}{2\pi} \frac{- \pi i v/g}{((\omega+i0^+)^2-v^2 p^2)((\omega-i0^+)^2-v^2 p^2)} \\ \times \left( \begin{array}{cc}
        \frac{2i\gamma v }{\pi g} p^2 & -((\omega+i0^+)^2-v^2 p^2) \\
        -((\omega-i0^+)^2-v^2 p^2) & 0
    \end{array} \right).
\end{multline}
Including this regularization, the frequency-integral can be performed. On the off-diagonals, the result is zero as the functions only have poles in either the upper or the lower half plane and we may close the contour in the other half plane. Therefore, the only contribution at equal times comes from the $cc$ correlation function which reads
\begin{equation}
    \langle \phi^{(k=0)*}_{c}(t,p) \phi^{(k=0)}_{c}(t,p') \rangle = 2 \pi \delta(p-p') \int \frac{d\omega}{2\pi} \frac{2 \gamma v^2 p^2/g^2}{(\omega^2-v^2 p^2-(0^+)^2)^2+4(0^+)^2 \omega^2}.
\end{equation}
The integrand now has 4 poles, $\omega=\pm 2i0^+ \pm pv$ i.e. 2 in both half planes. Therefore, we need to close the contour around two of the poles. We choose the upper half plane to find
\begin{multline}
    \int \frac{d\omega}{2\pi} \frac{1}{(\omega^2-v^2 p^2-(0^+)^2)^2+4(0^+)^2 \omega^2} = \ointctrclockwise_{z=pv+2i(0^+)} \frac{dz}{2\pi i} \frac{i}{(\omega^2-v^2 p^2-(0^+)^2)^2+4(0^+)^2 \omega^2}  \\+ \ointctrclockwise_{z=-pv+2i(0^+)} \frac{dz}{2\pi i} \frac{i}{(\omega^2-v^2 p^2-(0^+)^2)^2+4(0^+)^2 \omega^2}  
    = \frac{1}{8 (0^+) p^2 v^2}.
\end{multline}
This yields Eq.~\eqref{eq:InfiniteKeldysh}.
\section{Observables of the Sine-Gordon action} \label{app:BosoObservables}
Here we compute the Keldysh Greens function from the action~\eqref{eq:nearest_neighbor_action}.
To do so, remember that the fields are real in the original replica diagonal formulation. To take this into account when solving the path integral, we rewrite the (normalized) Keldysh partition function including real source terms $J^{(r)}_{\sigma}$ with the constraint (the $k=0$ mode has been integrated out such that we do not couple it to a source term) $\sum_r J^{(r)}_{\sigma}=0$
\begin{align}
    \mathcal{Z}[J] &= \int \mathcal{D}\phi e^{-\frac{1}{2} \sum_{k>0,\sigma}  \int_{\omega,p} \phi_{\sigma}^{(k)*} (-iG_\sigma^{(-1)}) \phi_{\sigma}^{(k)} +  \sum_{k>0,\sigma} \int_{x,t} J_\sigma^{(k)*}\phi_{\sigma}^{(k)}} \notag \\
    &= \prod_{k>0,\sigma} \int \mathcal{D}\phi e^{-\frac{1}{2} \int_{x,t} \phi_{\sigma}^{(k)*} (-iG_\sigma^{(-1)}) \phi_{\sigma}^{(k)} + \int_{x,t} J_\sigma^{(r)}\phi_{\sigma}^{(r)}} \notag \\
    &= \prod_{k,\sigma} \int \mathcal{D}\phi e^{-\frac{1}{2} \int_{x,t} \phi_{\sigma}^{(k)*} (-iG_\sigma^{(-1)}) \phi_{\sigma}^{(k)} + \int_{x,t} J_\sigma^{(r)}\phi_{\sigma}^{(r)}} \notag \\
    &= \int \mathcal{D}\phi e^{-\frac{1}{2} \sum_{k,\sigma} \int_{\omega,p} \phi_{\sigma}^{(k)*} (-iG_\sigma^{(-1)}) \phi_{\sigma}^{(k)} +  \sum_{k,\sigma} \int_{x,t} J_\sigma^{(k)*}\phi_{\sigma}^{(k)}} \notag \\
    &= \int \mathcal{D}\phi e^{-\frac{1}{2} \sum_{r,\sigma} \int_{\omega,p} \phi_{\sigma}^{(r)} (-iG_\sigma^{(-1)}) \phi_{\sigma}^{(r)} +  \sum_{r,\sigma} \int_{x,t} J_\sigma^{(r)}\phi_{\sigma}^{(r)}} \notag \\
    &= e^{\frac{1}{2} \sum_{r,\sigma} \int_{x,t} J_{\sigma}^{(r)} iG_\sigma J^{(r)}_\sigma}.
\end{align}
In between, we multiplied by $1$ reintroducing the field $\phi^{(k=0)}$ which has no physical meaning here but is just used to compute the integral. The important point is here, that the result is still a real Gaussian integral with the corresponding factor of $2$. We can read off the correlation function,
\begin{equation}
    \langle \phi_{\sigma}^{(k>0)*}(p,\omega) \phi_{\sigma'}^{(k'>0)}(p',\omega') \rangle = 2\pi \delta(\omega-\omega') 2\pi \delta(p-p') \delta_{\sigma \sigma'} \delta_{kk'} \frac{\pi i}{\sigma K_\sigma} \frac{1}{\frac{\omega^2}{\eta_\sigma} - p^2 \eta_\sigma}.
\end{equation}
We are interested in the correlations at equal times, i.e. 
\begin{equation}
    \langle \phi_{\sigma}^{(k>0)*}(t,p) \phi_{\sigma'}^{(k'>0)}(t,p') \rangle = 2\pi \delta(p-p') \delta_{\sigma \sigma'} \delta_{kk'} \int \frac{d\omega}{2\pi} \frac{\sigma \pi i\eta_\sigma/K_\sigma}{\omega^2 - p^2 \eta_\sigma^2}.
\end{equation}
The poles of the integral are at $\pm \sqrt{p^2 \eta_\sigma^2}$. As we see later, $\eta_\sigma$ is not renormalized and due to $\text{sgn} \Im \eta_\sigma = -\sigma$, we find that the $-\sigma$ solution is in the upper half plane. Therefore, when closing the contour in the upper half plane, we have to integrate counter-clockwise around the pole at $\omega = - \sqrt{p^2 \eta_\sigma^2}$. This yields for the integral
\begin{equation}
    \langle \phi_{\sigma}^{(k>0)*}(t,p) \phi_{\sigma'}^{(k'>0)}(t,p') \rangle = 2\pi \delta(p-p') \delta_{\sigma \sigma'} \delta_{kk'} \frac{\pi}{2 K_\sigma \vert p \vert}  .
\end{equation}
Therefore, we can read off the correlation function in momentum space, trivially taking the limit $R \to 1$, and with $c=\Re 1/K_+$ we obtain Eq.~\eqref{eq:Cofpres} from the main text.
\section{Integrating out the infinite temperature mode} \label{app:IntegratingOutK0}
The bosonized replica Keldysh partition function~\eqref{eq:ActionFromLindblad} of monitored free fermions takes the form
\begin{align}
    \mathcal{Z}_{R} &= \int \mathcal{D} \phi^{(k=0)} \mathcal{D} \phi^{(k>0)} e^{i\sum_{k>0} S_{0}^{(k)}[\phi^{(k)}] + i \Delta S[\phi]} e^{i S_{0}^{(k=0)}[\phi^{(k=0)}]} , 
\end{align}
where $\Delta S[\phi]$ contains the non-linear terms emerging due to the non-linear measurement operator $\hat{O}_2= m \cos{2 \hat{\phi}}$. In this section, we treat this term as a perturbation. The Gaussian part of the action can be separated into decoupled $k$ sectors $S_0^{(k)}[\phi^{(k)}]$. In App.~\ref{app:infititecorrelations}, we show that the $k=0$ sector has infinitely fluctuating fields due to the heating of the average mode. Because of that, the perturbative treatment of $\Delta S[\phi]$ simplifies drastically. We find
\begin{align}
    \mathcal{Z}_{R} &\approx \int \mathcal{D} \phi^{k>0} \mathcal{D} \phi^{k=0}  (1+i \Delta \Delta S[\phi]) e^{i S_{0}^{(k=0)}[\phi^{(k=0)}]}  e^{i\sum_{k>0} S_{0}^{k>0}[\phi^{(k)}]} \notag \\
    &\approx \int \mathcal{D} \phi^{k>0}  (1+i \langle \Delta S[\phi] \rangle_{S_{0}^{(k=0)}}) e^{i\sum_{k>0} S_{0}^{(k>0)}[\phi^{(k)}]} \notag \\
    &\approx \int \mathcal{D} \phi^{k>0} \exp{i \langle \Delta S[\phi] \rangle_{S_{0}^{(k=0)}} + i\sum_{k>0} S_{0}^{k>0}[\phi^{(k)}]},
\end{align}
where $\langle \dots \rangle_{S_{0}^{(k=0)}}$ denotes the expectation value over the infinite temperature state of the replica-averaged mode $k=0$, as derived above. Repeating this calculation to the next order yields the expansion
\begin{equation}
    S^{k>0}[\phi^{k>0}] = \sum_{k>0} S_{0}^{k>0}[\phi^{(k)}] + \langle \Delta S[\phi] \rangle_{S_{0}^{(k=0)}} -\frac{1}{2} \left( \langle (\Delta S[\phi])^2 \rangle_{S_{0}^{(k=0)}} - \langle \Delta S[\phi] \rangle_{S_{0}^{(k=0)}}^2 \right) + \dots
\end{equation}
Let us first just discuss the expansion to first order. All terms that appear in $\Delta S$ are of the form $\cos 2(\phi_\sigma^{(r)}-\phi_{\sigma'}^{(r')})$, where $\sigma,\sigma'=\pm$ and $r,r'$ arbitrary replica indices, and they are local in space and time. We use the Fourier transformation in replica space and that fluctuations of the $k=0$ mode diverge to obtain
\begin{align}
    \left\langle \cos \frac{2}{\sqrt{R}} (\phi_{\sigma}^{(k=0)}-\phi_{\sigma'}^{(k=0)}) \right\rangle_{S_{0}^{(k=0)}} &= e^{-\frac{2}{R} \left\langle \left( \phi_{\sigma}^{(k=0)}-\phi_{\sigma'}^{(k=0)}  \right)^2 \right\rangle_{S_{0}^{(k=0)}}} = \delta_{\sigma \sigma'}, \\
    \left\langle \sin \frac{2}{\sqrt{R}} (\phi_{\sigma}^{(k=0)}-\phi_{\sigma'}^{(k=0)}) \right\rangle_{S_{0}^{(k=0)}} &= 0.
\end{align}
Therefore, we can simplify
\begin{equation}
    \left\langle \cos 2(\phi_\sigma^{(r)}-\phi_{\sigma'}^{(r')}) \right\rangle_{S_{0}^{(k=0)}} = \delta_{\sigma \sigma'} \cos 2l(\phi_\sigma^{(r)}-\phi_{\sigma'}^{(r')}),
\end{equation}
which does not depend on the $k=0$ mode any more because this term just gives a constant whenever $r=r'$. This cancels out the Lindblad term entirely and due to the vanishing contour-coupling, the action drastically simplifies to
\begin{equation}
    \langle \Delta S[\phi] \rangle_{S_{0}^{(k=0)}} = -i\gamma m \int_{x,t} \sum_{\sigma=\pm} \sum_{r \neq r'} \cos 2(\phi_\sigma^{(r)}-\phi_{\sigma}^{(r')}) .
\end{equation}
Looking at the second order term, we realize that besides measure $0$ contributions from $x=y, t= \tau$ that vanish under integration, we find
\begin{multline}
    \left\langle \cos 2(\phi_\sigma^{(r)}(x,t)-\phi_{\sigma'}^{(r')}(x,t)) \cos 2(\phi_\rho^{(u)}(y, \tau)-\phi_{\rho'}^{(u')}(y, \tau)) \right\rangle_{S_{0}^{(k=0)}} \\ = \delta_{\sigma\sigma'} \delta_{\rho\rho'} \cos 2(\phi_\sigma^{(r)}(x,t)-\phi_{\sigma'}^{(r')}(x,t)) \cos 2(\phi_\rho^{(u)}(y, \tau)-\phi_{\rho'}^{(u')}(y, \tau)).
\end{multline}
Therefore
\begin{equation}
    \langle (\Delta S[\phi])^2 \rangle_{S_{0}^{(k=0)}} = \langle \Delta S[\phi] \rangle_{S_{0}^{(k=0)}}^2 ,   
\end{equation}
which means that the second order perturbative correction to the $k>0$ modes vanishes. The same structure repeats at higher order perturbation theory, which means that in the infinite temperature state of the Gaussian theory, we may integrate the $k=0$ mode out \textit{exactly}. This yields the action~\eqref{eq:relativereplicaaction}.
\section{Details on the renormalization of the Sine-Gordon model} \label{app:RGNearestNeighbor}
In this derivation, we drop the contour index since the path integral for the replica off-diagonal correlation functions~\eqref{eq:EuclidianAction} factorizes and the only difference between the two contours are the coupling that get an additional contour index. For the final RG step, the contour index becomes important again, as discussed in the main text. In order to renormalize the action~\eqref{eq:EuclidianAction}, we need the correlation functions
\begin{align}
    G_{>/<}(x,t) &=\left\langle \left( \phi^{(r)}_{>/<}(x,t)-\phi^{(r')}_{>/<}(x,t) \right) \left( \phi^{(r)}_{>/<}(0,0)-\phi^{(r')}_{>/<}(0,0) \right) \right\rangle_0 \notag \\
    &= \frac{1}{R} \sum_{k>0,k'>0} \left( e^{-i\frac{2\pi kr}{R}}-e^{-i\frac{2\pi kr'}{R}} \right) \left( e^{i\frac{2\pi k'r}{R}}-^{i\frac{2\pi k'r'}{R}} \right) \left\langle \phi^{(k)*}_{>/<}(x,t)  \phi^{(k')}_{>/<}(0,0)  \right\rangle_0 \notag \\
    &= \frac{1}{R} \sum_{k'>0} (2-e^{2\pi ik'(r-r')/R}-e^{-2\pi ik'(r-r')/R}) \left\langle \phi^{(k>0)*}_{>/<}(x,t)  \phi^{(k>0)}_{>/<}(0,0) \right\rangle_0 \notag \\
    &= \frac{1}{R} \sum_{k'=0}^{R-1} (2-e^{2\pi ik'(r-r')/R}-e^{-2\pi ik'(r-r')/R}) \left\langle \phi^{(k>0)*}_{>/<}(x,t)  \phi^{(k>0)}_{>/<}(0,0) \right\rangle_0 \notag \\
    &= 2(1-\delta_{r,r'}) \left\langle \phi^{(k>0)*}_{>/<}(x,t)  \phi^{(k>0)}_{>/<}(0,0)  \right\rangle_0 \notag \\
    &= 2(1-\delta_{r,r'}) \int_{>/<} e^{i(px+\omega t)} \int_{>/<} e^{i(p'0+\omega' 0)} \left\langle \phi^{(k>0)*}_{>/<}(p,\omega)  \phi^{(k>0)}_{>/<}(p',\omega')  \right\rangle_0 \notag \\
    &= 2(1-\delta_{r,r'}) \int_{>/<} \frac{\pi}{K} \frac{ e^{i(px+\omega t)}}{p^2 + \omega^2}. 
\end{align}
We used here that the Gaussian theory decouples $k$ modes and that it does not explicitly depend on $k$. Therefore, it does not matter, for which particular choice of $k>0$ the expectation value is computed, and we can add the $k=0$ term (which vanishes). By that mechanism, the explicit dependence on $R$ drops out and we obtain a finite contribution to the RG flow equations in that limit. Knowing that the correlation function obviously vanishes for $r=r'$, we leave out the Kronecker delta and evaluate the integral separately, writing only the leading order in $s$
\begin{align}
    G_>(x,t) &= \frac{2\pi}{K} \int_{b\Lambda < p <\Lambda} \frac{dp}{2\pi} \int_{-\infty}^{\infty} \frac{d\omega}{2\pi}  \frac{e^{i(px+\omega t)}}{p^2 +  \omega^2} \notag  \\ &= \frac{s}{\pi K} \cos \Lambda x \int_{-\infty}^\infty d\omega \frac{e^{i \omega t}}{ \omega^2 +  \Lambda^2} \notag \\
    &= \frac{s}{K} \cos \Lambda x e^{- \Lambda \vert t \vert} ,
\end{align}
and
\begin{align}
    G_<(x,t) &= \frac{2\pi}{K} \int_{-\Lambda}^\Lambda \frac{dp}{2\pi} \int_{-\infty}^{\infty} \frac{d\omega}{2\pi} e^{i(px+\omega t)} \frac{ e^{i(px+\omega t)}}{p^2 +  \omega^2} \notag  \\ &= \frac{1}{K} \int_0^\Lambda \frac{dp}{p} \cos px e^{- p \vert t \vert} \notag  \\
    &=  \frac{ \vert t \vert}{ t^2 +  x^2} \left( 1-\cos \Lambda x e^{- \Lambda \vert t \vert} \right) + \frac{x}{ x^2 + t^2} \sin \Lambda x  e^{- \Lambda \vert t \vert}.
\end{align}
This yields
\begin{align}
    \langle \Delta S \rangle_{0,>} &= i \lambda \int_{\mathbf{x}} \sum_{r \neq r'} \left\langle \cos 2 (\phi^{(r,r')}_< + \phi^{(r,r')}_>)  \right\rangle_{0,>} \notag \\
     &= i  \lambda \int_{\mathbf{x}} \sum_{r,r'} \cos 2  \phi^{(r,r')}_< e^{-2 \left\langle \left( \phi^{(r,r')}_> \right)^2 \right\rangle_{0,>}} \notag \\
     &= i \lambda \int_{\mathbf{x}} \sum_{r,r'} \cos 2 \phi^{(r,r')}_< e^{-2 G_>(0,0)} \notag  \\ & = i \lambda \int_{\mathbf{x}} \sum_{r,r'} \cos 2  \phi^{(r,r')}_< e^{-  \frac{2s}{K}}.
\end{align}
For the second order perturbative correction, consider
\begin{multline}
    -\frac{1}{2} \left( \langle \left(\Delta S\right)^2 \rangle_{0,>} - \left( \langle \Delta S \rangle_{0,>}  \right)^2 \right)
    =\frac{\lambda^2}{2} \sum_{r \neq r'} \sum_{u \neq u'} \int_{\mathbf{x},\mathbf{y}} \left\langle \cos 2 \phi^{(r,r')}(\mathbf{x}) \cos 2 \phi^{(u,u')}(\mathbf{y})   \right\rangle_{0,>}    \\
    -\frac{\lambda^2}{2} \sum_{r \neq r'} \sum_{u \neq u'} \int_{\mathbf{x},\mathbf{y}} \left\langle \cos 2 \phi^{(r,r')}(\mathbf{x}) \right\rangle_{0,>} \left\langle \cos 2 \phi^{(u,u')}(\mathbf{y})  \right\rangle_{0,>} .
\end{multline}
The locality of the model implies that no long-range coupled terms are generated, which allows us to expand in $\mathbf{x}-\mathbf{y}$ later on. In leading order, we simply generate an additional term $\Delta S^2$ which is less relevant than $\Delta S$ itself and therefore dropped. On the other hand, the spatial symmetry prohibits odd terms in $\mathbf{x}-\mathbf{y}$, and products of oscillating functions and derivative terms are less relevant than the derivative terms themselves. The only combinations where a non-oscillating term is generated at second order are $r=u,r'=u'$ and $r=u',r'=u$ which allows us to eliminate one of the sums which just generates a factor of $2$.
\begin{multline}
    -\frac{1}{2} \left( \langle \left(\Delta S\right)^2 \rangle_{0,>} - \left( \langle \Delta S \rangle_{0,>}  \right)^2 \right) \simeq
    \lambda^2 \sum_{r \neq r'} \int_{\mathbf{x},\mathbf{y}} \left\langle \cos 2 \phi^{(r,r')}(\mathbf{x}) \cos 2 \phi^{(r,r')}(\mathbf{y})   \right\rangle_{0,>}    \\
    -\lambda^2 \sum_{r \neq r'} \int_{\mathbf{x},\mathbf{y}} \left\langle \cos 2 \phi^{(r,r')}(\mathbf{x}) \right\rangle_{0,>} \left\langle \cos 2 \phi^{(r,r')}(\mathbf{y}) \right\rangle_{0,>} .
\end{multline}
When we expand the $\cos$ terms, we need to compute all combinations
\begin{align}
     \left\langle \cos 2 \phi^{(r,r')}_>(\mathbf{x}) \cos 2 \phi^{(r,r')}_>(\mathbf{y}) \right\rangle_{0,>} &= \frac{1}{2} \sum_{\sigma = \pm} \left( e^{-2\left\langle \left( \phi^{(r,r')}_>(\mathbf{x}) - \sigma \phi^{(r,r')}_>(\mathbf{y}) \right)^2 \right\rangle_{0,>}} \right) \\
     \left\langle \sin 2 \phi^{(r,r')}_>(\mathbf{x}) \sin 2 \phi^{(r,r')}_>(\mathbf{y}) \right\rangle_{0,>} &= \frac{1}{2} \sum_{\sigma = \pm} \sigma \left( e^{-2\left\langle \left( \phi^{(r,r')}_>(\mathbf{x}) - \sigma \phi^{(r,r')}_>(\mathbf{y}) \right)^2 \right\rangle_{0,>}} \right) \\
     \left\langle \sin 2 \phi^{(r,r')}_>(\mathbf{x}) \cos 2 \phi^{(r,r')}_>(\mathbf{y}) \right\rangle_{0,>} &= \left\langle \cos 2 \phi^{(r,r')}_>(\mathbf{x}) \sin 2 \phi^{(r,r')}_>(\mathbf{y}) \right\rangle_{0,>} = 0 .
\end{align}
Expanding and using translation-invariance yields
\begin{align}
    \left\langle \cos 2 \phi^{(r,r')}_>(\mathbf{x}) \cos 2 \phi^{(r,r')}_>(\mathbf{y}) \right\rangle_{0,>} &= \cosh 4 \left\langle \phi^{(r,r')}_>(\mathbf{x}) \phi^{(r,r')}_>(\mathbf{y}) \right\rangle_{0,>} e^{-4\left\langle \left( \phi^{(r,r')}_> \right)^2 \right\rangle_{0,>}} \\
    \left\langle \sin 2 \phi^{(r,r')}_>(\mathbf{x}) \sin 2 \phi^{(r,r')}_>(\mathbf{y}) \right\rangle_{0,>} &= \sinh 4 \left\langle \phi^{(r,r')}_>(\mathbf{x}) \phi^{(r,r')}_>(\mathbf{y}) \right\rangle_{0,>} e^{-4\left\langle \left( \phi^{(r,r')}_> \right)^2 \right\rangle_{0,>}}.
\end{align}
For the second order perturbative result this yields
\begin{equation}
    -\frac{1}{2} \left( \langle \left(\Delta S\right)^2 \rangle_{0,>} - \left( \langle \Delta S \rangle_{0,>}  \right)^2 \right) = \frac{\lambda^2}{2} \sum_{r \neq r'} \sum_{\sigma = \pm} \int_{\mathbf{x},\mathbf{y}} e^{-4G_>(\mathbf{0})} \left( e^{4\sigma G_>(\mathbf{x}-\mathbf{y})} -1 \right)  \cos 2 \left( \phi_<^{(r,r')}(\mathbf{x}) - \sigma \phi_<^{(r,r')}(\mathbf{y}) \right).
\end{equation}
Note that the fast mode correlation functions are $\mathcal{O}(s)$, both the local and the non-local one. Therefore, we can already at this point expand in $s$ which yields
\begin{equation}
    -\frac{1}{2} \left( \langle \left(\Delta S\right)^2 \rangle_{0,>} - \left( \langle \Delta S \rangle_{0,>}  \right)^2 \right) =  2 \lambda^2 \sum_{r \neq r'} \sum_{\sigma = \pm} \sigma \int_{\mathbf{x},\mathbf{y}} G_>(\mathbf{x}-\mathbf{y}) \cos 2 \left( \phi_<^{(r,r')}(\mathbf{x}) - \sigma \phi_<^{(r,r')}(\mathbf{y}) \right).
\end{equation}
By locality we know that the fast mode correlation function is peaked around $0$ and vanishes at large distances. Therefore it makes sense to expand $\sigma \phi_<^{(r,r')}(\mathbf{y}) \simeq \sigma \phi_<^{(r,r')}(\mathbf{x}) + (\mathbf{y}-\mathbf{x}) \nabla \phi_<^{(r,r')}(\mathbf{x}) $. The term with $\sigma=-$ is dominated by the zeroth order contribution which generates a term $\sim \cos 4\phi_<^{(r,r')}$ which is less relevant than $\cos 2 \phi_<^{(r,r')}$ and therefore ignored throughout the calculation. On the other hand, for $\sigma=+$, the constant contribution vanishes and we can expand the cosine in derivatives. This is slightly more complicated than one might assume at first glance because of the strong fluctuations of the field $\phi$ which makes the direct expansion in derivatives ill-defined. This can be cured using a normal ordering strategy~\cite{Giamarchi_2003},
\begin{align}
    \cos 2 \left( \phi_<^{(r,r')}(\mathbf{y}) - \phi_<^{(r,r')}(\mathbf{x}) \right) \notag  =& \frac{\cos 2 \left( \phi_<^{(r,r')}(\mathbf{y}) - \phi_<^{(r,r')}(\mathbf{x}) \right)}{\left\langle \cos 2 \left( \phi_<^{(r,r')}(\mathbf{y}) - \phi_<^{(r,r')}(\mathbf{x}) \right) \right\rangle_0} \left\langle \cos 2 \left( \phi_<^{(r,r')}(\mathbf{y}) - \phi_<^{(r,r')}(\mathbf{x}) \right) \right\rangle_0 \notag \\
    =& \frac{\cos 2 \left( \phi_<^{(r,r')}(\mathbf{y}) - \phi_<^{(r,r')}(\mathbf{x}) \right)}{\left\langle \cos 2 \left( \phi_<^{(r,r')}(\mathbf{y}) - \phi_<^{(r,r')}(\mathbf{x}) \right) \right\rangle_0} e^{-2 \left\langle \left( \phi_<^{(r,r')}(\mathbf{y}) - \phi_<^{(r,r')}(\mathbf{x}) \right)^2 \right\rangle_0} \notag \\ =&\left(  1 - 2\left( (\mathbf{y}-\mathbf{x})\nabla \phi_<^{(r,r')}(\mathbf{x}) \right)^2 + \dots \right) e^{-4\left\langle \phi^{(r,r')}_<(\mathbf{x})\left(\phi^{(r,r')}_<(\mathbf{x})-\phi^{(r,r')}_<(\mathbf{y})\right) \right\rangle_0} \notag \\ &= \left(  1 - 2\left( (\mathbf{y}-\mathbf{x})\nabla \phi_<^{(r,r')}(\mathbf{x}) \right)^2 + \dots \right) e^{-4(G_<(\mathbf{0})-G_<(\mathbf{y}-\mathbf{x}))} 
\end{align}
This means that after we absorbed the divergent fluctuations using the Gaussian expectation value, we can safely expand~\cite{Giamarchi_2003} in derivatives, which cures the integral appearing later. The constant term can be absorbed into the definition of the path integral while the derivative term generates an additional contribution to the Gaussian part of the action and renormalizes $K$. We find 
\begin{equation}
    -\frac{1}{2} \left( \langle \left(\Delta S\right)^2 \rangle_{0,>} - \left( \langle \Delta S \rangle_{0,>}  \right)^2 \right) = \\ -4 \lambda^2 \sum_{r \neq r'} \int_{\mathbf{x},\mathbf{x'}} G_>(\mathbf{x}') e^{-4F_<(\mathbf{x}')} \left( \mathbf{x'} \nabla \phi_<^{(r,r')}(\mathbf{x}) \right)^2. 
\end{equation}
We used the translation-invariance to decouple the two integrals and used the functions
\begin{align}
    G_>(x,t) = \frac{s}{K} \cos \Lambda x e^{-\Lambda \vert t \vert},
\end{align}
and
\begin{align}
    F_<(x,t) &= G_>(0,0)-G_>(x,t) = \frac{2\pi}{K} \int_< \frac{ 1-e^{i(px+\omega t)}}{p^2+\omega^2} \notag \\ &= \frac{1}{\pi K}  \int_0^\Lambda dp \int_{-\infty}^\infty d\omega \frac{1- \cos px e^{i\omega t}}{p^2+ \omega^2} \notag \\ &= \frac{1}{K} \int_0^\Lambda \frac{dp}{p} \left( 1-\cos px e^{- p \vert t \vert} \right).
\end{align}
Since both $G_>$ and $F_<$ are symmetric under inversion $t \to -t$ or $x \to -x$, we only need to compute the integral for the $\partial_x^2$ and the $\partial_t^2$ term separately. Let us start with the spatial term,
\begin{align}
    &\int_{-\infty}^\infty dx \int_{-\infty}^\infty dt x^2 G_>(x,t) e^{-4F_<(x,t)} \notag \\ = & \frac{s}{K} \int_{-\infty}^\infty dx \int_{-\infty}^\infty dt x^2  \cos \Lambda x e^{-\Lambda \vert t \vert} e^{-\frac{4}{K} \int_0^\Lambda \frac{dp}{p} \left( 1-\cos px e^{- p \vert t \vert} \right)} \notag \\ =& \frac{s}{K \Lambda^4} \int_{-\infty}^\infty dx \int_{-\infty}^\infty dt x^2 \cos x e^{- \vert t \vert} e^{-\frac{4}{K} \int_0^1 \frac{dp}{p}(1-\cos px) e^{- p \vert t \vert}}.
\end{align}
In general, this is a function of $K$. As we are mainly interested in the vicinity of the BKT point at $K=1$, we plug this into the integrals to get the leading contribution. The remaining integral can then be numerically evaluated, and we can do the same for the temporal integral, which yields
\begin{equation}
    \int_{-\infty}^\infty dx \int_{-\infty}^\infty dt x^2 G_>(x,t) e^{-4F_<(x,t)} \simeq \frac{Is}{\Lambda^4  K},
\end{equation}
where 
\begin{equation}
    I = 4 \int_0^\infty dx \int_0^\infty dt x^2 \cos x e^{- t} e^{-4 \int_0^1 \frac{dp}{p}(1-\cos px) e^{- p t}} \approx 0.07805.
\end{equation}
replacing $x^2 \to t^2$ yields the same numerical value up to the fifth digit. We conclude Eq.~\eqref{eq:secondOrderPert} from the main text.